\documentclass[a4paper,11pt]{article}
\pdfoutput=1
\usepackage{jheppub}
\usepackage[T1]{fontenc}
\usepackage{lmodern}
\usepackage{mathtools,booktabs,tabularx,multirow,etoolbox}
\usepackage{tikz}
\usetikzlibrary{decorations.pathmorphing,decorations.markings,decorations.pathreplacing}
\newcommand{\Sop}{\mathbb{S}}
\newcommand{\Cop}{\mathbb{C}}
\numberwithin{equation}{section}
\usepackage{microtype,flafter,placeins,needspace}

\allowdisplaybreaks[1]
\renewcommand{\afterTocRuleSpace}{\clearpage}
\let\papertableofcontents\tableofcontents
\renewcommand{\tableofcontents}{\begingroup\linespread{1}\selectfont\papertableofcontents\endgroup}
\makeatletter
\patchcmd{\l@section}{1.0em \@plus\p@}{0.25em \@plus\p@}{}{\PackageError{manuscript}{Cannot set contents spacing}{}}
\makeatother
\title{\boldmath NNLO QCD corrections to direct \(J/\psi\) production with open charm at B factories}
\hypersetup{bookmarksdepth=3,pdftitle={NNLO QCD corrections to direct J/psi production with open charm at B factories},pdfauthor={Xiangyu Chang, Chao Zeng, Bin Gong, Jian-Xiong Wang, Cong Li}}
\author[a,b,\dagger]{Xiangyu Chang,}
\author[a,b,\dagger]{Chao Zeng,\note[$\dagger$]{These authors contributed equally to this work and share first authorship.}}
\author[a,b,*]{Bin Gong,}
\author[a,b,*]{Jian-Xiong Wang,\note[*]{Corresponding authors.}}
\author[c]{and Cong Li}
\affiliation[a]{Institute of High Energy Physics, Chinese Academy of Sciences, Beijing 100049, China}
\affiliation[b]{University of Chinese Academy of Sciences, Beijing 100049, China}
\affiliation[c]{School of Physical Science and Technology, Southwest University, Chongqing 400700, China}
\emailAdd{changxy@ihep.ac.cn}
\emailAdd{zengchao@ihep.ac.cn}
\emailAdd{twain@ihep.ac.cn}
\emailAdd{jxwang@ihep.ac.cn}
\emailAdd{lc312321@163.com}
\abstract{We calculate the next-to-next-to-leading-order (NNLO) QCD corrections to direct colour-singlet \(J/\psi\) production with open charm at B factories, using nested soft-collinear subtraction for massive partons.
At \(\sqrt{s}=10.6\) GeV, \(m_c=1.5\) GeV and \(\mu_R=2m_c\), we obtain an approximate NNLO cross section of \(0.324^{+0.041}_{-0.066}\) pb, with scale uncertainties from \(m_c\leq\mu_R\leq4m_c\).
The total cross section is \(18.4\%\) above NLO and exhibits good perturbative convergence through NNLO, with reduced scale dependence.
An NLO-based mass extrapolation to \(m_c=1.4\) GeV, combined with estimated \(\psi(2S)\) feed-down, raises the rate to about \(0.59\) pb.
The NNLO correction brings the absolute momentum spectrum near its peak closer to Belle's prompt open-charm data, but shifts the angular shape further away from Belle's prompt charm-associated data.
The difference in angular shape motivates more precise measurements at Belle~II.}
\begin{document}
\maketitle
\raggedbottom
\Needspace{6\baselineskip}
\section{Introduction}\label{introduction}

Associated production of \(J/\psi\) and an open charm pair in electron-positron annihilation provides a test of heavy-quarkonium production mechanisms.
In nonrelativistic QCD (NRQCD), the production cross section is factorised into perturbative short-distance coefficients and non-perturbative long-distance matrix elements describing quarkonium formation \cite{ref1}.
Early colour-singlet calculations established the contributions from associated open charm and two-gluon production~\cite{cho_leibovich,yuan_qiao_chao}, with subsequent studies examining the charm-associated rate and differential distributions~\cite{liu_he_chao,hagiwara_inclusive}.

Measurements of prompt charmonium production by BaBar and Belle~\cite{babar_prompt,belle_prompt}, followed by Belle's observation of a large double-charm component~\cite{belle_double_charm}, exposed discrepancies with leading-order (LO) predictions.
Zhang and Chao calculated the next-to-leading-order (NLO) QCD correction to the total rate \cite{ref2}; Gong and Wang performed an independent calculation and studied differential distributions \cite{ref3}.
For commonly used inputs, these calculations found an enhancement by a factor of approximately 1.7--1.8 over LO \cite{ref2,ref3}.
Belle reported a charm-associated prompt cross section of \(0.74\pm0.08\,{}^{+0.09}_{-0.08}\) pb, including open-charm and double-charmonium final states, together with momentum and angular distributions \cite{ref4}.
The NLO production-angle distribution, however, differed from the shape measured by Belle~\cite{ref3}.
Further studies have examined relativistic corrections~\cite{he_double_charm_v2} and renormalisation-scale setting using the principle of maximum conformality~\cite{ref5}.

The complementary \(J/\psi+gg\) channel is important for interpreting Belle's sample without an associated charm pair~\cite{ref4}.
Its NLO QCD corrections were calculated independently in refs.~\cite{ma_gg_nlo,gong_gg_nlo}, and colour-singlet relativistic corrections were obtained in refs.~\cite{he_gg_v2,jia_gg_v2}.
Further developments include endpoint resummation~\cite{leibovich_endpoint}, NLO colour-octet contributions~\cite{zhang_octet_nlo}, initial-state radiation~\cite{shao_isr}, and the recent mixed \(O(\alpha_s v^2)\) correction to the colour-singlet rate and angular distribution~\cite{li_gg_asv2}.

NLO QCD predictions for these inclusive production channels have been available for more than a decade~\cite{ref2,ref3,ma_gg_nlo,gong_gg_nlo}.
More recently, several exclusive processes have been studied at next-to-next-to-leading order (NNLO).
For exclusive \(e^+e^-\to J/\psi+\eta_c\), Feng et al.~\cite{feng_jia_nnlo} first obtained NNLO predictions, followed by Huang et al.~\cite{ref15}; analytic two-loop amplitudes are now available as asymptotic expansions in \(m_c^2/s\)~\cite{chen_analytic_nnlo}.
NNLO corrections to exclusive double-\(J/\psi\) production have been studied in refs.~\cite{ref17,ref16}.

Large higher-order corrections in exclusive quarkonium processes have also motivated alternative perturbative prescriptions.
An amplitude-level prescription retains the complete modulus square of the truncated amplitude and has been proposed and studied for \(J/\psi\to3\gamma\) and exclusive charmonium production~\cite{jpsi_3gamma_nnlo,exclusive_reanalysis,amplitude_positivity}.
Zhan et al. have applied this prescription at NLO to \(\gamma\gamma\to J/\psi+\gamma\) in ultraperipheral collisions~\cite{zhan_upc_amplitude}.
Separately, Maxia et al. have proposed a soft-improved matching scheme that includes finite soft contributions in the NNLO renormalisation of colour-singlet S-wave NRQCD matrix elements~\cite{soft_improved_matching}.

In the inclusive case, NNLO results are available for the hadronic decays of \(\eta_c\) and \(\eta_b\), whose leading partonic channel is \(\eta_Q\to gg\)~\cite{eta_hadronic_nnlo}.
For \(J/\psi\) production with an open charm pair, the sizeable NLO correction makes it important to test the perturbative stability of both the total rate and differential distributions at NNLO.
This process already has a three-body Born final state containing massive coloured particles.
Besides the two-loop virtual terms, the NNLO correction involves two gluons or a massless quark pair in the double-real channels, and the one-loop correction to \(J/\psi+c\bar c+g\) in the real-virtual contribution.
Their infrared singularities must be cancelled while retaining a consistent definition of the measured observables across the different final-state multiplicities.

General subtraction methods organise this cancellation locally in phase space.
At NLO, the FKS~\cite{fks} and dipole~\cite{catani_seymour} methods provide standard constructions, with extensions to massive partons~\cite{massive_dipoles} and NRQCD production~\cite{nrqcd_dipoles}.
NNLO approaches include antenna subtraction~\cite{antenna}, sector-improved residue subtraction~\cite{stripper}, CoLoRFulNNLO~\cite{colorful} and local analytic sector subtraction~\cite{local_sector}, as well as nested soft-collinear subtraction~\cite{ref20}.
Fully differential massive-pair production from a colourless initial state has been treated within the latter scheme~\cite{ref6} and CoLoRFulNNLO~\cite{massive_colorful}.

In this work we present the first NNLO QCD calculation of direct colour-singlet \(J/\psi\) production with an open charm pair in electron-positron annihilation through a virtual photon.
The real-emission contributions are treated with the massive-parton formulation of nested soft-collinear subtraction~\cite{ref6}.
The heavy-quark mass removes collinear singularities involving the open charm quarks, but overlapping soft and light-parton collinear limits remain in the double-real channels.
Soft limits are subtracted before the remaining collinear limits, and the integrated counterterms are combined with the virtual contributions on the corresponding reduced phase spaces.
For the double-soft terms, we use the analytic integrals for back-to-back massive emitters of ref.~\cite{ref7}; their extension to arbitrary relative angles has recently been studied in ref.~\cite{arbitrary_double_soft}.

We combine the real and virtual corrections into finite phase-space integrals, with analytic cancellation of infrared poles on the single-real and Born phase spaces.
The subtraction mappings keep the measured quarkonium momentum fixed, allowing us to retain the differential information.
We present approximate predictions at \(\sqrt{s}=10.6\ {\rm GeV}\) for the total cross section and the production-angle and momentum distributions, including their renormalisation-scale dependence.

Section~\ref{theoretical-framework} defines the process, observables and renormalisation conventions.
Section~\ref{combining-real-and-virtual-corrections} explains how the NLO and NNLO real and virtual corrections are separated and recombined into finite contributions.
The numerical implementation and the phenomenological results are discussed separately in sections~\ref{numerical-setup-and-checks} and~\ref{phenomenological-results}, followed by the conclusions in section~\ref{conclusions}.
Appendix~\ref{angular-reconstruction-derivation} derives the angular reconstruction, and appendix~\ref{finite-contributions-and-poles} gives the integrated kernels and pole cancellations.
The local counterterms and their integrals, phase-space mappings and auxiliary analytic functions are given in appendices~\ref{real-emission-subtraction}, \ref{common-parametrisations-of-the-real-and-subtraction-terms} and~\ref{integrated-kernels-and-one-loop-soft-currents}, respectively.

\FloatBarrier

\Needspace{12\baselineskip}
\section{Theoretical framework}\label{theoretical-framework}

\Needspace{5\baselineskip}
\subsection{Process definition and perturbative expansion}\label{process-and-expansion}

The process studied here is direct \(J/\psi\) production with open charm in electron-positron annihilation through a virtual photon,
\begin{equation}
\begin{gathered}
e^-(p_-)+e^+(p_+)\longrightarrow\gamma^*(Q)
\longrightarrow J/\psi(P)+c(p_1)+\bar c(p_2)+X_{\rm light},\\
Q=p_-+p_+,\qquad Q^2=s.
\end{gathered}
\label{eq:2.1}
\end{equation}
The quarkonium is produced directly in the colour-singlet \(c\bar c[{}^3S_1^{[1]}]\) state.
At leading order in the heavy-quark relative velocity \(v\), the charm quark and antiquark forming the \(J/\psi\) each carry momentum \(P/2\), with \(P^2=M_\psi^2=4m_c^2\).
The momenta \(p_1\) and \(p_2\) denote the open charm quark and antiquark, respectively.
The incoming beams are unpolarised and the electron mass is set to zero.
The light-parton final states are \(X_{\rm light}=\varnothing,g,gg,q\bar q\), with \(q=u,d,s\).

NRQCD factorisation gives~\cite{ref1}
\begin{equation}
d\sigma
 =d\widehat\sigma\bigl[e^+e^-\to c\bar c[{}^3S_1^{[1]}]+c\bar c+X_{\rm light}\bigr]
   \langle\mathcal O_1\rangle_{J/\psi}(\mu_\Lambda).
\label{eq:nrqcd-factorisation}
\end{equation}
Here \(d\widehat\sigma\) is the perturbative short-distance coefficient, and \(\langle\mathcal O_1\rangle_{J/\psi}\) is the non-perturbative colour-singlet matrix element in the single-polarisation convention, related to the wave function by \(\lvert R(0)\rvert^2=(2\pi/3)\langle\mathcal O_1\rangle_{J/\psi}\).

We denote the contribution of order \(\alpha^2\alpha_s^{2+n}\) to the physical cross section \(\sigma\), including this matrix element, by \(\sigma^{(n)}\), with \(n=0,1,2\).
The fixed-order cross sections are \[
\begin{aligned}
\sigma_{\rm LO}&=\sigma^{(0)},\\
\sigma_{\rm NLO}&=\sigma^{(0)}+\sigma^{(1)},\\
\sigma_{\rm NNLO}&=\sigma^{(0)}+\sigma^{(1)}+\sigma^{(2)}.
\end{aligned}
\]

\Needspace{5\baselineskip}
\subsection{Angular reconstruction from the virtual-photon tensor}\label{total-rate-production-angle-and-energy-bins}

We obtain the total cross sections \(\sigma\) and \(\sigma_L\) with the respective photon-polarisation tensors
\begin{equation}
\begin{gathered}
\Pi^{\mu\nu}=-g^{\mu\nu}+\frac{Q^\mu Q^\nu}{s},
\qquad \Pi_L^{\mu\nu}=e_L^\mu e_L^\nu,\\
e_L^\mu=\frac{P^\mu-(P\cdot Q)Q^\mu/s}
 {\sqrt{(P\cdot Q)^2/s-M_\psi^2}}.
\end{gathered}
\label{eq:photon-projectors}
\end{equation}
Let \(\theta\) be the angle between the \(J/\psi\) and incoming-electron momenta in the centre-of-mass frame.
The angular distribution is then
\begin{equation}
\begin{aligned}
\frac{d\sigma}{d\cos\theta}
 &=\frac38\left[(\sigma+\sigma_L)
       +(\sigma-3\sigma_L)\cos^2\theta\right],\\
\alpha_\theta&=\frac{\sigma-3\sigma_L}{\sigma+\sigma_L}.
\end{aligned}
\label{eq:2.4}
\end{equation}
The derivation is given in appendix~\ref{angular-reconstruction-derivation}.

\Needspace{5\baselineskip}
\subsection{Renormalisation conventions}\label{renormalisation-conventions}

We use dimensional regularisation with \(d=4-2\epsilon\).
The charm-quark mass and field are renormalised in the on-shell scheme, and the strong coupling in the \(\overline{\rm MS}\) scheme with four active flavours.
The three light flavours \(u,d,s\) are treated as massless.

At relative order \(\alpha_s^2\), the infrared pole associated with the colour-singlet \(c\bar c[{}^3S_1^{[1]}]\) pair is absorbed into the \(\overline{\rm MS}\)-renormalised NRQCD matrix element~\cite{nrqcd_vector_matching_cm,nrqcd_vector_matching,nrqcd_current_matching}.
Subtracting the term in \eqref{eq:nrqcd-counterterm} from the ultraviolet-renormalised QCD amplitude implements this matching and introduces the factorisation scale \(\mu_\Lambda\).
We fix \(\mu_\Lambda=m_c\), independently of the QCD renormalisation scale \(\mu_R\).
The short-distance amplitudes used below include this NRQCD matching; the remaining QCD infrared poles cancel in the cross section after adding the real-emission contributions.

\FloatBarrier

\Needspace{6\baselineskip}
\section{Organisation of the NLO and NNLO calculation}\label{combining-real-and-virtual-corrections}

\begin{figure}[!htbp]
\centering
\begingroup
\tikzset{
  fermion/.style={draw=.,line width=.45pt,postaction={decorate},
    decoration={markings,mark=at position .5 with {\arrow[scale=.7]{stealth}}}},
  outgoing fermion/.style={draw=.,line width=.45pt,postaction={decorate},
    decoration={markings,mark=at position .76 with {\arrow[scale=.7]{stealth}}}},
  gluon/.style={draw=.,line width=.45pt,decorate,
    decoration={coil,amplitude=1.6pt,segment length=3.8pt,aspect=1}},
  photon/.style={draw=.,line width=.45pt,decorate,
    decoration={snake,amplitude=1.6pt,segment length=6.5pt}},
}
\newcommand{\diagramframe}{%
  \path[use as bounding box] (-2.45,-2.25) rectangle (3.65,1.95);
  \coordinate (photon) at (0,0);
  \coordinate (split) at (1.18,.3);
  \coordinate (charm) at (2.85,-.95);
  \coordinate (anticharm) at (2.7,-1.65);
  \path (photon) -- coordinate[pos=.15] (leftone)
    coordinate[pos=.43] (lefttwo) (anticharm);
  \path (split) -- coordinate[pos=.12] (rightone)
    coordinate[pos=.18] (emitone) coordinate[pos=.32] (emitpair)
    coordinate[pos=.46] (righttwo) (charm);
  \draw[fermion] (-2.05,1.0) -- (-1.15,0);
  \draw[fermion] (-1.15,0) -- (-2.05,-1.0);
  \node[above left,inner sep=2pt] at (-2.05,1.0) {\(e^-\)};
  \node[below left,inner sep=2pt] at (-2.05,-1.0) {\(e^+\)};
  \draw[photon] (-1.15,0) -- (photon);
  \node[above,inner sep=3pt] at (-.575,0) {\(\gamma^*\)};
  \draw[fermion] (anticharm) -- (photon);
  \draw[outgoing fermion] (photon) -- (2.5,1.5);
  \draw[fermion] (2.68,1.2) -- (split);
  \draw[outgoing fermion] (split) -- (charm);
  \draw[gluon] (.55,.33) -- (split);
  \draw[draw=.,fill=gray!25,line width=.45pt]
    (2.59,1.35) ellipse [x radius=.175cm,y radius=.075cm,rotate=120.96];
  \node[above right,inner sep=3pt] at (2.59,1.5) {\(J/\psi\)};
  \node[right,inner sep=3pt] at (anticharm) {\(\bar c\)};
  \node[right,inner sep=3pt] at (charm) {\(c\)};
}
\begin{tabular}{@{}c@{\hspace{.4cm}}c@{}}
\begin{tikzpicture}[font=\small]
  \diagramframe
  \draw[gluon] (leftone) -- (rightone);
  \draw[gluon] (lefttwo) -- (righttwo);
  \node at (.6,-2.0) {(a) VV};
\end{tikzpicture}
&
\begin{tikzpicture}[font=\small]
  \diagramframe
  \draw[gluon] (lefttwo) -- (righttwo);
  \draw[gluon] (emitone) -- (3.25,.4);
  \node[right,inner sep=3pt] at (3.25,.4) {\(g\)};
  \node at (.6,-2.0) {(b) RV};
\end{tikzpicture}
\\[2pt]
\begin{tikzpicture}[font=\small]
  \diagramframe
  \draw[gluon] (emitone) -- (3.25,.5);
  \draw[gluon] (righttwo) -- (3.25,-.1);
  \node[right,inner sep=3pt] at (3.25,.5) {\(g\)};
  \node[right,inner sep=3pt] at (3.25,-.1) {\(g\)};
  \node at (.6,-2.0) {(c) RR \((gg)\)};
\end{tikzpicture}
&
\begin{tikzpicture}[font=\small]
  \diagramframe
  \draw[gluon] (emitpair) -- (2.6,.15);
  \draw[fermion] (2.6,.15) -- (3.25,.55);
  \draw[fermion] (3.25,-.25) -- (2.6,.15);
  \node[right,inner sep=3pt] at (3.25,.55) {\(q\)};
  \node[right,inner sep=3pt] at (3.25,-.25) {\(\bar q\)};
  \node at (.6,-2.0) {(d) RR \((q\bar q)\)};
\end{tikzpicture}
\end{tabular}
\endgroup
\caption{Representative NNLO diagrams for \(e^+e^-\to J/\psi+c\bar c+X\): (a) two-loop, (b) one-loop real-emission, and (c,d) double-real tree diagrams.
The shaded oval denotes the \(c\bar c[{}^3S_1^{[1]}]\) projection; \(q=u,d,s\).
VV also includes the squared one-loop amplitude.}\label{fig:representative-diagrams}
\end{figure}
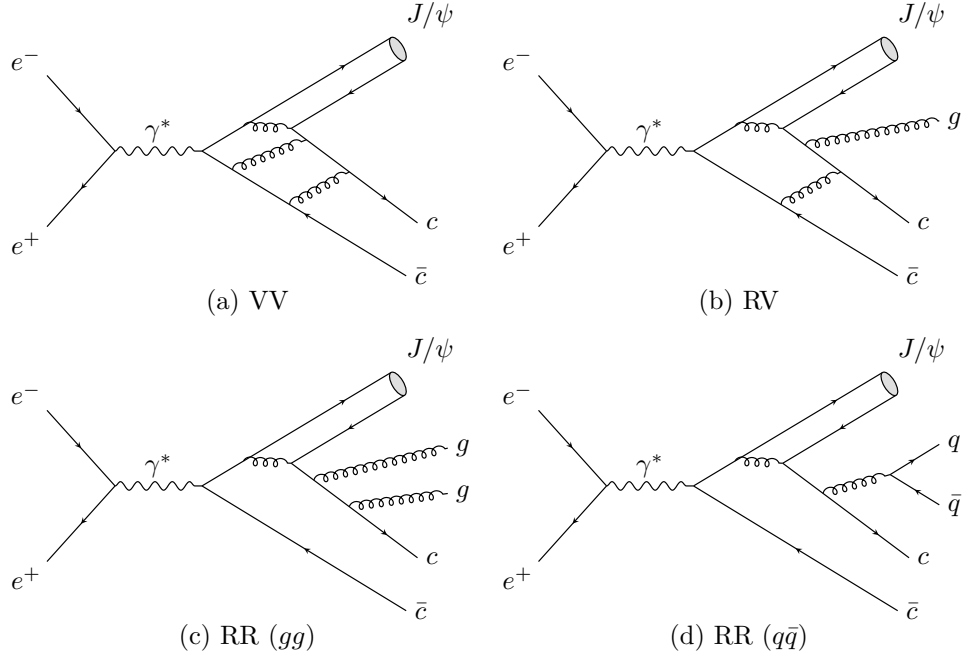

Let \(\mathcal M_{c\bar c+X}^{(\ell)}\) be the \(\ell\)-loop short-distance amplitude coefficient for \(\gamma^*\to c\bar c[{}^3S_1^{[1]}]+c\bar c+X\), and \(d\Phi_{c\bar c+X}\) the phase space for the corresponding \(J/\psi+c\bar c+X\) final state.
The common \(J/\psi\) is understood in these subscripts.
With \(a_s=\alpha_s/(4\pi)\), the renormalised amplitude is \(\mathcal M=\mathcal M^{(0)}+a_s\mathcal M^{(1)}+a_s^2\mathcal M^{(2)}+\cdots\).
We define the perturbative contributions to the squared matrix element by
\begin{equation}
\begin{aligned}
F_{c\bar c+X}^{(0)}&=|\mathcal M_{c\bar c+X}^{(0)}|^2,\\
F_{c\bar c+X}^{(1)}&=2a_s\operatorname{Re}
 [\mathcal M_{c\bar c+X}^{(0)*}\mathcal M_{c\bar c+X}^{(1)}],\\
F_{c\bar c+X}^{(2)}&=a_s^2\left(2\operatorname{Re}
 [\mathcal M_{c\bar c+X}^{(0)*}\mathcal M_{c\bar c+X}^{(2)}]
 +|\mathcal M_{c\bar c+X}^{(1)}|^2\right).
\end{aligned}
\label{eq:partonic-matrix-elements}
\end{equation}
Final-state spin and colour sums and the photon projection in \eqref{eq:photon-projectors} are implicit.
The phase-space integrals of these short-distance terms give the physical cross sections after multiplication by
\[
\mathcal N=\frac{\alpha\lvert R(0)\rvert^2}{6m_cs^2}.
\]
This factor includes the initial-state flux and the NRQCD matrix element through \(\lvert R(0)\rvert^2\); the couplings of the hadronic current remain in the amplitudes.
The Born term and the dimensionally regulated higher-order contributions are
\begin{equation}
\begin{aligned}
\sigma^{(0)}&=\underbrace{\mathcal N\int\!\bigl[d\Phi_{c\bar c}F_{c\bar c}^{(0)}\bigr]_{\epsilon^0}}_{B},\\[3pt]
\sigma^{(1)}&=\underbrace{\mathcal N\int d\Phi_{c\bar c+g}F_{c\bar c+g}^{(0)}}_{R}
 +\underbrace{\mathcal N\int d\Phi_{c\bar c}F_{c\bar c}^{(1)}}_{V},\\[3pt]
\sigma^{(2)}&=\underbrace{\mathcal N\sum_{ab}\int d\Phi_{c\bar c+ab}F_{c\bar c+ab}^{(0)}}_{\mathrm{RR}}
 +\underbrace{\mathcal N\int d\Phi_{c\bar c+g}F_{c\bar c+g}^{(1)}}_{\mathrm{RV}}
 +\underbrace{\mathcal N\int d\Phi_{c\bar c}F_{c\bar c}^{(2)}}_{\mathrm{VV}}.
\end{aligned}
\label{eq:unsubtracted-orders}
\end{equation}
Here \(ab=gg,u\bar u,d\bar d,s\bar s\), and the \(gg\) contribution includes the identical-particle factor \(1/2!\).
Representative NNLO diagrams are shown in figure~\ref{fig:representative-diagrams}.

We use nested soft-collinear subtraction for massive partons~\cite{ref6} to turn these sums into separately finite integrals.
Local soft and collinear counterterms are subtracted from the real-emission integrands and integrated analytically over the unresolved variables.
Their integrals cancel the explicit infrared poles of the virtual corrections at fixed lower-multiplicity kinematics.
In each finite contribution below, \([\cdot]_{\epsilon^0}\) extracts the \(\epsilon^0\) coefficient of the complete integrand, including its phase space, after pole cancellation and local subtraction.
The superscript \({\rm fin}\) on \(F^{(1,2)}\) denotes the expressions in \eqref{eq:partonic-matrix-elements} with the loop amplitudes replaced by their infrared-finite remainders, defined in appendix~\ref{virtual-amplitudes-and-infrared-matching}.

\Needspace{5\baselineskip}
\subsection{NLO: real--virtual matching}\label{nlo-contributions}

At NLO, soft subtraction and infrared factorisation split \(R\) and \(V\) into finite remainders, labelled by the subscript \(\mathrm F\), and terms whose poles cancel in their sum.
The terms in the boxed column sum to the finite contribution \(U\):
\begin{equation}
\begin{array}{@{}c@{\;}c@{\;}c@{\;}c@{\;}c@{}}
\begin{array}{@{}r@{}}R\\[3pt]V\end{array}
&\begin{array}{@{}c@{}}\longrightarrow\\[3pt]\longrightarrow\end{array}
&\begin{array}{@{}l@{}}R_{\mathrm F}\\[3pt]V_{\mathrm F}\end{array}
&\begin{array}{@{}c@{}}+\\[3pt]+\end{array}
&\boxed{\begin{array}{c}R_U\\[3pt]V_U\end{array}}\\[-1pt]
&&&&\downarrow\\[-2pt]
&&&&U
\end{array}
\label{eq:nlo-separation}
\end{equation}

The real-emission contribution \(R\) has only a soft-gluon singularity; \(k\) denotes the emitted gluon momentum.
The operator \(\Sop_k\) extracts the leading soft behaviour of the squared matrix element and phase space at fixed quarkonium momentum \(P\).
Subtracting this limit gives
\begin{equation}
R_{\mathrm F}=\mathcal N\int\!\left[(1-\Sop_k)
 \left\{d\Phi_{c\bar c+g}F_{c\bar c+g}^{(0)}\right\}\right]_{\epsilon^0}.
\label{eq:nlo-local-definition}
\end{equation}
The soft projection removes \(k\) from the recoil constraint, leaving Born kinematics for the heavy pair.
Integrating over the unresolved gluon at fixed Born momenta defines the integrated counterterm \(F_{R,U}\):
\begin{equation}
d\Phi_{c\bar c}F_{R,U}
 =\int_{\rm unres}\!\Sop_k
 \left\{d\Phi_{c\bar c+g}F_{c\bar c+g}^{(0)}\right\}.
\label{eq:nlo-integrated-definition}
\end{equation}
Here and below, \(\int_{\rm unres}\) leaves the reduced phase-space variables unintegrated.
The unresolved phase space and the explicit result for \(F_{R,U}\) are given in appendices~\ref{appendix-subtraction-conventions} and~\ref{jpsicbar-cg-at-tree-level}.

The virtual interference separates into its finite remainder and infrared term,
\begin{equation}
F_{c\bar c}^{(1)}=F_{c\bar c}^{(1),{\rm fin}}
 +F_{V,U}.
\label{eq:nlo-virtual-definition}
\end{equation}
The regulated Born-phase-space integrals of \(F_{R,U}\) and \(F_{V,U}\), multiplied by \(\mathcal N\), define \(R_U\) and \(V_U\) in \eqref{eq:nlo-separation}.
Their poles cancel at fixed Born kinematics, giving
\begin{equation}
\begin{aligned}
U&=\mathcal N\int\!\left[d\Phi_{c\bar c}
 \bigl(F_{R,U}+F_{V,U}\bigr)\right]_{\epsilon^0},\\
V_{\mathrm F}&=\mathcal N\int\!\left[d\Phi_{c\bar c}F_{c\bar c}^{(1),{\rm fin}}\right]_{\epsilon^0}.
\end{aligned}
\label{eq:4.2}
\end{equation}
The explicit infrared terms and their cancellation are given in appendix~\ref{nlo-fr-pole-cancellation}.
Thus the NLO correction is
\begin{equation}
\sigma^{(1)}
 =\underbrace{R_{\mathrm F}}_{c\bar c+g}
  +\underbrace{V_{\mathrm F}+U}_{c\bar c}.
\label{eq:nlo-finite-overview}
\end{equation}
The braces label the remaining phase-space integrations.

\Needspace{5\baselineskip}
\subsection{NNLO: double-real, real-virtual and double-virtual matching}\label{nnlo-contributions}

At NNLO, the same separation into finite remainders and terms to be combined gives
\begin{equation}
\begin{array}{@{}c@{\;}c@{\;}c@{\;}c@{\;}c@{\;}c@{\;}c@{\;}c@{\;}c@{}}
\begin{array}{@{}r@{}}\mathrm{RR}\\[3pt]\mathrm{RV}\\[3pt]\mathrm{VV}\end{array}
&\begin{array}{@{}c@{}}\longrightarrow\\[3pt]\longrightarrow\\[3pt]\longrightarrow\end{array}
&\begin{array}{@{}l@{}}\mathrm{RR}_{\mathrm F}\\[3pt]\mathrm{RV}_{\mathrm F}\\[3pt]\mathrm{VV}_{\mathrm F}\end{array}
&\begin{array}{@{}c@{}}+\\[3pt]+\\[3pt]\phantom{+}\end{array}
&\boxed{\begin{array}{c}\mathrm{RR}_{\mathrm{SU}}\\[3pt]\mathrm{RV}_{\mathrm{SU}}\\[3pt]\phantom{\mathrm{VV}_{\mathrm{SU}}}\end{array}}
&\begin{array}{@{}c@{}}\phantom{+}\\[3pt]+\\[3pt]+\end{array}
&\boxed{\begin{array}{c}\phantom{\mathrm{RR}_{\mathrm{FR}}}\\[3pt]\mathrm{RV}_{\mathrm{FR}}\\[3pt]\mathrm{VV}_{\mathrm{FR}}\end{array}}
&\begin{array}{@{}c@{}}+\\[3pt]+\\[3pt]+\end{array}
&\boxed{\begin{array}{c}\mathrm{RR}_{\mathrm{DU}}\\[3pt]\mathrm{RV}_{\mathrm{DU}}\\[3pt]\mathrm{VV}_{\mathrm{DU}}\end{array}}\\[-1pt]
&&&&\downarrow&&\downarrow&&\downarrow\\[-2pt]
&&&&\mathrm{SU}&&\mathrm{FR}&&\mathrm{DU}
\end{array}
\label{eq:nnlo-separation}
\end{equation}
The \(\mathrm F\) terms are the locally subtracted real contributions and the finite double-virtual contribution.
Each boxed column sums to the finite contribution below it: the single-unresolved contribution \(\mathrm{SU}\), the terms proportional to the finite one-loop interference \(\mathrm{FR}\), and the Born-proportional double-unresolved contribution \(\mathrm{DU}\).

\paragraph{Locally subtracted real contributions.}
For the two light-parton momenta \(k_3,k_4\), \(\Sop_{34}\) extracts the double-soft limit and \(\Cop\equiv\Cop_{34}\) the pair-collinear limit.
In the \(gg\) channel we order the energies as \(E_4^*<E_3^*\) in the rest frame of \(K=Q-P\), so the remaining single-soft limit is \(\Sop_4\).
The finite double-real contributions are
\begin{equation}
\begin{aligned}
\mathrm{RR}_{\mathrm F}(q\bar q)
 &=\mathcal N\sum_{q=u,d,s}\int\!\Bigl[(1-\Cop)(1-\Sop_{34})
 \bigl\{d\Phi_{c\bar c+q\bar q}F_{c\bar c+q\bar q}^{(0)}\bigr\}\Bigr]_{\epsilon^0},\\[3pt]
\mathrm{RR}_{\mathrm F}(gg)
 &=\mathcal N\int\!\Bigl[(1-\Cop)(1-\Sop_4)(1-\Sop_{34})
 \bigl\{d\Phi_{c\bar c+gg}F_{c\bar c+gg}^{(0)}\bigr\}\Bigr]_{\epsilon^0}.
\end{aligned}
\label{eq:finite-rr}
\end{equation}
The operators act from right to left on the complete expressions in braces.
The \(gg\) integral includes both energy-ordered sectors.
The local limits, overlap terms and integration domains are specified in appendices~\ref{jpsicbar-cqbar-q-at-tree-level} and~\ref{jpsicbar-cgg-at-tree-level}.

For RV, infrared factorisation first gives
\begin{equation}
F_{c\bar c+g}^{(1)}=F_{c\bar c+g}^{(1),{\rm fin}}
 +F_{\mathrm{RV,SU}}.
\label{eq:rv-factorisation-overview}
\end{equation}
The finite remainder still becomes singular as the gluon momentum vanishes.
Subtracting its soft limit gives
\begin{equation}
\mathrm{RV}_{\mathrm F}=\mathcal N\int\!\left[(1-\Sop_k)
 \left\{d\Phi_{c\bar c+g}F_{c\bar c+g}^{(1),{\rm fin}}\right\}\right]_{\epsilon^0}.
\label{eq:finite-rv}
\end{equation}
The infrared term \(F_{\mathrm{RV,SU}}\) will combine with the singly integrated RR counterterms below.
Its explicit expression is given in appendix~\ref{virtual-amplitudes-and-infrared-matching}, and the soft counterterms of both RV terms in appendix~\ref{local-soft-subtraction}.

\paragraph{Single-unresolved contribution.}
The single-unresolved terms removed in \eqref{eq:finite-rr} are generated by \(\Cop(1-\Sop_{34})\) for \(q\bar q\) and \([\Cop+(1-\Cop)\Sop_4](1-\Sop_{34})\) for \(gg\).
Integrating the pieces without \(\Sop_{34}\) defines \(F_{\mathrm{RR,SU}}\) on the one-gluon phase space:
\begin{equation}
\begin{aligned}
d\Phi_{c\bar c+g}\,F_{\mathrm{RR,SU}}
 ={}&\sum_{q=u,d,s}\int_{\rm unres}\!\Cop
   \bigl\{d\Phi_{c\bar c+q\bar q}F_{c\bar c+q\bar q}^{(0)}\bigr\}\\
 &+\int_{\rm unres}\!\bigl[\Cop+(1-\Cop)\Sop_4\bigr]
   \bigl\{d\Phi_{c\bar c+gg}F_{c\bar c+gg}^{(0)}\bigr\}.
\end{aligned}
\label{eq:rr-single-integrated-definition}
\end{equation}
The remaining momentum \(k\) is the massless parent in the collinear terms and the harder gluon in the single-soft term.

The pieces containing \(-\Sop_{34}\) subtract the overlap with the double-soft limit.
After one unresolved integration, these overlaps supply \(-\Sop_k\{d\Phi_{c\bar c+g}F_{\mathrm{RR,SU}}\}\), with the remaining momentum left unintegrated.
The part of the collinear--double-soft overlap above the kinematic energy limit for the parent gluon in the \(c\bar c+g\) phase space is instead integrated fully into \(\mathrm{RR}_{\mathrm{DU}}\).
Appendix~\ref{single-unresolved-terms} gives this domain split and the explicit relation \eqref{eq:rr-single-soft-origin}, including the nested single-soft overlap in \(gg\).

Combining \(F_{\mathrm{RR,SU}}\) and the RV infrared term \(F_{\mathrm{RV,SU}}\) with their remaining soft subtractions gives
\begin{equation}
\mathrm{SU}=\mathcal N\int\!\left[(1-\Sop_k)\left\{d\Phi_{c\bar c+g}
 \bigl(F_{\mathrm{RR,SU}}+F_{\mathrm{RV,SU}}\bigr)
 \right\}\right]_{\epsilon^0}.
\label{eq:finite-su}
\end{equation}
The two terms, each with this soft subtraction, define \(\mathrm{RR}_{\mathrm{SU}}\) and \(\mathrm{RV}_{\mathrm{SU}}\) in \eqref{eq:nnlo-separation}.
Their poles cancel both before and after the soft projection; the explicit expressions and both identities are given in appendix~\ref{single-unresolved-terms}.

\paragraph{Born-phase-space contributions.}
The doubly integrated RR counterterms define \(F_{\mathrm{RR,DU}}\) through
\begin{equation}
d\Phi_{c\bar c}\,F_{\mathrm{RR,DU}}
 =\sum_{ab}\int_{\rm unres}\!(\Sop_{34}-\Cop\Sop_>)
 \bigl\{d\Phi_{c\bar c+ab}F_{c\bar c+ab}^{(0)}\bigr\}.
\label{eq:rr-double-integrated-definition}
\end{equation}
Both unresolved partons are integrated here.
The symbol \(\Cop\Sop_>\) denotes the high-parent-energy part of the collinear--double-soft overlap described above.
Together, the RR parts of \(\mathrm{SU}\) and \(\mathrm{DU}\) restore all counterterms subtracted in \eqref{eq:finite-rr}.

The soft counterterms of the RV finite remainder and infrared term sum to the soft limit of the full RV interference.
Soft factorisation separates its integral into a term proportional to \(F_{c\bar c}^{(1),{\rm fin}}\), denoted by \(F_{\mathrm{RV,FR}}\), and a term proportional to \(F_{c\bar c}^{(0)}\), denoted by \(F_{\mathrm{RV,DU}}\):
\begin{equation}
\int_{\rm unres}\!\Sop_k
 \bigl\{d\Phi_{c\bar c+g}F_{c\bar c+g}^{(1)}\bigr\}
 =d\Phi_{c\bar c}\bigl(F_{\mathrm{RV,FR}}+F_{\mathrm{RV,DU}}\bigr).
\label{eq:rv-integrated-definition}
\end{equation}
Their explicit expressions are given in appendix~\ref{integrated-soft-counterterm}.
Infrared factorisation gives the corresponding decomposition of the double-virtual contribution,
\begin{equation}
F_{c\bar c}^{(2)}=F_{c\bar c}^{(2),{\rm fin}}
 +F_{\mathrm{VV,FR}}+F_{\mathrm{VV,DU}},
\label{eq:vv-factorisation-overview}
\end{equation}
where the \(\mathrm{FR}\) and \(\mathrm{DU}\) terms are proportional to \(F_{c\bar c}^{(1),{\rm fin}}\) and \(F_{c\bar c}^{(0)}\), respectively, as specified in appendix~\ref{virtual-amplitudes-and-infrared-matching}.
Combining these terms gives
\begin{align}
\mathrm{FR}&=\mathcal N\int\!\left[d\Phi_{c\bar c}
 \bigl(F_{\mathrm{RV,FR}}+F_{\mathrm{VV,FR}}\bigr)\right]_{\epsilon^0},
\label{eq:finite-fr}\\[3pt]
\mathrm{DU}&=\mathcal N\int\!\left[d\Phi_{c\bar c}
 \bigl(F_{\mathrm{RR,DU}}+F_{\mathrm{RV,DU}}+F_{\mathrm{VV,DU}}\bigr)\right]_{\epsilon^0}.
\label{eq:4.12}
\end{align}
The separate regulated integrals of these \(F_{X,Y}\) terms over the Born phase space, multiplied by \(\mathcal N\), define the corresponding \(X_Y\) contributions in \eqref{eq:nnlo-separation}.
In \(\mathrm{FR}\), the same integrated soft and virtual infrared factors as in \(U\) multiply the finite one-loop interference, so the poles cancel by the NLO identity.
Those in \(\mathrm{DU}\) cancel between the RR, RV and VV terms at fixed Born kinematics.
The explicit kernels and cancellations are given in appendices~\ref{terms-proportional-to-the-finite-one-loop-remainder} and~\ref{double-unresolved-terms}.
The finite double-virtual contribution is
\begin{equation}
\mathrm{VV}_{\mathrm F}=\mathcal N\int\!\left[d\Phi_{c\bar c}F_{c\bar c}^{(2),{\rm fin}}\right]_{\epsilon^0}.
\label{eq:hard-double-virtual-definition}
\end{equation}

The finite NNLO correction is therefore
\begin{equation}
\begin{aligned}
\sigma^{(2)}
 &=\underbrace{\mathrm{RR}_{\mathrm F}(gg)+\mathrm{RR}_{\mathrm F}(q\bar q)}
              _{c\bar c+gg,\;c\bar c+q\bar q}\\[3pt]
 &\quad+\underbrace{\mathrm{RV}_{\mathrm F}+\mathrm{SU}}_{c\bar c+g}
       +\underbrace{\mathrm{VV}_{\mathrm F}+\mathrm{FR}+\mathrm{DU}}_{c\bar c}.
\end{aligned}
\label{eq:4.1}
\end{equation}
The phase-space mappings in appendix~\ref{common-parametrisations-of-the-real-and-subtraction-terms} keep \(P\) fixed between each real term and its counterterms, so the same subtraction applies bin by bin in the quarkonium energy spectrum.

\FloatBarrier

\Needspace{6\baselineskip}
\section{Numerical implementation}\label{numerical-setup-and-checks}

\Needspace{4\baselineskip}
\subsection{Input parameters and approximation of the virtual terms}\label{input-parameters}

The input parameters used in the numerical calculation are listed in table~\ref{tab:1}.
We take \(\alpha_s^{(5)}(M_Z)=0.1180\), \(M_Z=91.1876\ {\rm GeV}\) and \(\overline m_b(\overline m_b)=4.18\ {\rm GeV}\)~\cite{ref22}.
The coupling is evolved with three-loop running and two-loop flavour matching at \(\mu_b=\overline m_b(\overline m_b)\)~\cite{ref23}.
This gives \(\alpha_s^{(4)}(2m_c)=0.2528\).

For the wave function, we adopt the central value \(\langle\mathcal O_1\rangle_{J/\psi}=0.440\ {\rm GeV}^3\) used in eq.~(3.2) of ref.~\cite{ref15}.
This value was extracted in ref.~\cite{ref21}, including NLO QCD and relativistic corrections.
We use the same running coupling and matrix element at all three perturbative orders to isolate the successive QCD corrections, as in ref.~\cite{ref15}.

\begin{table}[!htbp]
\centering
\small
\renewcommand{\arraystretch}{1.18}
\begin{tabularx}{\linewidth}{>{\raggedright\arraybackslash}p{.47\linewidth}X}
\toprule
Parameter & Value \\
\midrule
Centre-of-mass energy \(\sqrt{s}\) & \(10.6\ {\rm GeV}\) \\
Charm mass \(m_c\) & \(1.5\ {\rm GeV}\) \\
Quarkonium mass \(M_\psi\) & \(2m_c=3.0\ {\rm GeV}\) \\
Squared wave function at the origin \(\lvert R(0)\rvert^2\) & \(0.922\ {\rm GeV}^3\) \\
Electromagnetic coupling \(\alpha\) & \(1/137\) \\
Renormalisation scale \(\mu_R\) & \(2m_c\) \\
Strong coupling \(\alpha_s^{(4)}(2m_c)\) & \(0.2528\) \\
NRQCD factorisation scale \(\mu_\Lambda\) & \(m_c\) \\
\bottomrule
\end{tabularx}
\caption{Central input parameters.}\label{tab:1}
\end{table}
\FloatBarrier

For the finite virtual contributions \(V_{\mathrm F}\) and \(\mathrm{VV}_{\mathrm F}\), we use the results at \(\sqrt{s_V}=10.5\ {\rm GeV}\) to approximate their values at \(\sqrt{s}=10.6\ {\rm GeV}\).
The choice \(s_V/m_c^2=49\) simplifies the two-loop integration-by-parts reduction.
The remaining contributions are evaluated at \(\sqrt{s}\).
This replacement is applied to the finite terms after analytic infrared cancellation.
The NLO cross section agrees with the independent calculation of ref.~\cite{ref3} within \(0.5\%\) for matched inputs (section~\ref{benchmarks-and-numerical-uncertainties}), providing a lower-order check of the approximation used here.

\Needspace{4\baselineskip}
\subsection{Matrix elements and phase-space integration}\label{virtual-quadrature-and-binning}

\paragraph{Virtual contributions.}
We reduce the loop integrals to master integrals with Kira 3.1 \cite{ref14} and evaluate them with AMFlow \cite{ref11}.
Values at successive phase-space points are obtained with AMFlow's internal differential-equation solver.
The finite virtual contributions \(V_{\mathrm F}\) and \(\mathrm{VV}_{\mathrm F}\) are integrated over phase space using fixed-grid Gauss--Kronrod quadrature.

\paragraph{Real radiation and integrated counterterms.}
The real-emission squared matrix elements and one-loop--tree interference are obtained with an in-house program written in FORM~\cite{form}.
We simplify their rational expressions with the \texttt{pfd.lib} library~\cite{pfd} in Singular~\cite{singular}.
Common subexpression elimination allows repeated numerator and denominator factors to be evaluated once and reused, reducing the cost of the phase-space integration.
The tree-level squared matrix elements agree with independent results generated with FDC~\cite{fdc}.
We evaluate the scalar one-loop integrals in the real-emission calculation using the dimension-changing transformation (DCT)~\cite{ref19}.
The implementation uses Arb/Acb, a C library for arbitrary-precision real and complex ball arithmetic~\cite{ref13}, to evaluate the integrals and loop interferences with adjustable precision and propagated numerical error bounds.
The phase-space integrals of the Born, real-emission and integrated-counterterm contributions are evaluated with \mbox{VEGAS+~\cite{ref12}}.

\paragraph{Energy and momentum spectra.}
We first compute the quarkonium energy spectrum in bins of \(E_\psi=Q\cdot P/\sqrt{s}\).
To obtain \(d\sigma/d\lvert\boldsymbol P\rvert\), we convert the bin boundaries to momentum and divide each bin cross section by its momentum width.
The binning and interpolation are specified in appendix~\ref{numerical-integration-and-spectra}.

\Needspace{4\baselineskip}
\subsection{Numerical checks and uncertainty estimates}\label{benchmarks-and-numerical-uncertainties}

\paragraph{Lower-order benchmarks.}
To compare with Gong and Wang \cite{ref3}, we use their input parameters, including \(\alpha_s=0.259\) and \(\lvert R(0)\rvert^2=0.944\ {\rm GeV}^3\).
We then obtain LO and NLO cross sections of \(0.1711\) and \(0.2967\ {\rm pb}\), compared with their \(0.171\) and \(0.298\ {\rm pb}\), respectively.
The LO results agree to the quoted precision, and our NLO result is lower by \(0.43\%\).
The NLO angular distribution also agrees with their prediction for direct \(J/\psi+c\bar c\) production to the accuracy that can be read from the published plot.

\paragraph{Monte Carlo uncertainties.}
We estimate the combined Monte Carlo uncertainty conservatively by adding the standard errors of the individual contributions linearly.
At the central scale, this gives \(0.00042\ {\rm pb}\) for the NNLO total cross section, or \(0.13\%\) of the total and \(0.83\%\) of the NNLO correction, with \(\mathrm{RV}_{\mathrm F}\) providing the largest contribution.
The corresponding estimate for \(\sigma_L\) is \(0.000055\ {\rm pb}\).

\paragraph{Virtual-integration errors.}
We estimate the phase-space integration errors in \(V_{\mathrm F}\) and \(\mathrm{VV}_{\mathrm F}\) from the embedded differences between the 7-point Gauss and 15-point Kronrod rules~\cite{quadpack}.
Combining the inner and outer estimates and summing over the separately integrated terms gives \(1.2\times10^{-3}\ {\rm pb}\) for the NNLO total cross section at the central scale.
The individual estimates are included in table~\ref{tab:4}, with their definition and sum given in appendix~\ref{numerical-integration-and-spectra}.

\paragraph{Spectrum checks.}
At NNLO, the relative Monte Carlo uncertainty estimates per energy bin have a median of \(1.4\%\) and a maximum of \(3.0\%\).
The integrated spectrum agrees with the total cross section within the conservative Monte Carlo uncertainty estimates.

To check the sensitivity to interpolation, we replace the cubic spline for \(V_{\mathrm F}\) and \(\mathrm{VV}_{\mathrm F}\) by linear interpolation.
The NNLO cross section in the lowest energy bin, \(E_\psi=3.00\)--\(3.10\ {\rm GeV}\), changes by \(3.3\%\).
The lower input energy of the virtual terms also shifts their momentum endpoint relative to the other contributions (appendix~\ref{numerical-integration-and-spectra}).

\FloatBarrier

\Needspace{6\baselineskip}
\section{Phenomenological results}\label{phenomenological-results}

The central scale is \(\mu_R=2m_c\); the quoted scale uncertainties and theory bands are the envelopes obtained by varying \(m_c\leq\mu_R\leq4m_c\), with all other inputs fixed, including \(\mu_\Lambda=m_c\).
We vary both \(\alpha_s^{(4)}(\mu_R)\) and the explicit scale logarithms, retaining the four-flavour scheme throughout this interval.

\Needspace{5\baselineskip}
\subsection{Experimental samples and comparison procedure}\label{experimental-inputs-and-comparison-conventions}

Our prediction describes direct colour-singlet \(J/\psi\) production with an open charm pair.
For the total rate and production angle, we compare with Belle's charm-associated prompt sample, which includes both open-charm and double-charmonium production, with \(J/\psi\) produced directly or through higher-charmonium decays (feed-down)~\cite{ref4}.
For the momentum spectrum, the main comparison uses Belle's prompt open-charm sample, which excludes double charmonium but retains feed-down.

We digitise the experimental points and error bars from Belle's published plots.
For the angular distribution, we convert the efficiency-corrected yields to cross sections using Belle's measured total rate; the plotted errors are obtained by applying the same conversion factor to the yield errors.

\FloatBarrier

\Needspace{5\baselineskip}
\subsection{Integrated cross sections and NNLO corrections}\label{total-cross-section-and-the-nnlo-correction}

Table~\ref{tab:3} gives the fixed-order results.
Writing \(\alpha_s=\alpha_s^{(4)}(\mu_R)\) and \(L_R=\ln[\mu_R/(2m_c)]\), we express the NNLO results as corrections to the LO cross sections:
\begin{equation}
\begin{aligned}
\sigma_{\rm NNLO}
 &\simeq\sigma_{\rm LO}\Bigl[1+(2.837+2.621\,L_R)\alpha_s
 +(4.945+12.49\,L_R+5.165\,L_R^2)\alpha_s^2\Bigr],\\[4pt]
\sigma_{L,\rm NNLO}
 &\simeq\sigma_{L,\rm LO}\Bigl[1+(2.195+2.650\,L_R)\alpha_s
 +(1.258+10.017\,L_R+5.241\,L_R^2)\alpha_s^2\Bigr],
\end{aligned}
\label{eq:scale-polynomials}
\end{equation}
where \(\sigma_{\rm LO}\simeq2.4893\,\alpha_s^2\ {\rm pb}\) and \(\sigma_{L,\rm LO}\simeq0.38866\,\alpha_s^2\ {\rm pb}\).

At the central scale, the NLO cross section is 1.72 times the LO value; the NNLO correction is \begin{equation}
\sigma^{(2)}\simeq0.0503\ {\rm pb},\qquad
\frac{\sigma^{(2)}}{\sigma_{\rm NLO}}\simeq18.4\%.
\label{eq:5.1}
\end{equation}

\begin{table}[!htbp]
\centering
\small
\renewcommand{\arraystretch}{1.18}
\begin{tabular}{lrrr}
\toprule
Order & \(\sigma\ [{\rm pb}]\) & \(\sigma_L\ [{\rm pb}]\) & \(\alpha_\theta\) \\
\midrule
LO & \(0.159^{+0.14}_{-0.059}\) & \(0.0248^{+0.022}_{-0.0092}\) & \(0.460\) \\
NLO & \(0.273^{+0.13}_{-0.080}\) & \(0.0386^{+0.014}_{-0.010}\) & \(0.505^{+0.036}_{-0.014}\) \\
NNLO & \(0.324^{+0.041}_{-0.066}\) & \(0.0406^{+0.0007}_{-0.0058}\) & \(0.554^{+0.096}_{-0.031}\) \\
\bottomrule
\end{tabular}
\caption{Results for \(\sigma\), \(\sigma_L\) and \(\alpha_\theta\) at \(\mu_R=2m_c\), with scale uncertainties from \(m_c\leq\mu_R\leq4m_c\).
The LO angular coefficient is scale independent.}\label{tab:3}
\end{table}

\begin{figure}[!htbp]
\centering
\includegraphics[width=0.82\linewidth]{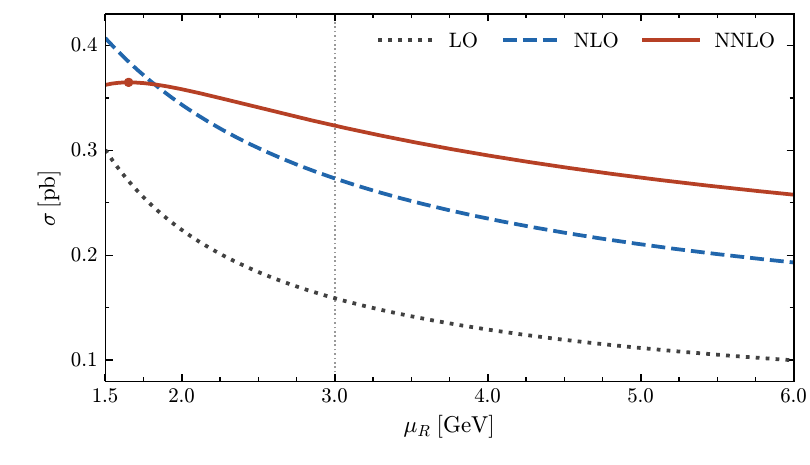}
\caption{Renormalisation-scale dependence of the direct \(J/\psi+c\bar c+X\) cross section at LO, NLO and NNLO, with \(m_c=1.5\ {\rm GeV}\).
The dot marks the NNLO maximum; the vertical line indicates the central scale \(\mu_R=2m_c\).}\label{fig:scale-dependence}
\end{figure}

Figure~\ref{fig:scale-dependence} shows the renormalisation-scale dependence of the LO, NLO and NNLO total cross sections.
The relative scale variation decreases from \({}^{+48.9\%}_{-29.3\%}\) at NLO to \({}^{+12.7\%}_{-20.3\%}\) at NNLO.
The NNLO cross section reaches its maximum at \(\mu_R\simeq1.65\ {\rm GeV}\).
At this lower scale, we find
\[
(\sigma_{\rm LO},\sigma_{\rm NLO},\sigma_{\rm NNLO})\big|_{\mu_R=1.65\,\mathrm{GeV}}
 \simeq(0.271,\,0.385,\,0.365)\ {\rm pb}.
\]
Relative to NLO, the NNLO cross section is \(5.2\%\) lower at this scale but \(18.4\%\) higher at the central scale.

Table~\ref{tab:4} gives the Born and finite higher-order contributions.
For \(\mathrm{VV}_{\mathrm F}\), we list the finite two-loop--tree interference \(\mathrm{VV}_{20}\) and squared one-loop finite remainder \(\mathrm{VV}_{11}\) separately.

\begin{table}[!htbp]
\centering
\small
\renewcommand{\arraystretch}{1.18}
\begin{tabular}{clrrc}
\toprule
Order & Contribution & \(\sigma\ [{\rm fb}]\) & \(\sigma_L\ [{\rm fb}]\) & Method \\
\midrule
\(\sigma^{(0)}\) & \(B\) & +159.14(2) & +24.846(3) & MC \\
\midrule
\multirow{3}{*}{\(\sigma^{(1)}\)} & \(R_{\mathrm{F}}\) & -51.05(3) & -8.314(5) & MC \\
 & \(U\) & -32.70(2) & -4.963(6) & MC \\
 & \(V_{\mathrm{F}}\) & \(+197.886\pm0.078\) & \(+27.068\pm0.020\) & GK \\
\midrule
\multirow{8}{*}{\(\sigma^{(2)}\)} & \(\mathrm{RR}_{\mathrm{F}}(gg)\) & -18.75(1) & -3.216(2) & MC \\
 & \(\mathrm{RR}_{\mathrm{F}}(q\bar q)\), \(u,d,s\) & -0.565(5) & -0.0777(9) & MC \\
 & \(\mathrm{RV}_{\mathrm{F}}\) & -23.1(3) & -2.94(3) & MC \\
 & \(\mathrm{SU}\) & -29.84(3) & -4.296(6) & MC \\
 & \(\mathrm{FR}\) & -40.70(3) & -5.343(4) & MC \\
 & \(\mathrm{DU}\) & +1.9282(3) & +0.30227(4) & MC \\
 & \(\mathrm{VV}_{20}\) & \(+84.26\pm0.79\) & \(+7.261\pm0.033\) & GK \\
 & \(\mathrm{VV}_{11}\) & \(+77.07\pm0.32\) & \(+10.309\pm0.014\) & GK \\
\bottomrule
\end{tabular}
\caption{Finite contributions at \(\mu_R=2m_c\), in fb.
For MC (VEGAS+), parentheses give statistical errors in the last quoted digit.
For GK (fixed-grid Gauss--Kronrod), the errors following \(\pm\) are the embedded estimates of \eqref{eq:virtual-gk-error}.}\label{tab:4}
\end{table}

Belle's charm-associated prompt cross section is \mbox{\(0.74\pm0.08\ ({\rm stat.})\,{}^{+0.09}_{-0.08}\ ({\rm syst.})\ {\rm pb}\)}~\cite{ref4}.
The NNLO result in table~\ref{tab:3} gives the direct colour-singlet open-charm contribution to this rate.
A new NNLO calculation at \(m_c=1.4\ {\rm GeV}\) would require several months with our available computational resources.
We therefore estimate the effect of this lower mass using the NLO mass dependence obtained by Gong and Wang~\cite{ref3}.
We form the ratio of their NLO cross sections at the same \(\mu_R\) and with a common \(\lvert R(0)\rvert^2\), so that the wave-function factor cancels.\footnote{Table~I of ref.~\cite{ref3} uses \(\lvert R(0)\rvert^2\propto m_c^2\) and \(\mu_R=2m_c\), so both inputs differ between its mass values. We remove this wave-function dependence and use eq.~(16) of that reference to evaluate both masses at a common scale with our coupling.}
We assume that this mass ratio also applies at NNLO:
\begin{equation}
\begin{aligned}
K_m(\mu_R)
 &=\frac{\sigma_{\rm NLO}(m_c=1.4\ {\rm GeV};\mu_R)}
         {\sigma_{\rm NLO}(m_c=1.5\ {\rm GeV};\mu_R)},\\[3pt]
\sigma_{\rm NNLO}^{\rm est}(1.4\ {\rm GeV};\mu_R)
 &=K_m(\mu_R)\,\sigma_{\rm NNLO}(1.5\ {\rm GeV};\mu_R).
\end{aligned}
\label{eq:mass-rescaling}
\end{equation}
The NLO results give \(K_m\simeq1.40\) at both \(\mu_R=3\) and \(1.65\ {\rm GeV}\), so this extrapolation enhances the NNLO total rate by about \(40\%\).

For the open-charm channel, production of \(\psi(2S)\) followed by its inclusive decay to \(J/\psi\) is estimated to add about \(29\%\) of the direct rate~\cite{ref3}.
Table~\ref{tab:mass-feeddown} summarises the effects of the lower mass and this feed-down estimate.
Together they raise the estimated open-charm rate to about \(0.59\ {\rm pb}\) at \(\mu_R=3\ {\rm GeV}\) and \(0.66\ {\rm pb}\) at \(\mu_R=1.65\ {\rm GeV}\), accounting for a larger fraction of the measured prompt rate.
Belle's measured rate also includes double-charmonium production.

\begin{table}[!htbp]
\centering
\small
\renewcommand{\arraystretch}{1.18}
\begin{tabular}{crrrr}
\toprule
 & \multicolumn{2}{c}{\(m_c=1.5\ {\rm GeV}\)}
 & \multicolumn{2}{c}{\(m_c=1.4\ {\rm GeV}\) (estimated)} \\
\cmidrule(lr){2-3}\cmidrule(lr){4-5}
\(\mu_R\ [{\rm GeV}]\) & Direct & With feed-down & Direct & With feed-down \\
\midrule
3.00 & 0.324 & 0.42 & 0.45 & 0.59 \\
1.65 & 0.365 & 0.47 & 0.51 & 0.66 \\
\bottomrule
\end{tabular}
\caption{NNLO open-charm cross sections in pb. The \(m_c=1.4\ {\rm GeV}\) entries use the NLO mass rescaling in \eqref{eq:mass-rescaling}; \(\psi(2S)\) feed-down is estimated as \(29\%\) of direct production~\cite{ref3}.}\label{tab:mass-feeddown}
\end{table}

To illustrate these effects on the distributions, we also show the NNLO result multiplied by \(1.29\) and by \(1.29K_m(\mu_R)\).
These estimates rescale the differential cross sections while retaining the calculated \(m_c=1.5\ {\rm GeV}\) shapes and kinematics.
Their scale bands are obtained by varying \(\mu_R\) in the distribution and \(K_m\) together over the same \(1.5\)--\(6\ {\rm GeV}\) interval.

\Needspace{5\baselineskip}
\subsection{Production-angle distribution}\label{production-angle-distribution}

For each fixed \(\cos\theta\), we evaluate the angular distribution in \eqref{eq:2.4} with \(\sigma\) and \(\sigma_L\) at the same \(\mu_R\).
Its minimum and maximum over \(m_c\leq\mu_R\leq4m_c\) define the scale band.
At \(\mu_R=2m_c\), the NNLO production-angle distribution is \begin{equation}
\frac{d\sigma_{\rm NNLO}}{d\cos\theta}
\simeq\left(0.1366+0.0756\cos^2\theta\right){\rm pb}.
\label{eq:5.2}
\end{equation}
The central angular coefficient \(\alpha_\theta\) increases from 0.505 at NLO to 0.554 at NNLO, enhancing forward and backward production relative to the central region.
Its NNLO scale range is \(0.523\leq\alpha_\theta\leq0.649\); unlike the total cross section, its scale uncertainty does not decrease at this order.

\begin{figure}[!htbp]
\centering
\includegraphics[width=1.00\linewidth]{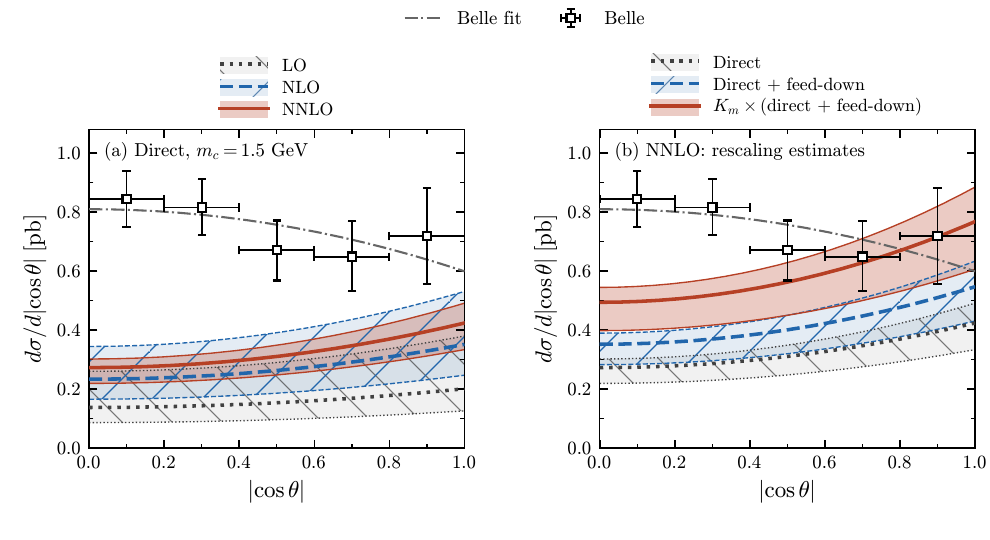}
\caption{Production-angle distributions compared with Belle prompt charm-associated data~\cite{ref4}: (a) direct production at \(m_c=1.5\ {\rm GeV}\); (b) NNLO with the indicated overall rescalings for feed-down and the lower mass.
Thick lines show \(\mu_R=3\ {\rm GeV}\); bands show scale variation.
The dash-dotted curve is the Belle fit.}\label{fig:2}
\end{figure}

Figure~\ref{fig:2} compares the angular distribution in \(|\cos\theta|\) with Belle's prompt charm-associated data~\cite{ref4}.
Belle measures \(\alpha_\theta=-0.26^{+0.24}_{-0.22}\) (stat.), favouring central production, whereas the direct colour-singlet result favours forward and backward production throughout the NNLO scale band.
The rescaling estimates in the right panel raise the rate but leave \(\alpha_\theta\) unchanged.
The experimental coefficient describes the combined prompt sample; assessing this difference requires the angular distributions and relative rates of open-charm feed-down and double-charmonium production.

\Needspace{5\baselineskip}
\subsection{Momentum distribution}\label{momentum-distribution}

For the energy spectrum, we vary \(\mu_R\) in the running coupling and the explicit scale logarithms of each bin cross section.
The minimum and maximum over \(m_c\leq\mu_R\leq4m_c\) give its scale uncertainty.
We then convert the bin cross sections and their scale envelopes to the momentum spectrum.

\begin{figure}[!htbp]
\centering
\includegraphics[width=1.00\linewidth]{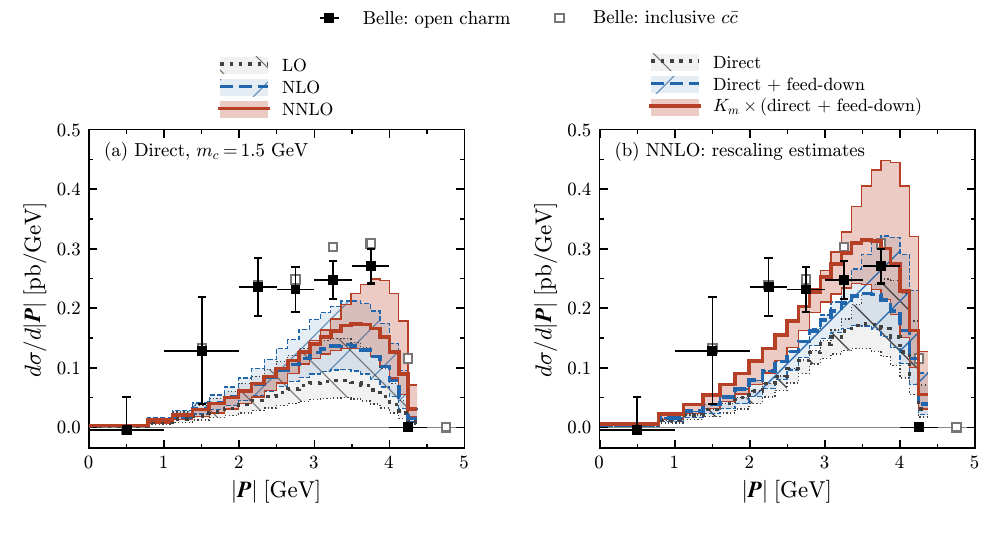}
\caption{Momentum spectra: (a) direct production at \(m_c=1.5\ {\rm GeV}\); (b) NNLO with the same overall rescalings as figure~\ref{fig:2}, keeping the calculated spectral shape.
Filled squares are Belle prompt open-charm data; open squares also include double charmonium~\cite{ref4}.
Thick lines show \(\mu_R=3\ {\rm GeV}\); bands show scale variation.}\label{fig:3}
\end{figure}

Figure~\ref{fig:3} compares the \(J/\psi\) momentum spectrum with Belle's prompt open-charm data, digitised from figure~3 of ref.~\cite{ref4}.\footnote{We read the ordinate as \({\rm fb}/(0.5\,{\rm GeV}/c)\), taking the printed nb to be a typographical error. The charm-associated spectrum then integrates to \(0.736\ {\rm pb}\), consistent with the published \(0.74\ {\rm pb}\). The first lower error bar is clipped in the original plot; no uncertainty is given for the last point.}
The digitised open-charm spectrum integrates to about \(0.62\ {\rm pb}\).

At the central scale, the NNLO correction raises the direct cross section near the peak, bringing it closer to Belle's prompt open-charm data.
The combined mass and feed-down rescaling brings the integrated rate close to the measured open-charm rate, but leaves the direct-production shape unchanged.
This shape still gives a larger fraction of high-momentum \(J/\psi\) mesons than Belle's prompt spectrum, which includes feed-down.
A prediction for the prompt spectrum therefore requires adding the feed-down momentum spectrum to the direct contribution, using the production spectra of higher charmonia with open charm and the kinematics of their decays to \(J/\psi\).

\Needspace{6\baselineskip}
\section{Conclusions}\label{conclusions}

We have calculated the NNLO QCD corrections to direct colour-singlet \(J/\psi\) production with an open charm pair in electron-positron annihilation.
The real-radiation contributions are treated with the nested soft-collinear subtraction scheme for massive partons.
We have organised the real and virtual corrections into finite contributions and demonstrated the infrared pole cancellation on the single-real and Born phase spaces.
The phase-space mappings keep the quarkonium momentum fixed in the real terms and their counterterms, allowing us to calculate differential distributions.

For \(m_c=1.5\ {\rm GeV}\), the approximate NNLO cross section at \(\sqrt{s}=10.6\ {\rm GeV}\) is \(0.324^{+0.041}_{-0.066}\ {\rm pb}\), with the uncertainty obtained by varying \(m_c\leq\mu_R\leq4m_c\) at fixed \(\mu_\Lambda=m_c\).
The central value at \(\mu_R=2m_c\) is \(18.4\%\) above NLO, and its relative scale variation is reduced to \({}^{+12.7\%}_{-20.3\%}\).
The NNLO correction is less than half the NLO correction; together with the reduced scale dependence, this shows good perturbative convergence of the total cross section through NNLO.

Assuming that the NLO mass dependence persists at NNLO, lowering \(m_c\) to \(1.4\ {\rm GeV}\) and including the estimated \(\psi(2S)\) feed-down gives an open-charm rate of about \(0.59\ {\rm pb}\) at \(\mu_R=3\ {\rm GeV}\) and \(0.66\ {\rm pb}\) at \(\mu_R=1.65\ {\rm GeV}\).
The mass and feed-down estimates increase the differential cross sections without changing their calculated shapes.

At the central scale, the NNLO correction raises the absolute momentum spectrum near the peak, bringing it closer to Belle's prompt open-charm data.
The predicted spectrum still gives a larger fraction of high-momentum \(J/\psi\) mesons than the measured spectrum.

By contrast, the NNLO correction shifts the angular shape further away from Belle's prompt charm-associated data.
At \(\mu_R=2m_c\), \(\alpha_\theta\) increases from \(0.505\) at NLO to \(0.554\) at NNLO, strengthening the preference for forward and backward production.
Belle's measured value, \(\alpha_\theta=-0.26\), instead favours production at central angles.
The NNLO scale uncertainty in \(\alpha_\theta\), \({}^{+0.096}_{-0.031}\), is larger than at NLO.
The difference in angular shape motivates more precise measurements at Belle~II that separate direct open-charm production from feed-down and double charmonium.

\acknowledgments
This work was supported by the National Natural Science Foundation of China under Grant No.~12135013.
Generative AI tools assisted with code development, analytic and numerical checks, and manuscript preparation.

\FloatBarrier

\appendix

\Needspace{6\baselineskip}
\section{Derivation of the angular reconstruction}\label{angular-reconstruction-derivation}

We write the quarkonium energy as \(E_\psi=Q\cdot P/\sqrt{s}\).
In the centre-of-mass frame, define
\[
Q=(\sqrt{s},\boldsymbol 0),\qquad
\boldsymbol n=\frac{\boldsymbol P}{|\boldsymbol P|},\qquad
\boldsymbol n_-=\frac{\boldsymbol p_-}{|\boldsymbol p_-|},\qquad
\cos\theta=\boldsymbol n\cdot\boldsymbol n_-.
\]
At fixed quarkonium momentum \(P\), we integrate inclusively over the open charm pair and additional radiation and sum over final-state spins and colours.
After infrared subtraction and pole cancellation, let \(H^{\mu\nu}(E_\psi,\boldsymbol n)\) denote the symmetric part of the finite four-dimensional hadronic tensor, normalised by
\[
\begin{gathered}
\frac{d\sigma}{dE_\psi\,d\Omega_{\boldsymbol n}}
 =\frac{3}{4\pi s}L_{\mu\nu}H^{\mu\nu},\\
L^{\mu\nu}=p_-^\mu p_+^\nu+p_+^\mu p_-^\nu-\frac{s}{2}g^{\mu\nu}.
\end{gathered}
\]
Here \(L^{\mu\nu}\) is the spin-averaged leptonic tensor for massless incoming leptons.
The projectors in \eqref{eq:photon-projectors} give the covariant contractions
\[
H=\Pi_{\mu\nu}H^{\mu\nu},\qquad
H_L=\Pi_{L,\mu\nu}H^{\mu\nu},\qquad
Q\cdot e_L=0,\quad e_L^2=-1.
\]
Since both projectors depend only on \(Q,P\), these contractions commute with the integration over unobserved particles.
The replacement \(\Pi^{\mu\nu}\to\Pi_L^{\mu\nu}\) can therefore be made directly in the squared matrix elements and interferences at each phase-space point.

In the centre-of-mass frame, current conservation, \(Q_\mu H^{\mu\nu}=Q_\nu H^{\mu\nu}=0\), leaves only spatial components.
The inclusive integration leaves \(\boldsymbol n\) as the only spatial direction in the hadronic tensor.
Rotational symmetry about this direction gives the two-structure decomposition used in appendix~A of ref.~\cite{ref3},
\begin{equation}
\begin{aligned}
H^{ij}(E_\psi,\boldsymbol n)
 &=\frac{H_T(E_\psi)}{2}(\delta^{ij}-n^in^j)
   +H_L(E_\psi)n^in^j,\\
H(E_\psi)&=\delta_{ij}H^{ij}=H_T(E_\psi)+H_L(E_\psi),
\qquad H_L(E_\psi)=n_i n_jH^{ij}.
\end{aligned}
\label{eq:angular-tensor}
\end{equation}
The contributions from the two photon polarisations perpendicular to \(\boldsymbol P\) sum to \(H_T\).
With \(e_L^\mu=(0,\boldsymbol n)\), the covariant projectors reduce to the spatial contractions in \eqref{eq:angular-tensor}.

Using \(L^{ij}=\tfrac{s}{2}(\delta^{ij}-n_-^in_-^j)\), contraction with \eqref{eq:angular-tensor} and integration over the quarkonium azimuthal angle give
\begin{equation}
\frac{d^2\sigma}{dE_\psi\,d\cos\theta}
 =\frac38(1+\cos^2\theta)H_T(E_\psi)
  +\frac34(1-\cos^2\theta)H_L(E_\psi).
\label{eq:angular-transverse-weights}
\end{equation}
Both angular weights integrate to unity over \(-1\leq\cos\theta\leq1\), so \(H_T+H_L=d\sigma/dE_\psi=H\).
Substituting \(H_T=H-H_L\) gives
\begin{equation}
\frac{d^2\sigma}{dE_\psi\,d\cos\theta}
 =\frac38(1+\cos^2\theta)H(E_\psi)
  +\frac38(1-3\cos^2\theta)H_L(E_\psi).
\label{eq:2.3}
\end{equation}
The two projected cross sections are
\[
\sigma=\int dE_\psi\,H(E_\psi),\qquad
\sigma_L=\int dE_\psi\,H_L(E_\psi).
\]
Integrating \eqref{eq:2.3} over \(E_\psi\) therefore yields \eqref{eq:2.4}.

\FloatBarrier

\Needspace{6\baselineskip}
\section{Finite contributions and infrared pole cancellation}\label{finite-contributions-and-poles}\label{analytic-pole-cancellation}

This appendix gives the integrated kernels and pole cancellations for the finite contributions defined in section~\ref{combining-real-and-virtual-corrections}.
The local counterterms and their unresolved integrals are derived in appendix~\ref{real-emission-subtraction}.

\subsection{Kinematics and phase-space conventions}\label{appendix-subtraction-conventions}
We define the recoil momentum, its invariant mass and four-velocity, and the heavy-quark speed at Born kinematics by \begin{equation}
K=Q-P,\qquad W=\sqrt{K^2},\qquad u_K=K/W,\qquad
\beta=\sqrt{1-\frac{4m_c^2}{K^2}}.
\label{eq:3.1}
\end{equation} A superscript \(*\) labels energies, angles and momentum components in the rest frame of \(K\), with \(E_k^*=u_K\cdot k\).

For an emitted massless parton with momentum \(k\), the recoiling system contains the heavy-quark pair.
Hence \((K-k)^2=K^2-2WE_k^*\geq4m_c^2\), which gives \begin{equation}
E_k^*\leq E_0^*(K),\qquad
E_0^*(K)=\frac{K^2-4m_c^2}{2W}=\frac{W\beta^2}{2}.
\label{eq:3.2}
\end{equation} We include this kinematic bound in the massless one-particle phase-space element,
\begin{equation}
[dk]_K=\frac{d^d k}{(2\pi)^{d-1}}\delta_+(k^2)
             \theta(E_0^*-u_K\cdot k).
\label{eq:3.3}
\end{equation} Here \(\delta_+(k^2)=\theta(k^0)\delta(k^2)\). In a soft counterterm, the unresolved momenta are removed from the momentum-conservation constraint and integrated independently of the reduced phase space.

To separate the spectator kinematics from the coloured final state \(X\) of total momentum \(K\), we write \begin{equation}
d\Phi_{J/\psi+X}(Q)
=\frac{dK^2}{2\pi}\,d\Phi_2(Q;P,K)\,d\Phi_X(K).
\label{eq:3.4}
\end{equation} Here \(d\Phi_X(K)\) contains only the coloured final state.
For the full \(J/\psi+X\) phase spaces, we use the channel notation of section~\ref{combining-real-and-virtual-corrections} and abbreviate
\[
d\Phi_B=d\Phi_{c\bar c},\qquad
d\Phi_R=d\Phi_{c\bar c+g},\qquad
d\Phi_{\mathrm{RR}}^{ab}=d\Phi_{c\bar c+ab}.
\]
The soft and collinear projections act on the squared matrix elements and interferences together with their phase space, at fixed \(P\).
Thus \(K\), the recoil frame and \(E_0^*(K)\) are common to all terms, which enter the same quarkonium energy bin.

All counterterms use the dimensional factor \(\mathcal C_\epsilon=(\mu_R^2e^{\gamma_E}/4\pi)^\epsilon\), where \(\gamma_E\) is the Euler--Mascheroni constant.

\Needspace{5\baselineskip}
\subsection{Virtual amplitudes and infrared factors}\label{virtual-amplitudes-and-infrared-matching}

The colour factors are \(C_F=(N_c^2-1)/(2N_c)=4/3\), \(C_A=N_c=3\) and \(T_F=1/2\), with \(n_l=3\) massless flavours and \(n_f=4\) active flavours in the coupling.
The one-loop beta-function coefficients and the charm-quark contribution \(\beta_{0,Q}\) are
\begin{equation}
\begin{gathered}
\beta_0^{(n_f)}=\frac{11}{3}C_A-\frac43T_F n_f,
\qquad \beta_{0,Q}=-\frac43T_F=-\frac23,\\
\beta_0^{(3)}=9,\qquad
\beta_0^{(4)}=\beta_0^{(3)}+\beta_{0,Q}=\frac{25}{3}.
\end{gathered}
\label{eq:2.2}
\end{equation}
Coupling renormalisation uses \(\beta_0^{(4)}\), while the infrared factors involve \(\beta_0^{(3)}\).
With the four-flavour coupling used throughout, the charm-loop terms proportional to \(\beta_{0,Q}\) appear in \(Z_2\), \eqref{eq:2.6}, and in the gluon-field renormalisation coefficient \(Z_A^{(1)}\), \eqref{eq:3.38}~\cite{ref6}.

\paragraph{NRQCD matching.}
The colour-singlet vector-current renormalisation gives the NRQCD subtraction term~\cite{nrqcd_vector_matching_cm,nrqcd_vector_matching,nrqcd_current_matching}
\begin{equation}
\mathcal M_{\rm NRQCD}
 =-\frac{\alpha_s^2}{4\epsilon}
   \left(\frac{C_F^2}{3}+\frac{C_FC_A}{2}\right)
   \left(\frac{\mu_\Lambda^2e^{\gamma_E}}{4\pi\mu_R^2}\right)^{-2\epsilon}
   \mathcal M^{(0)}.
\label{eq:nrqcd-counterterm}
\end{equation}
Through NNLO, the matched amplitude is \(\mathcal M=\mathcal M_{\rm QCD,ren}-\mathcal M_{\rm NRQCD}\), where \(\mathcal M_{\rm QCD,ren}\) is the ultraviolet-renormalised QCD amplitude.
This subtraction removes the two-loop pole associated with the bound-state pair before the remaining QCD infrared poles are separated below.
The coefficient \(C_F^2/3+C_FC_A/2=70/27\) is independent of the open-charm kinematics.

\paragraph{Infrared factors and finite remainders.}
We separate the remaining QCD infrared poles from the finite remainders following ref.~\cite{ref6}, section~2.3.
For the Born channel, we abbreviate \(\mathcal M_{c\bar c}^{(\ell)}\) to \(\mathcal M^{(\ell)}\).
The infrared-finite coefficients are \begin{equation}
\begin{aligned}
\mathcal M_{\rm fin}^{(1)}&=\mathcal M^{(1)}-Z_1\mathcal M^{(0)},\\
\mathcal M_{\rm fin}^{(2)}&=\mathcal M^{(2)}
 -Z_1\mathcal M_{\rm fin}^{(1)}-Z_2\mathcal M^{(0)}.
\end{aligned}
\label{eq:2.5}
\end{equation}

In terms of the heavy-pair anomalous-dimension coefficients \(\Gamma_n=C_F\gamma_{{\rm cusp},Q}^{(n)}(v_{12})+2\gamma_Q^{(n)}\), with \(v_{12}=\sqrt{1-m_c^4/(p_1\cdot p_2)^2}\), the scalar infrared factors are~\cite{becher_neubert,ferroglia_massive,ref6} \begin{equation}
\begin{aligned}
Z_1&=\frac{\Gamma_0}{2\epsilon},\\
Z_2&=\frac{\Gamma_0(\Gamma_0-2\beta_0^{(3)})}{8\epsilon^2}
     +\frac{\Gamma_1}{4\epsilon}
     +\frac{\beta_{0,Q}\Gamma_0}{2\epsilon}
                          \ln\frac{\mu_R^2}{m_c^2}.
\end{aligned}
\label{eq:2.6}
\end{equation}
The anomalous dimensions are expanded as \(\gamma_i(\alpha_s)=\sum_{n\geq0}a_s^{n+1}\gamma_i^{(n)}\).
For the outgoing massive pair, \begin{equation}
\begin{aligned}
\gamma_{{\rm cusp},Q}^{(0)}(v)
 &=\frac4v\left[\frac12\ln\frac{1+v}{1-v}-i\pi\right],
 \qquad \gamma_Q^{(0)}=-2C_F,\\
\gamma_Q^{(1)}
 &=C_FC_A\left(\frac{2\pi^2}{3}-\frac{98}{9}-4\zeta_3\right)
   +\frac{40}{9}C_FT_Fn_l.
\end{aligned}
\label{eq:2.7}
\end{equation} The two-loop massive cusp coefficient \(\gamma_{{\rm cusp},Q}^{(1)}(v)\), including its time-like imaginary part, is given in appendix~\ref{two-loop-massive-cusp-anomalous-dimension}. Here and below, \(\zeta_n=\zeta(n)\) denotes a Riemann zeta value.

\paragraph{Born and finite virtual contributions.}
We abbreviate \(F_{c\bar c}^{(0)}\), \(F_{c\bar c}^{(1),{\rm fin}}\) and \(F_{c\bar c}^{(2),{\rm fin}}\) as \(F_B\), \(F_V^{\rm fin}\) and \(F_{\mathrm{VV}}^{\rm fin}\), respectively:
\begin{equation}
\begin{aligned}
F_B&=|\mathcal M^{(0)}|^2,\\
F_V^{\rm fin}&=2a_s\operatorname{Re}
             [\mathcal M^{(0)*}\mathcal M_{\rm fin}^{(1)}],\\
F_{\mathrm{VV}}^{\rm fin}
 &=a_s^2\left(2\operatorname{Re}
             [\mathcal M^{(0)*}\mathcal M_{\rm fin}^{(2)}]
                     +|\mathcal M_{\rm fin}^{(1)}|^2\right).
\end{aligned}
\label{eq:2.8}
\end{equation}

The finite double-virtual contribution consists of
\begin{equation}
\begin{gathered}
\mathrm{VV}_{\mathrm F}=\mathrm{VV}_{20}+\mathrm{VV}_{11},\\
\mathrm{VV}_{20}=2a_s^2\mathcal N\int\!\left[d\Phi_B\,
 \operatorname{Re}(\mathcal M^{(0)*}\mathcal M_{\rm fin}^{(2)})\right]_{\epsilon^0},
\qquad
\mathrm{VV}_{11}=a_s^2\mathcal N\int\!\left[d\Phi_B\,|\mathcal M_{\rm fin}^{(1)}|^2\right]_{\epsilon^0}.
\end{gathered}
\label{eq:2.9}
\end{equation}
The two terms are the finite two-loop--tree interference and squared one-loop finite remainder.

The infrared terms generated by \eqref{eq:2.5} and \eqref{eq:2.6} are \begin{equation}
\begin{aligned}
F_{V,U}&=2a_s\operatorname{Re}Z_1\,F_B,\\
F_{\mathrm{VV,FR}}&=2a_s\operatorname{Re}Z_1\,F_V^{\rm fin},\\
F_{\mathrm{VV,DU}}&=a_s^2
 \left(2\operatorname{Re}Z_2+|Z_1|^2\right)F_B.
\end{aligned}
\label{eq:2.10}
\end{equation}

\paragraph{Real-virtual contribution.}
For \(J/\psi+c\bar c+g\), we write \(F_R=F_{c\bar c+g}^{(0)}\), \(F_{\mathrm{RV}}=F_{c\bar c+g}^{(1)}\) and \(F_{\mathrm{RV}}^{\rm fin}=F_{c\bar c+g}^{(1),{\rm fin}}\).
The renormalised one-loop--tree interference separates into an explicit infrared-pole term and a finite remainder at resolved kinematics,
\begin{equation}
F_{\mathrm{RV}}=2a_s\operatorname{Re}Z_{c\bar cg}^{(1)}F_R
                         +F_{\mathrm{RV}}^{\rm fin}.
\label{eq:3.34}
\end{equation}
The first term is \(F_{\mathrm{RV,SU}}\) in \eqref{eq:rv-factorisation-overview}, with the one-loop infrared factor~\cite{ref6}
\begin{equation}
\begin{aligned}
Z_{c\bar cg}^{(1)}
&=-\frac{C_A\gamma_{\rm cusp}^{(0)}}{4\epsilon^2}
+\frac1{2\epsilon}\biggl[\gamma_g^{(0)}+2\gamma_Q^{(0)}
+\left(C_F-\frac{C_A}{2}\right)\gamma_{{\rm cusp},Q}^{(0)}(v_{12})\\
&\qquad\qquad
-\frac{C_A\gamma_{\rm cusp}^{(0)}}2
\left(\ln\frac{m_c\mu_R}{2p_1\cdot k}
     +\ln\frac{m_c\mu_R}{2p_2\cdot k}+2i\pi\right)\biggr],
\end{aligned}
\label{eq:3.35}
\end{equation} where \(\gamma_{\rm cusp}^{(0)}=4\) and \(\gamma_g^{(0)}=-\beta_0^{(3)}\).

\Needspace{5\baselineskip}
\subsection{NLO contributions and pole cancellation}\label{nlo-fr-pole-cancellation}

In the soft counterterm, omitting \(k\) from the momentum-conservation constraint factorises the phase space as \(d\Phi_B[dk]_K\).
The locally subtracted integrand is integrable in four dimensions, giving
\begin{equation}
R_{\mathrm F}
 =\mathcal N\int\!\left[d\Phi_R(k)\,F_R-d\Phi_B[dk]_K\,\Sop_kF_R\right]_{\epsilon^0}.
\label{eq:finite-nlo-real-virtual}
\end{equation}
The soft counterterm is specified in appendix~\ref{jpsicbar-cg-at-tree-level}, and the common phase-space parametrisation of the two terms in appendix~\ref{jpsicbar-cg}.
Integrating the counterterm in \(d\) dimensions at fixed Born kinematics defines the soft kernel \(\mathcal I_{\mathrm S}\) given in \eqref{eq:A.8},
\[
F_{R,U}=\int[dk]_K\,\Sop_kF_R=a_sC_F\mathcal I_{\mathrm S}F_B.
\]
At Born kinematics, \(v_{12}=2\beta/(1+\beta^2)\).
Using \(A(\beta)\) from \eqref{eq:3.7} and the infrared factor \eqref{eq:2.6}, the integrated NLO soft counterterm and the one-loop infrared term have poles
\begin{equation}
\begin{aligned}
C_F\mathcal I_{\mathrm S}&=\frac{C_FA(\beta)}{\epsilon}+O(\epsilon^0),\\[2pt]
2\operatorname{Re}Z_1
 &=\frac{\operatorname{Re}\Gamma_0}{\epsilon}
  =-\frac{C_FA(\beta)}{\epsilon}.
\end{aligned}
\label{eq:4.5}
\end{equation}
The pole part, denoted by \([\cdot]_{\rm poles}\), therefore satisfies
\begin{equation}
\bigl[C_F\mathcal I_{\mathrm S}+2\operatorname{Re}Z_1\bigr]_{\rm poles}=0.
\label{eq:4.6}
\end{equation}

\Needspace{5\baselineskip}
\subsection{NNLO contributions and pole cancellation}\label{nnlo-integral-definitions}

The pole-cancellation identities below hold separately for the two photon projections.

\subsubsection{Subtracted real contributions}\label{nnlo-subtracted-real-and-hard-virtual}

For the two double-real channels we abbreviate \(F_{c\bar c+q\bar q}^{(0)}\) and \(F_{c\bar c+gg}^{(0)}\) as \(F_{q\bar q}\) and \(F_{gg}\).
The counterterms in \eqref{eq:finite-rr} are specified in appendices~\ref{jpsicbar-cqbar-q-at-tree-level} and~\ref{jpsicbar-cgg-at-tree-level}; their phase-space parametrisations are given in appendices~\ref{jpsicbar-cqbar-q} and~\ref{jpsicbar-cgg}, respectively.
The integral for \(\mathrm{RV}_{\mathrm F}\) follows from \eqref{eq:finite-nlo-real-virtual} by replacing \(F_R\) with \(F_{\mathrm{RV}}^{\rm fin}\).
Its soft counterterm is constructed in appendix~\ref{local-soft-subtraction}, and its numerical integral uses the single-gluon parametrisation of appendix~\ref{jpsicbar-cg}.

\Needspace{4\baselineskip}
\subsubsection{Single-unresolved contribution (SU)}\label{single-unresolved-terms}\label{su-pole-cancellation}

The integrated counterterm defined in \eqref{eq:rr-single-integrated-definition} is \(F_{\mathrm{RR,SU}}=I_{\mathrm{RR,SU}}F_R\), with
\begin{equation}
I_{\mathrm{RR,SU}}(k,\epsilon)
 =g_s^2\mathcal K_{\Sop_4}
       +\mathcal C_{gg}(E_k^*,\epsilon)
       +n_l\mathcal C_{q\bar q}(E_k^*,\epsilon)
       -\mathcal C_{\Cop\Sop_4}(E_k^*,\epsilon).
\label{eq:4.8}
\end{equation}
Here \(g_s^2\mathcal K_{\Sop_4}\) is the integrated single-soft kernel, and \(\mathcal C_{gg}\), \(\mathcal C_{q\bar q}\) are the collinear kernels.
The kernel \(\mathcal C_{\Cop\Sop_4}\) subtracts the overlap of the gluon single-soft and collinear terms.
Their explicit integrals are given in \eqref{eq:3.15}, \eqref{eq:3.25}, \eqref{eq:3.28} and \eqref{eq:3.33}.
The remaining gluon momentum \(k\) is the massless parent momentum \(p\) for a collinear splitting, or \(k_3\) for single-soft emission.

Adding the RV infrared term and subtracting the remaining soft limit gives \(\mathrm{SU}\) in \eqref{eq:finite-su}.
The operator \(\Sop_k\) acts on \(d\Phi_R[I_{\mathrm{RR,SU}}+2a_s\operatorname{Re}Z_{c\bar cg}^{(1)}]F_R\) and gives a counterterm on \(d\Phi_B[dk]_K\).

In the double-soft counterterms, the separate bounds \(E_3^*,E_4^*<E_0^*\) allow the parent energy \(E^*=E_3^*+E_4^*\) to exceed \(E_0^*\).
We therefore split the double-soft--collinear overlap \(\Cop\Sop_{34}\) at \(E^*=E_0^*\).
The low-energy part, \(\Cop\Sop_<\), supplies the collinear part of the soft subtraction in \eqref{eq:finite-su}: only the splitting variables are integrated, with the parent momentum left unintegrated in \(d\Phi_B[dp]_K\).
The high-energy part, \(\Cop\Sop_>\), is fully integrated and subtracted from the double-soft contribution in \(\mathrm{DU}\) on \(d\Phi_B\).
The partially and fully integrated expressions are given in \eqref{eq:3.18} and \eqref{eq:3.30}.
With this domain split, the RR decomposition can be written as the operator identities
\[
\begin{aligned}
1={}&(1-\Cop)(1-\Sop_{34})
 +(\Cop-\Cop\Sop_<)+(\Sop_{34}-\Cop\Sop_>),
 &&q\bar q,\\[3pt]
1={}&(1-\Cop)(1-\Sop_4)(1-\Sop_{34})\\
 &+\bigl[\Cop-\Cop\Sop_<+(1-\Cop)\Sop_4(1-\Sop_{34})\bigr]
 +(\Sop_{34}-\Cop\Sop_>),
 &&gg.
\end{aligned}
\]
Each identity acts on the squared matrix element and phase space of the corresponding double-real channel.
The three groups give \(\mathrm{RR}_{\mathrm F}\), \(\mathrm{RR}_{\mathrm{SU}}\) and \(\mathrm{RR}_{\mathrm{DU}}\), respectively.
The negative overlaps in the middle group yield \(-\Sop_k\{d\Phi_R F_{\mathrm{RR,SU}}\}\) after one unresolved integration:
\begin{equation}
\begin{aligned}
\Sop_k\!\left\{d\Phi_R F_{\mathrm{RR,SU}}\right\}
 ={}&\sum_{ab}\int_{\rm unres}\!\Cop\Sop_<
   \bigl\{d\Phi_{\mathrm{RR}}^{ab}F_{ab}\bigr\}\\
 &+\int_{\rm unres}\!(1-\Cop)\Sop_4\Sop_{34}
   \bigl\{d\Phi_{\mathrm{RR}}^{gg}F_{gg}\bigr\}.
\end{aligned}
\label{eq:rr-single-soft-origin}
\end{equation}
The first term leaves the parent momentum unintegrated on \(d\Phi_B[dp]_K\); the second leaves the harder-gluon momentum on \(d\Phi_B[dk_3]_K\).

The pole parts of the integrated RR kernels are
\begin{equation}
\begin{gathered}
\begin{aligned}
\bigl[g_s^2\mathcal K_{\Sop_4}\bigr]_{\rm poles}
 &=a_s\Biggl\{\frac{2C_A}{\epsilon^2}
  +\frac1\epsilon\biggl[
       4C_F-\left(C_F-\frac{C_A}{2}\right)
         \operatorname{Re}\gamma_{{\rm cusp},Q}^{(0)}(v_{12})\\
 &\qquad
       +2C_A\ln\frac{m_c^2\mu_R^2}
                   {4(p_1\cdot k)(p_2\cdot k)}
       \biggr]\Biggr\},
\end{aligned}\\[3pt]
\bigl[\mathcal C_{gg}-\mathcal C_{\Cop\Sop_4}\bigr]_{\rm poles}
 =\frac{a_s}{\epsilon}\frac{11C_A}{3},
\qquad
\bigl[n_l\mathcal C_{q\bar q}\bigr]_{\rm poles}
 =-\frac{a_s}{\epsilon}\frac{4T_Fn_l}{3}.
\end{gathered}
\label{eq:4.9}
\end{equation}
Using the infrared factor in \eqref{eq:3.35}, we obtain
\begin{equation}
\bigl[I_{\mathrm{RR,SU}}\bigr]_{\rm poles}
 =\bigl[g_s^2\mathcal K_{\Sop_4}\bigr]_{\rm poles}
   +\frac{a_s\beta_0^{(3)}}{\epsilon}
 =-2a_s\operatorname{Re}Z_{c\bar cg}^{(1)}.
\label{eq:4.10}
\end{equation}

For the remaining soft subtraction, we define \(I_{\mathrm{RR,SU}}^{\mathrm S}(k,\epsilon)\) by
\[
\Sop_k\!\left[I_{\mathrm{RR,SU}}F_R\right]
 =I_{\mathrm{RR,SU}}^{\mathrm S}\,\Sop_kF_R.
\]
The kernel has the form of \eqref{eq:4.8} with Born heavy-quark momenta, \(p_1+p_2=K\) and \(v_{12}=2\beta/(1+\beta^2)\), while \(k\) remains an integration variable on \(d\Phi_B[dk]_K\).
Its poles are
\begin{equation}
\begin{aligned}
\bigl[I_{\mathrm{RR,SU}}^{\mathrm S}\bigr]_{\rm poles}
 &=a_s\Biggl\{\frac{2C_A}{\epsilon^2}
 +\frac1\epsilon\biggl[\beta_0^{(3)}
       +\left(C_F-\frac{C_A}{2}\right)A(\beta)\\
 &\qquad
       +2C_A\left(1+\ln\frac{m_c^2\mu_R^2}
                            {4(p_1\cdot k)(p_2\cdot k)}\right)
       \biggr]\Biggr\},
\end{aligned}
\label{eq:su-rr-soft-poles}
\end{equation}

The soft limit of the RV infrared term follows from one-loop soft factorisation~\cite{ref6},
\begin{equation}
\Sop_k\!\left[2a_s\operatorname{Re}Z_{c\bar cg}^{(1)}F_R\right]
 =a_s\Bigl\{2\operatorname{Re}Z_1
   +\bigl[R_{12}^{(1)}(k)+Z_{\alpha_s}^{(1)}+Z_A^{(1)}\bigr]_{\rm poles}
   \Bigr\}\,\Sop_kF_R.
\label{eq:su-rv-soft-counterterm}
\end{equation}
Here \(R_{12}^{(1)}\) is the massive one-loop soft-current correction, and \(Z_{\alpha_s}^{(1)}\), \(Z_A^{(1)}\) are the coupling and gluon-field renormalisation coefficients in \eqref{eq:3.38}.
At the same Born kinematics, their pole parts are
\begin{equation}
\begin{aligned}
\bigl[R_{12}^{(1)}(k)\bigr]_{\rm poles}
 &=-\frac{2C_A}{\epsilon^2}
   -\frac{2C_A}{\epsilon}\left[1-\frac{A(\beta)}4
       +\ln\frac{m_c^2\mu_R^2}{4(p_1\cdot k)(p_2\cdot k)}\right],\\[3pt]
\bigl[Z_{\alpha_s}^{(1)}+Z_A^{(1)}\bigr]_{\rm poles}
 &=-\frac{\beta_0^{(3)}}{\epsilon}.
\end{aligned}
\label{eq:su-rv-soft-poles}
\end{equation}

Together with \(2\operatorname{Re}Z_1=-C_FA(\beta)/\epsilon\), these give
\[
\bigl[I_{\mathrm{RR,SU}}^{\mathrm S}\bigr]_{\rm poles}
 +a_s\Bigl\{2\operatorname{Re}Z_1
   +\bigl[R_{12}^{(1)}(k)+Z_{\alpha_s}^{(1)}+Z_A^{(1)}\bigr]_{\rm poles}
   \Bigr\}=0.
\]
Thus the poles cancel separately on \(d\Phi_R(k)\) and \(d\Phi_B[dk]_K\).
The complete soft-subtracted combination consequently obeys
\begin{equation}
\Bigl[(1-\Sop_k)\bigl\{d\Phi_R(k)\,
      \bigl[I_{\mathrm{RR,SU}}+2a_s\operatorname{Re}Z_{c\bar cg}^{(1)}\bigr]
       F_R\bigr\}\Bigr]_{\rm poles}=0.
\label{eq:4.11}
\end{equation}

\Needspace{4\baselineskip}
\subsubsection{Finite-remainder contribution (FR)}\label{terms-proportional-to-the-finite-one-loop-remainder}

The part of the integrated RV soft counterterm \eqref{eq:3.40} proportional to \(F_V^{\rm fin}\) is \(F_{\mathrm{RV,FR}}=a_sC_F\mathcal I_{\mathrm S}F_V^{\rm fin}\).
Combining it with \(F_{\mathrm{VV,FR}}=2a_s\operatorname{Re}Z_1F_V^{\rm fin}\) in \eqref{eq:2.10} gives \(\mathrm{FR}\) in \eqref{eq:finite-fr}.
The poles cancel by the same identity \eqref{eq:4.6}.

\Needspace{4\baselineskip}
\subsubsection{Double-unresolved contribution (DU)}\label{double-unresolved-terms}\label{du-pole-cancellation}

For the Born-proportional terms in \(\mathrm{DU}\), we define the kernels by
\[
\begin{aligned}
F_{\mathrm{RR,DU}}&=a_s^2 I_{\mathrm{RR,DU}}F_B,\\
F_{\mathrm{RV,DU}}&=a_s^2\bigl(I_{\mathrm{RV,cur}}+I_{\mathrm{RV},Z}\bigr)F_B,\\
F_{\mathrm{VV,DU}}&=a_s^2 I_{\mathrm{VV,DU}}F_B.
\end{aligned}
\]
Integrating the double-soft counterterms minus the high-energy overlaps, \(\Sop_{34}-\Cop\Sop_>\), over both unresolved partons gives the RR kernel
\begin{equation}
\begin{aligned}
I_{\mathrm{RR,DU}}
 ={}&\frac{C_F^2}{2}\mathcal I_{\mathrm S}^2
       -8C_FC_A\mathcal A_S(f_{AA}^{gg}-f_{AB}^{gg})\\[3pt]
 &+16C_FT_Fn_l\mathcal A_S
                  (f_{AA}^{q\bar q}-f_{AB}^{q\bar q})
       -\frac1{a_s^2}
       \bigl(\mathcal K_{\Cop\Sop,g}
       +n_l\mathcal K_{\Cop\Sop,q}\bigr).
\end{aligned}
\label{eq:4.13}
\end{equation}
The first term describes independent soft-gluon emissions; the functions \(f_{ij}^{gg}\) and \(f_{ij}^{q\bar q}\) describe correlated double-soft emission, with normalisation \(\mathcal A_S\).
The labels \(AA\) and \(AB\) denote equal and distinct emitters, respectively.
The kernels \(\mathcal K_{\Cop\Sop,g}\) and \(\mathcal K_{\Cop\Sop,q}\) are the integrals of the overlaps over \(E^*>E_0^*\).
These terms are given in \eqref{eq:3.11}, \eqref{eq:3.18}, \eqref{eq:3.23} and \eqref{eq:3.30}.
The RV soft-current and renormalisation terms in \eqref{eq:3.40}, together with the VV infrared terms in \eqref{eq:2.10}, give
\begin{equation}
\begin{aligned}
I_{\mathrm{RV,cur}}
 &=4C_FC_Ae^{2\epsilon L_S}
       \sum_{n=-3}^{0}\epsilon^n\mathcal R_n(\beta)+O(\epsilon),\\
I_{\mathrm{RV},Z}
 &=C_F\mathcal I_{\mathrm S}
       \bigl[Z_{\alpha_s}^{(1)}+Z_A^{(1)}
                         +2\operatorname{Re}Z_1\bigr],\\
I_{\mathrm{VV,DU}}
 &=2\operatorname{Re}Z_2+|Z_1|^2.
\end{aligned}
\label{eq:4.14}
\end{equation}

The coefficients \(\mathcal R_n\) are defined by the integrated one-loop soft current in \eqref{eq:3.39} and given explicitly in appendix~\ref{integrated-one-loop-soft-current-coefficients}.
Using the single-soft coefficients \(A(\beta)\) and \(B_s(\beta)\) in \eqref{eq:3.7}, we extract the poles with
\begin{equation}
\begin{aligned}
C_F\mathcal I_{\mathrm S}
 &=-\frac{\operatorname{Re}\Gamma_0}{\epsilon}
       +C_F\bigl[A(\beta)L_S+B_s(\beta)\bigr]+O(\epsilon),\\[2pt]
Z_{\alpha_s}^{(1)}+Z_A^{(1)}
 &=-\frac{\beta_0^{(3)}}{\epsilon}
       +\beta_{0,Q}\ln\frac{\mu_R^2}{m_c^2}+O(\epsilon).
\end{aligned}
\label{eq:4.15}
\end{equation}
Expanding \(I_{\mathrm{RR,DU}}\) and \(I_{\mathrm{RV,cur}}\) gives \cite{ref6,ref7,ref10}
\begin{equation}
\begin{gathered}
\bigl[I_{\mathrm{RR,DU}}\bigr]_{\epsilon^{-3}}
 =-C_A\operatorname{Re}\Gamma_0,
\qquad
\bigl[I_{\mathrm{RV,cur}}\bigr]_{\epsilon^{-3}}
 =+C_A\operatorname{Re}\Gamma_0,\\[4pt]
\begin{aligned}
\bigl[I_{\mathrm{RR,DU}}+I_{\mathrm{RV,cur}}\bigr]_{\rm poles}
 &=\frac{\operatorname{Re}\Gamma_0
              (\operatorname{Re}\Gamma_0-\beta_0^{(3)})}{2\epsilon^2}
       -\frac{\operatorname{Re}\Gamma_1}{2\epsilon}\\[3pt]
 &\quad-\frac{C_F(\operatorname{Re}\Gamma_0-\beta_0^{(3)})
                  [A(\beta)L_S+B_s(\beta)]}{\epsilon}.
\end{aligned}
\end{gathered}
\label{eq:4.16}
\end{equation}

Using \eqref{eq:4.15} and \eqref{eq:4.5} gives
\begin{equation}
\begin{aligned}
\bigl[I_{\mathrm{RV},Z}\bigr]_{\rm poles}
 ={}&-\frac{\operatorname{Re}\Gamma_0
                (\operatorname{Re}\Gamma_0-\beta_0^{(3)})}{\epsilon^2}
 -\frac{\beta_{0,Q}\operatorname{Re}\Gamma_0}{\epsilon}
                         \ln\frac{\mu_R^2}{m_c^2}\\[3pt]
 &+\frac{C_F(\operatorname{Re}\Gamma_0-\beta_0^{(3)})
                   [A(\beta)L_S+B_s(\beta)]}{\epsilon}.
\end{aligned}
\label{eq:4.17}
\end{equation}
From \eqref{eq:2.6},
\begin{equation}
I_{\mathrm{VV,DU}}
 =\frac{\operatorname{Re}\Gamma_0
                (\operatorname{Re}\Gamma_0-\beta_0^{(3)})}{2\epsilon^2}
       +\frac{\operatorname{Re}\Gamma_1}{2\epsilon}
 +\frac{\beta_{0,Q}\operatorname{Re}\Gamma_0}{\epsilon}
                         \ln\frac{\mu_R^2}{m_c^2}.
\label{eq:4.18}
\end{equation}
Combining \eqref{eq:4.16}--\eqref{eq:4.18} gives
\begin{equation}
\bigl[I_{\mathrm{RR,DU}}+I_{\mathrm{RV,cur}}
       +I_{\mathrm{RV},Z}+I_{\mathrm{VV,DU}}\bigr]_{\rm poles}=0.
\label{eq:4.19}
\end{equation}

\FloatBarrier

\Needspace{6\baselineskip}
\section{Local and integrated subtraction terms}\label{real-emission-subtraction}

The kinematics, unresolved-energy bounds and phase-space notation are defined in appendix~\ref{appendix-subtraction-conventions}.
In the locally subtracted combinations, denoted by the superscript \({\rm sub}\), each squared matrix element or interference is integrated with its corresponding full or projected phase space.

\Needspace{5\baselineskip}
\subsection{Single-real contribution (R)}\label{jpsicbar-cg-at-tree-level}

The squared matrix element \(F_R\) is singular only when the gluon momentum \(k\) becomes soft; the charm mass regulates the heavy-quark collinear limits.

\paragraph{Local soft subtraction.}
The soft counterterm uses \(d\Phi_B[dk]_K\), with \(0<E_k^*<E_0^*\) and Born heavy momenta satisfying \(p_1+p_2=K\).
The real-emission contribution is integrated over \(d\Phi_R\), with \(p_1+p_2=K-k\).
Soft-gluon emission from the two massive quarks is described by the eikonal factor
\begin{equation}
\mathcal S_{12}(k)=
\frac{2p_1\cdot p_2}{(p_1\cdot k)(p_2\cdot k)}
-\frac{m_c^2}{(p_1\cdot k)^2}
-\frac{m_c^2}{(p_2\cdot k)^2}.
\label{eq:3.5}
\end{equation}
The corresponding soft limit is
\[
\Sop_kF_R=g_s^2\mathcal C_\epsilon C_F\mathcal S_{12}(k)F_B.
\]
The local subtraction is abbreviated as \(F_R^{\rm sub}=F_R-\Sop_kF_R\).
The finite integral \(R_{\mathrm F}\) is expressed in common phase-space variables in \eqref{eq:B.12}.

\paragraph{Integrated soft counterterm.}
Integration over the unresolved gluon gives \begin{equation}
F_{R,U}=\int[dk]_K\,\Sop_k F_R
=\frac{g_s^2C_F}{(4\pi)^2}
\left[\frac{A(\beta)}{\epsilon}
      +A(\beta)\ln\frac{\mu_R^2}{4(E_0^*)^2}
      +B_s(\beta)+O(\epsilon)\right]F_B,
\label{eq:3.6}
\end{equation} where \begin{equation}
\begin{aligned}
A(\beta)&=4+\frac{2(1+\beta^2)}{\beta}
                    \ln\frac{1-\beta}{1+\beta},\\
B_s(\beta)&=-\frac4\beta\ln\frac{1-\beta}{1+\beta}
-\frac{2(1+\beta^2)}{\beta}
 \left[\operatorname{Li}_2\!\left(\frac{2\beta}{1+\beta}\right)
       -\operatorname{Li}_2\!\left(\frac{-2\beta}{1-\beta}\right)\right].
\end{aligned}
\label{eq:3.7}
\end{equation}

At NNLO, the soft integral is needed beyond \(O(\epsilon^0)\).
Writing \(L_S=\ln[\mu_R^2/(4(E_0^*)^2)]\), the kernel defined by \(F_{R,U}=a_sC_F\mathcal I_{\mathrm S}F_B\) and its angular function are
\begin{align}
\mathcal I_{\mathrm S}(\beta,\epsilon,L_S)&=-\frac{4e^{\epsilon(\gamma_E+L_S)}\Gamma(1-\epsilon)}
          {\epsilon\Gamma(2-2\epsilon)}h(\beta,\epsilon),
\label{eq:A.8}\\*[3pt]
h(\beta,\epsilon)&=(1+\beta^2-2\epsilon)
 {}_2F_1\!\left(1,\frac12;\frac32-\epsilon;\beta^2\right)
                    -(1-2\epsilon).
\label{eq:A.3}
\end{align}
The integrated counterterm combines with the virtual infrared term to give \(U\) in appendix~\ref{nlo-fr-pole-cancellation}; the same kernel enters \(\mathrm{FR}\) in appendix~\ref{terms-proportional-to-the-finite-one-loop-remainder}.

\Needspace{5\baselineskip}
\subsection{Double-real contribution (RR)}\label{double-real-contribution}

\Needspace{5\baselineskip}
\subsubsection{\texorpdfstring{Light-quark-pair emission: \(q\bar q\)}{Light-quark-pair emission: q qbar}}\label{jpsicbar-cqbar-q-at-tree-level}

For an additional \(q(k_3)\bar q(k_4)\) pair of a given flavour, the squared matrix element \(F_{q\bar q}\) has double-soft (\(\Sop_{34}\)) and pair-collinear (\(\Cop\)) singularities.
They are subtracted through \begin{equation}
\begin{aligned}
F_{q\bar q}^{\rm sub}&=(1-\Cop)(1-\Sop_{34})F_{q\bar q}\\
 &=F_{q\bar q}-\Sop_{34}F_{q\bar q}-\Cop F_{q\bar q}+\Cop\Sop_{34}F_{q\bar q}.
\end{aligned}
\label{eq:3.8}
\end{equation}

\Needspace{4\baselineskip}
\paragraph{Double-soft subtraction.}\label{double-soft-subtraction}

The factorised phase space is \(d\Phi_B[dk_3]_K[dk_4]_K\), with independent energies \(0<E_3^*<E_0^*\) and \(0<E_4^*<E_0^*\) and both directions integrated over their full solid angles.
The heavy quarks have back-to-back Born momenta in the \(K\) rest frame.

The local double-soft counterterm is \begin{equation}
\Sop_{34}F_{q\bar q}=g_s^4\mathcal C_\epsilon^2 C_FT_F
     \bigl(\mathcal I_{11}-\mathcal I_{12}
                    -\mathcal I_{21}+\mathcal I_{22}\bigr)F_B,
\label{eq:3.9}
\end{equation} with \begin{equation}
\begin{gathered}
\mathcal I_{ij}=
\frac{(p_i\cdot k_3)(p_j\cdot k_4)
     +(p_i\cdot k_4)(p_j\cdot k_3)
     -(p_i\cdot p_j)(k_3\cdot k_4)}
     {(k_3\cdot k_4)^2(p_i\cdot k_{34})(p_j\cdot k_{34})},\\
k_{34}=k_3+k_4.
\end{gathered}
\label{eq:3.10}
\end{equation}

Integrating over both unresolved momenta gives \cite{ref7} \begin{equation}
\int[dk_3]_K[dk_4]_K\,\Sop_{34}F_{q\bar q}
=\frac{g_s^4C_FT_F}{(4\pi)^4}
       16\mathcal A_S(f_{AA}^{q\bar q}-f_{AB}^{q\bar q})F_B.
\label{eq:3.11}
\end{equation} The common double-soft normalisation is
\begin{equation}
\begin{aligned}
\mathcal A_S(\epsilon,L_S)
 &=\frac{e^{2\epsilon(\gamma_E+L_S)}\Gamma^2(1-\epsilon)}
     {\Gamma^2(2-2\epsilon)}.
\end{aligned}
\label{eq:double-soft-normalisation}
\end{equation}
The functions \(f_{ij}^{q\bar q}\) are expanded through \(O(\epsilon^0)\) in appendix~\ref{double-soft-functions}.

\Needspace{4\baselineskip}
\paragraph{Collinear subtraction.}\label{pair-collinear-subtraction}

We define \begin{equation}
E^*=E_3^*+E_4^*,\qquad z=\frac{E_3^*}{E^*},\qquad
\eta=\frac{1-\cos\theta_{34}^*}{2}.
\label{eq:3.12}
\end{equation} The collinear projection replaces \(k_3+k_4\) in the recoil by the massless parent \(p=k_3/z\): \[
\begin{aligned}
&\frac{dK^2}{2\pi}\,d\Phi_2(Q;P,K)\,
d\Phi_2(K-k_3-k_4;p_1,p_2)\,[dk_3]_K[dk_4]_K\\
&\quad\xrightarrow{k_3+k_4\mapsto p}\quad
\frac{dK^2}{2\pi}\,d\Phi_2(Q;P,K)\,
d\Phi_2(K-p;p_1,p_2)\,[dk_3]_K[dk_4]_K.
\end{aligned}
\] The heavy-pair threshold in \(d\Phi_2(K-p;p_1,p_2)\) requires
\((K-p)^2=W^2-2WE^*\geq4m_c^2\), hence \(E^*\leq E_0^*\).

Taking the spatial direction of \(p\) as the polar axis in the \(K\) frame, let \(d\Omega_\perp^*\) be the solid-angle element for the transverse direction of \(k_4\), on the unit sphere in \(d-2\) dimensions.
In four dimensions, \(d\Omega_\perp^*=d\phi_\perp^*\) with \(0\leq\phi_\perp^*<2\pi\).
The energy and angular changes of variables give \[
\begin{gathered}
dE_3^*\,dE_4^*=E^*\,dE^*\,dz,\\
d\Omega_4^*
 =(\sin\theta_{34}^*)^{1-2\epsilon}
       d\theta_{34}^*\,d\Omega_\perp^*
 =2^{1-2\epsilon}[\eta(1-\eta)]^{-\epsilon}
       d\eta\,d\Omega_\perp^*.
\end{gathered}
\] Following ref.~\cite{ref6}, appendix D.1, the collinear projection sets the smooth factor \((1-\eta)^{-\epsilon}\) to one and retains \(\eta^{-\epsilon}\).
Factoring out \([dp]_K\) gives \(d\Phi_R(p)d\Phi_{\Cop,\rm unres}\), with \(0<z,\eta<1\) and the full transverse angular range at fixed \(p\):
\begin{equation}
\begin{aligned}
d\Phi_{\Cop,\rm unres}
 &=\frac{(E^*)^{2-2\epsilon}[z(1-z)]^{1-2\epsilon}}
        {2(2\pi)^{3-2\epsilon}}\,dz\,
 2^{1-2\epsilon}\eta^{-\epsilon}d\eta\,d\Omega_\perp^*,\\
\int d\Omega_\perp^*
 &=\frac{2\pi^{1-\epsilon}}{\Gamma(1-\epsilon)}.
\end{aligned}
\label{eq:3.13}
\end{equation}

With \(n=2u_K-p/E^*\) and \(g_\perp^{\mu\nu}=g^{\mu\nu}-(p^\mu n^\nu+n^\mu p^\nu)/(p\cdot n)\), the component of \(k_4\) transverse to the parent direction is \(k_\perp^\mu=g_\perp^\mu{}_{\nu}k_4^\nu\).
Its squared norm is
\[
k_\perp^2=-(E_4^*)^2\sin^2\theta_{34}^*
          =-4(1-z)^2(E^*)^2\eta(1-\eta).
\]
The spin-correlated splitting tensor and local counterterm are \begin{equation}
\begin{aligned}
P_{q\bar q}^{\mu\nu}
 &=T_F\left[-g_\perp^{\mu\nu}
       +4z(1-z)\frac{k_\perp^\mu k_\perp^\nu}{k_\perp^2}\right],\\
\Cop\,F_{q\bar q}
 &=\frac{g_s^2\mathcal C_\epsilon}{k_3\cdot k_4}
         P_{q\bar q}^{\mu\nu}F_{R,\mu\nu}.
\end{aligned}
\label{eq:3.14}
\end{equation} Here \(F_{R,\mu\nu}\) is the spin-correlated squared matrix element for the reduced single-real process, with the parent-gluon polarisation indices left uncontracted.
The polarisation sum gives \(F_R(p)=-g_\perp^{\mu\nu}F_{R,\mu\nu}\).
The scalar product is \(k_3\cdot k_4=2z(1-z)(E^*)^2\eta\).

At fixed reduced kinematics, \(F_{R,\mu\nu}\) is independent of the transverse direction.
Rotational symmetry gives the normalised angular average \cite{ref6}
\[
\left\langle\frac{k_\perp^\mu k_\perp^\nu}{k_\perp^2}\right\rangle_\perp
\equiv\frac{\displaystyle\int d\Omega_\perp^*\,
                         k_\perp^\mu k_\perp^\nu/k_\perp^2}
             {\displaystyle\int d\Omega_\perp^*}
=\frac{g_\perp^{\mu\nu}}{d-2},
\]
so \(\langle P_{q\bar q}^{\mu\nu}F_{R,\mu\nu}\rangle_\perp
=T_F[1-2z(1-z)/(1-\epsilon)]F_R(p)\).
Including the solid-angle volume in \eqref{eq:3.13} and integrating over \(z\) and \(\eta\) gives, with \(L_C=\ln[\mu_R^2/(4(E^*)^2)]\), \begin{equation}
\begin{gathered}
\int d\Phi_{\Cop,\rm unres}\,\Cop F_{q\bar q}
 =\mathcal C_{q\bar q}(E^*,\epsilon)F_R(p),\\
\begin{aligned}
\mathcal C_{q\bar q}(E^*,\epsilon)
={}&\frac{g_s^2T_F}{(4\pi)^2}
\biggl[-\frac{4}{3\epsilon}-\frac43L_C-\frac{46}{9}\\
&\quad+\epsilon\left(-\frac23L_C^2-\frac{46}{9}L_C
                       -\frac{562}{27}+6\zeta_2\right)
 +O(\epsilon^2)\biggr].
\end{aligned}
\end{gathered}
\label{eq:3.15}
\end{equation} The expression valid to all orders in \(\epsilon\), needed for the subsequent soft integration, is~\cite{ref6}
\begin{equation}
\begin{aligned}
\mathcal C_{q\bar q}(E^*,\epsilon)
={}&\frac{g_s^2T_F}{(4\pi)^2}
 \left(\frac{\mu_R^2}{4(E^*)^2}\right)^\epsilon
 \frac{2[\epsilon(10-8\epsilon)-4]}{\epsilon}
 \frac{e^{\gamma_E\epsilon}\Gamma(1-2\epsilon)\Gamma(2-2\epsilon)}
 {\Gamma(4-4\epsilon)\Gamma(2-\epsilon)}.
\end{aligned}
\label{eq:A.1}
\end{equation}

\Needspace{4\baselineskip}
\paragraph{Soft-collinear overlap and its integration domain.}\label{soft-collinear-overlap}

To construct the overlap \(\Cop\Sop_{34}\), we first apply the double-soft projection,
\[
\begin{aligned}
&\frac{dK^2}{2\pi}\,d\Phi_2(Q;P,K)\,
d\Phi_2(K-k_3-k_4;p_1,p_2)\,[dk_3]_K[dk_4]_K\\
&\quad\xrightarrow{\Sop_{34}}\quad d\Phi_B[dk_3]_K[dk_4]_K,
\end{aligned}
\] so the reduced heavy momenta satisfy \(p_1+p_2=K\).
We use \(E_3^*=zE^*\), \(E_4^*=(1-z)E^*\) and \(p=k_3/z\), with \(dE_3^*dE_4^*=E^*dE^*dz\), and define the one-particle phase-space element for the parent as \(d\Pi(p)=d^dp\,\delta_+(p^2)/(2\pi)^{d-1}\).
The same collinear angular projection \((1-\eta)^{-\epsilon}\to1\) then gives \[
\begin{aligned}
d\Phi_B[dk_3]_K[dk_4]_K
 &\xrightarrow{\Cop}d\Phi_B\,d\Phi_{\Cop\Sop,\rm unres},\\
d\Phi_{\Cop\Sop,\rm unres}
 &=d\Pi(p)\,d\Phi_{\Cop,\rm unres}\,
   \theta(E_0^*-zE^*)\theta(E_0^*-(1-z)E^*).
\end{aligned}
\] Here \(d\Phi_{\Cop,\rm unres}\) is given in \eqref{eq:3.13}.
The Born phase space is independent of \(p\), so the parent-energy range follows from the bounds on the two daughter energies:
\begin{equation}
0<E^*<E_{\max}^*(z),\qquad
E_{\max}^*(z)=\min\!\left(\frac{E_0^*}{z},\frac{E_0^*}{1-z}\right)
=\frac{E_0^*}{\max(z,1-z)}.
\label{eq:3.16}
\end{equation} The remaining integrations run over \(0<z<1\), \(0<\eta<1\) and the full parent and transverse angular ranges.
We denote the parts of \(\Cop\Sop_{34}\) below and above \(E^*=E_0^*\) by \(\Cop\Sop_<\) and \(\Cop\Sop_>\), respectively.

With the eikonal current \(j^\mu(p)\) and the splitting tensor \eqref{eq:3.14}, the local overlap is \begin{equation}
\begin{aligned}
j^\mu(p)&=\frac{p_1^\mu}{p_1\cdot p}
                -\frac{p_2^\mu}{p_2\cdot p},\\
\Cop\Sop_{34}\,F_{q\bar q}
 &=\frac{g_s^4\mathcal C_\epsilon^2 C_F}
                  {k_3\cdot k_4}
           P_{q\bar q}^{\mu\nu}j_\mu(p)j_\nu(p)F_B.
\end{aligned}
\label{eq:3.17}
\end{equation}

The current obeys \(p\cdot j=0\), so \(-g_\perp^{\mu\nu}j_\mu j_\nu=-j^2=\mathcal S_{12}(p)\).
The same transverse average therefore gives
\(\langle P_{q\bar q}^{\mu\nu}j_\mu j_\nu\rangle_\perp
=T_F[1-2z(1-z)/(1-\epsilon)]\mathcal S_{12}(p)\).
For \(\Cop\Sop_<\), we integrate the splitting variables at fixed \(p\), with \(0<E^*<E_0^*\); for \(\Cop\Sop_>\), we also integrate the parent energy and direction:
\begin{equation}
\begin{gathered}
\begin{aligned}
\int d\Phi_{\Cop,\rm unres}\,\Cop\Sop_< F_{q\bar q}
 &=g_s^2\mathcal C_\epsilon C_F\,
   \mathcal C_{q\bar q}(E^*,\epsilon)\mathcal S_{12}(p)F_B,\\
\int d\Phi_{\Cop\Sop,\rm unres}\,\Cop\Sop_> F_{q\bar q}
 &=\mathcal K_{\Cop\Sop,q}F_B,
\end{aligned}\\
\mathcal K_{\Cop\Sop,q}
=\frac{g_s^4C_FT_F}{(4\pi)^4}
 \frac{4e^{2\epsilon(\gamma_E+L_S)}}
      {\epsilon^2\Gamma(2-2\epsilon)}h(\beta,\epsilon)
      \bigl(\mathcal W_{q,m}-\mathcal W_{q,0}\bigr).
\end{gathered}
\label{eq:3.18}
\end{equation} Here \(\mathcal C_{q\bar q}\) is the exact collinear kernel in \eqref{eq:A.1}.
The angular function \(h(\beta,\epsilon)\) is defined in \eqref{eq:A.3}.
The low- and high-energy overlap terms enter \(\mathrm{SU}\) and \(\mathrm{DU}\), respectively, as detailed in appendix~\ref{single-unresolved-terms}.

\paragraph{Overlap weights.}\label{energy-weights-for-the-double-soft-collinear-overlap}
We use the complete and incomplete beta functions
\[
\mathrm B_x(a,b)=\int_0^x dt\,t^{a-1}(1-t)^{b-1},\qquad
\mathrm B(a,b)=\mathrm B_1(a,b),
\]
with analytic continuation where required.
After factoring out \(T_F\), the quark-pair splitting weight is
\[
w_q(z,\epsilon)=[z(1-z)]^{-2\epsilon}
 \left[1-\frac{2z(1-z)}{1-\epsilon}\right].
\]
The two weights in \eqref{eq:3.18} are
\begin{equation}
\begin{aligned}
\mathcal W_{q,0}&=\int_0^1 dz\,w_q(z,\epsilon)\\
 &=\mathrm B(1-2\epsilon,1-2\epsilon)
 -\frac{2}{1-\epsilon}\mathrm B(2-2\epsilon,2-2\epsilon),\\
\mathcal W_{q,m}&=\int_0^1 dz\,w_q(z,\epsilon)\max(z,1-z)^{4\epsilon}\\
 &=2\left[\mathrm B_{1/2}(1-2\epsilon,1+2\epsilon)
 -\frac{2}{1-\epsilon}\mathrm B_{1/2}(2-2\epsilon,2+2\epsilon)\right].
\end{aligned}
\label{eq:A.4}
\end{equation}
Their difference is
\begin{equation}
\mathcal W_{q,m}-\mathcal W_{q,0}
 =\epsilon\left(\frac83\ln2-\frac{23}{9}\right)
 +\epsilon^2\left(\frac{16}{3}\zeta_2-\frac43\ln2-\frac{245}{27}\right)
 +O(\epsilon^3).
\label{eq:A.5}
\end{equation}

\FloatBarrier

\Needspace{5\baselineskip}
\subsubsection{\texorpdfstring{Two-gluon emission: \(gg\)}{Two-gluon emission: gg}}\label{jpsicbar-cgg-at-tree-level}

For the two gluons \(g(k_3)g(k_4)\) in the ordered sector \(E_3^*>E_4^*\), the squared matrix element \(F_{gg}\) has double-soft (\(\Sop_{34}\)), single-soft (\(\Sop_4\)) and pair-collinear (\(\Cop\)) singularities.
The nested subtraction is \begin{equation}
\begin{aligned}
F_{gg}^{\rm sub}
 &=(1-\Cop)(1-\Sop_4)(1-\Sop_{34})F_{gg}\\
 &=F_{gg}-\Sop_{34}F_{gg}-\Sop_4F_{gg}-\Cop F_{gg}\\
 &\quad+\Sop_4\Sop_{34}F_{gg}+\Cop\Sop_{34}F_{gg}
       +\Cop\Sop_4F_{gg}-\Cop\Sop_4\Sop_{34}F_{gg}.
\end{aligned}
\label{eq:3.19}
\end{equation} The factor \(2\) restoring the two energy-ordered sectors cancels the identical-gluon symmetry factor \(1/2!\) in all expressions below.
Appendix~\ref{jpsicbar-cgg} gives the common coordinates, recoils and Jacobians for all eight terms in \eqref{eq:3.19}.

\Needspace{4\baselineskip}
\paragraph{Double-soft subtraction.}\label{double-soft-subtraction-1}

The double-soft projection removes both gluons from the heavy-pair recoil: \[
\begin{aligned}
d\Phi_{\mathrm{RR}}^{gg}
 &=\frac{dK^2}{2\pi}\,d\Phi_2(Q;P,K)\,
   d\Phi_2(K-k_3-k_4;p_1,p_2)
   [dk_3]_K[dk_4]_K\theta(E_3^*-E_4^*)\\
 &\xrightarrow{\Sop_{34}}
   d\Phi_B[dk_3]_K[dk_4]_K\theta(E_3^*-E_4^*).
\end{aligned}
\] The unresolved energies obey \(0<E_4^*<E_3^*<E_0^*\), with both directions integrated over their full solid angles.

The massive double-soft current of ref.~\cite{ref6} gives \begin{equation}
\begin{gathered}
\Sop_{34}F_{gg}
 =g_s^4\mathcal C_\epsilon^2
 \Bigl[C_F^2\mathcal S_{12}(k_3)\mathcal S_{12}(k_4)
 -C_AC_F\bigl(\mathcal S^{gg}_{11}-\mathcal S^{gg}_{12}
                   -\mathcal S^{gg}_{21}+\mathcal S^{gg}_{22}\bigr)\Bigr]F_B,\\[3pt]
\mathcal S^{gg}_{ij}=\mathcal S^{(0)}_{ij}
              +m_c^2\mathcal S^{(m)}_{ij}+m_c^2\mathcal S^{(m)}_{ji}.
\end{gathered}
\label{eq:3.20}
\end{equation} To specify the correlated part, set \(k_{34}=k_3+k_4\) and \begin{equation}
\begin{aligned}
A_{ij}&=(p_i\cdot k_3)(p_j\cdot k_4)+(p_i\cdot k_4)(p_j\cdot k_3),\\
B_{ij}&=(p_i\cdot k_{34})(p_j\cdot k_{34}),\\
C_{ij}&=(p_i\cdot k_3)(p_j\cdot k_4)(p_i\cdot k_4)(p_j\cdot k_3).
\end{aligned}
\label{eq:3.21}
\end{equation} Then \begin{equation}
\begin{aligned}
\mathcal S^{(0)}_{ij}={}&
\frac{(1-\epsilon)A_{ij}}{(k_3\cdot k_4)^2B_{ij}}
-\frac{(p_i\cdot p_j)^2}{2C_{ij}}\left(2-\frac{A_{ij}}{B_{ij}}\right)\\
&+\frac1{k_3\cdot k_4}\biggl[
\frac{p_i\cdot p_j}{(p_i\cdot k_3)(p_j\cdot k_4)}
+\frac{p_i\cdot p_j}{(p_i\cdot k_4)(p_j\cdot k_3)}
-\frac{p_i\cdot p_j}{B_{ij}}\left(2+\frac{A_{ij}^2}{2C_{ij}}\right)\biggr],\\[4pt]
\mathcal S^{(m)}_{ij}={}&
-\frac1{4(p_i\cdot k_3)(p_i\cdot k_4)(k_3\cdot k_4)}
+\frac{p_i\cdot p_j}{2C_{ij}}\frac{p_j\cdot k_{34}}{p_i\cdot k_{34}}\\
&-\frac1{2(k_3\cdot k_4)B_{ij}}\left[
\frac{(p_j\cdot k_3)^2}{(p_i\cdot k_3)(p_j\cdot k_4)}
+\frac{(p_j\cdot k_4)^2}{(p_i\cdot k_4)(p_j\cdot k_3)}\right].
\end{aligned}
\label{eq:3.22}
\end{equation}

Integrating the double-soft counterterm over the ordered gluon phase space at fixed Born kinematics gives \cite{ref7} \begin{equation}
\int[dk_3]_K[dk_4]_K\,\theta(E_3^*-E_4^*)\Sop_{34}F_{gg}
 =\frac{g_s^4}{(4\pi)^4}
 \biggl[\frac{C_F^2}{2}\mathcal I_{\mathrm S}^2
 -8C_AC_F\mathcal A_S\bigl(f_{AA}^{gg}-f_{AB}^{gg}\bigr)\biggr]F_B.
\label{eq:3.23}
\end{equation} The functions \(\mathcal I_{\mathrm S}\) and \(\mathcal A_S\) are defined in \eqref{eq:A.8} and \eqref{eq:double-soft-normalisation}; the coefficients of \(f_{ij}^{gg}\) are given in appendix~\ref{gluon-emission}.

\Needspace{4\baselineskip}
\paragraph{Single-soft subtraction and double-soft overlap.}\label{single-soft-subtraction-and-its-double-soft-overlap}

The two soft projections give \[
\begin{aligned}
d\Phi_{\mathrm{RR}}^{gg}
 &\xrightarrow{\Sop_4}
 d\Phi_R(k_3)[dk_4]_K\theta(E_3^*-E_4^*),\\
d\Phi_{\mathrm{RR}}^{gg}
 &\xrightarrow{\Sop_4\Sop_{34}}
 d\Phi_B[dk_3]_K[dk_4]_K\theta(E_3^*-E_4^*).
\end{aligned}
\] At fixed \(k_3\), both terms use the same unresolved-gluon phase space, \[
[dk_4]_K\theta(E_3^*-E_4^*)
=\frac{(E_4^*)^{1-2\epsilon}}{2(2\pi)^{3-2\epsilon}}
  dE_4^*\,d\Omega_4^*,\qquad 0<E_4^*<E_3^*<E_0^*,
\] with \(d\Omega_4^*\) covering the full solid angle.

The local single-soft term and its double-soft overlap are \begin{equation}
\begin{gathered}
\begin{aligned}
\Sop_4F_{gg}&=g_s^2\mathcal C_\epsilon\,\mathcal D_4^g F_R(k_3),\\
\Sop_4\Sop_{34}F_{gg}&=g_s^4\mathcal C_\epsilon^2\,
               \mathcal D_4^g C_F\mathcal S_{12}(k_3)F_B,
\end{aligned}\\[4pt]
\mathcal D_4^g=C_F\mathcal S_{12}(k_4)
+C_A\biggl[
\frac{p_1\cdot k_3}{(p_1\cdot k_4)(k_3\cdot k_4)}
+\frac{p_2\cdot k_3}{(p_2\cdot k_4)(k_3\cdot k_4)}
-\frac{p_1\cdot p_2}{(p_1\cdot k_4)(p_2\cdot k_4)}\biggr].
\end{gathered}
\label{eq:3.24}
\end{equation} The momenta in each counterterm are evaluated on its corresponding phase space.

Integrating over the softer-gluon momentum at fixed \(k_3\) gives \begin{equation}
\begin{gathered}
\begin{aligned}
\int[dk_4]_K\theta(E_3^*-E_4^*)\,\Sop_4F_{gg}
 &=g_s^2\mathcal K_{\Sop_4}F_R(k_3),\\
\int[dk_4]_K\theta(E_3^*-E_4^*)\,\Sop_4\Sop_{34}F_{gg}
 &=g_s^4\mathcal C_\epsilon C_F\mathcal K_{\Sop_4}
                  \mathcal S_{12}(k_3)F_B,
\end{aligned}\\
\begin{aligned}
\mathcal K_{\Sop_4}={}&C_F[2\mathcal J(p_1,p_2)-\mathcal J(p_1,p_1)
                                      -\mathcal J(p_2,p_2)]\\
&+C_A[\mathcal J(p_1,k_3)+\mathcal J(p_2,k_3)-\mathcal J(p_1,p_2)] .
\end{aligned}
\end{gathered}
\label{eq:3.25}
\end{equation} Here \(\mathcal J(a,b)\equiv\mathcal J(a,b;E_3^*)\) is defined in appendix~\ref{single-soft-insertion-with-a-surviving-gluon}.
The remaining integrations are over \(d\Phi_R(k_3)\) and \(d\Phi_B[dk_3]_K\), respectively; the corresponding terms are combined in appendix~\ref{single-unresolved-terms}.

\Needspace{4\baselineskip}
\paragraph{Collinear subtraction.}\label{pair-collinear-subtraction-1}

In this ordered channel we define the energy fraction of the softer gluon by \(z=E_4^*/E^*\), with \(E^*=E_3^*+E_4^*\), so that \(0<z<1/2\).
The collinear projection replaces \(k_3+k_4\) in the recoil by the massless parent \(p=k_3/(1-z)\): \[
d\Phi_{\mathrm{RR}}^{gg}
 \xrightarrow{k_3+k_4\mapsto p}
 \frac{dK^2}{2\pi}\,d\Phi_2(Q;P,K)\,d\Phi_2(K-p;p_1,p_2)
 [dk_3]_K[dk_4]_K\theta(E_3^*-E_4^*).
\] With \(E_3^*=(1-z)E^*\), \(E_4^*=zE^*\) and \(\eta=(1-\cos\theta_{34}^*)/2\), the energy Jacobian and angular projection are the same as in appendix~\ref{jpsicbar-cqbar-q-at-tree-level}.
Factoring out \([dp]_K\) gives \(d\Phi_R(p)d\Phi_{\Cop,\rm unres}^{gg}\), with
\[
d\Phi_{\Cop,\rm unres}^{gg}
=d\Phi_{\Cop,\rm unres}\,\theta\!\left(\frac12-z\right).
\] Thus \eqref{eq:3.13} is integrated over \(0<z<1/2\), \(0<\eta<1\) and the full transverse solid angle at fixed \(p\).

Using the transverse projector of appendix~\ref{jpsicbar-cqbar-q-at-tree-level}, we have \(k_\perp^\mu=g_\perp^\mu{}_{\nu}k_4^\nu\) and \(k_\perp^2=-4z^2(E^*)^2\eta(1-\eta)\).
The splitting tensor is \begin{equation}
P_{gg}^{\mu\nu}=2C_A\left[
 -g_\perp^{\mu\nu}\left(\frac{z}{1-z}+\frac{1-z}{z}\right)
 -2(1-\epsilon)z(1-z)
                  \frac{k_\perp^\mu k_\perp^\nu}{k_\perp^2}\right],
\label{eq:3.26}
\end{equation} and the local counterterm is \begin{equation}
\begin{aligned}
\Cop F_{gg}&=\frac{g_s^2\mathcal C_\epsilon}{k_3\cdot k_4}
                     P_{gg}^{\mu\nu}F_{R,\mu\nu}(p),\\
k_3\cdot k_4&=2z(1-z)(E^*)^2\eta .
\end{aligned}
\label{eq:3.27}
\end{equation}

Transverse averaging gives
\[
\langle P_{gg}^{\mu\nu}F_{R,\mu\nu}(p)\rangle_\perp
=2C_A\left[\frac{z}{1-z}+\frac{1-z}{z}+z(1-z)\right]F_R(p).
\]
Integrating over the ordered splitting variables gives \begin{equation}
\begin{gathered}
\int d\Phi_{\Cop,\rm unres}^{gg}\,\Cop F_{gg}
 =\mathcal C_{gg}(E^*,\epsilon)F_R(p),\\
\mathcal C_{gg}(E^*,\epsilon)
 =\frac{g_s^2C_A}{(4\pi)^2}
 \biggl[\frac{2}{\epsilon^2}+\frac{2L_C+11/3}{\epsilon}
 +L_C^2+\frac{11}{3}L_C+\frac{134}{9}-9\zeta_2
 +O(\epsilon)\biggr].
\end{gathered}
\label{eq:3.28}
\end{equation} The higher-order terms in \(\epsilon\) follow from
\begin{equation}
\begin{aligned}
\mathcal C_{gg}(E^*,\epsilon)
={}&\frac{g_s^2C_A}{(4\pi)^2}
 \left(\frac{\mu_R^2}{4(E^*)^2}\right)^\epsilon
 \frac{6(2-3\epsilon)}{\epsilon^2}
 \frac{e^{\gamma_E\epsilon}\Gamma^2(2-2\epsilon)}
 {\Gamma(4-4\epsilon)\Gamma(1-\epsilon)}.
\end{aligned}
\label{eq:A.2}
\end{equation}

\Needspace{4\baselineskip}
\paragraph{Double-soft--collinear overlap.}\label{double-soft-collinear-overlap}

Starting from the double-soft phase space, the same change of variables and angular projection give \[
\begin{aligned}
d\Phi_B[dk_3]_K[dk_4]_K\theta(E_3^*-E_4^*)
 &\xrightarrow{\Cop}d\Phi_B\,d\Phi_{\Cop\Sop,\rm unres}^{gg},\\
d\Phi_{\Cop\Sop,\rm unres}^{gg}
 &=d\Pi(p)\,d\Phi_{\Cop,\rm unres}^{gg}
                  \theta(E_0^*-(1-z)E^*).
\end{aligned}
\]
The reduced heavy momenta satisfy \(p_1+p_2=K\).
The ordered daughter-energy bounds give \(0<z<1/2\) and \mbox{\(0<E^*<E_{\max}^*(z)=E_0^*/(1-z)\)}, with \(0<\eta<1\) and the full parent and transverse angular ranges.
The split at \(E^*=E_0^*\) defines \(\Cop\Sop_<\) and \(\Cop\Sop_>\) for this channel.

With the eikonal current \(j^\mu(p)\) in \eqref{eq:3.17}, the local overlap is \begin{equation}
\Cop\Sop_{34}F_{gg}
 =\frac{g_s^4\mathcal C_\epsilon^2 C_F}{k_3\cdot k_4}
        P_{gg}^{\mu\nu}j_\mu(p)j_\nu(p)F_B.
\label{eq:3.29}
\end{equation}

Using \(-g_\perp^{\mu\nu}j_\mu j_\nu=\mathcal S_{12}(p)\), the transverse average gives
\[
\langle P_{gg}^{\mu\nu}j_\mu j_\nu\rangle_\perp
=2C_A\left[\frac{z}{1-z}+\frac{1-z}{z}+z(1-z)\right]\mathcal S_{12}(p).
\]
As in the quark-pair channel, the splitting variables are integrated at fixed \(p\) for \(\Cop\Sop_<\), with \(0<E^*<E_0^*\), while \(\Cop\Sop_>\) is fully integrated:
\begin{equation}
\begin{gathered}
\begin{aligned}
\int d\Phi_{\Cop,\rm unres}^{gg}\,\Cop\Sop_<F_{gg}
 &=g_s^2\mathcal C_\epsilon C_F\,
   \mathcal C_{gg}(E^*,\epsilon)\mathcal S_{12}(p)F_B,\\
\int d\Phi_{\Cop\Sop,\rm unres}^{gg}\,\Cop\Sop_>F_{gg}
 &=\mathcal K_{\Cop\Sop,g}F_B,
\end{aligned}\\
\mathcal K_{\Cop\Sop,g}
=\frac{g_s^4C_FC_A}{(4\pi)^4}
 \frac{4e^{2\epsilon(\gamma_E+L_S)}}
      {\epsilon^2\Gamma(2-2\epsilon)}h(\beta,\epsilon)
      \bigl(\mathcal W_{g,m}-\mathcal W_{g,0}\bigr).
\end{gathered}
\label{eq:3.30}
\end{equation} Here \(\mathcal C_{gg}\) and \(h\) are given in \eqref{eq:A.2} and \eqref{eq:A.3}.

For the ordered gluon pair, we factor out \(C_A\) and define
\[
w_g(z,\epsilon)=2[z(1-z)]^{-2\epsilon}
 \left[\frac{z}{1-z}+\frac{1-z}{z}+z(1-z)\right].
\]
The weights in \eqref{eq:3.30} are
\begin{equation}
\begin{aligned}
\mathcal W_{g,0}={}&\int_0^{1/2}dz\,w_g(z,\epsilon)\\
={}&2\mathrm B(2-2\epsilon,-2\epsilon)
                     +\mathrm B(2-2\epsilon,2-2\epsilon),\\
\mathcal W_{g,m}={}&\int_0^{1/2}dz\,w_g(z,\epsilon)(1-z)^{4\epsilon}\\
={}&2\bigl[\mathrm B_{1/2}(2-2\epsilon,2\epsilon)
                  +\mathrm B_{1/2}(-2\epsilon,2+2\epsilon)
                  +\mathrm B_{1/2}(2-2\epsilon,2+2\epsilon)\bigr].
\end{aligned}
\label{eq:A.6}
\end{equation}
Their difference is
\begin{equation}
\mathcal W_{g,m}-\mathcal W_{g,0}
 =\epsilon\left(\frac{131}{18}-\frac{22}{3}\ln2-4\zeta_2\right)
 +\epsilon^2\left(\frac{790}{27}-\frac{44}{3}\zeta_2-16\zeta_3\right)
 +O(\epsilon^3).
\label{eq:A.7}
\end{equation}

\Needspace{4\baselineskip}
\paragraph{Single-soft--collinear and iterated overlaps.}\label{single-soft-collinear-and-triple-overlap}

For these counterterms, we keep the momentum \(k_3\) fixed, as in the single-soft subtraction of \eqref{eq:3.24}.
The change of variables \[
E_4^*=\frac{z}{1-z}E_3^*,\qquad
dE_4^*=\frac{E_3^*}{(1-z)^2}dz,\qquad 0<z<\frac12
\] and the angular projection \((1-\eta)^{-\epsilon}\to1\) give \[
\begin{aligned}
d\Phi_R(k_3)[dk_4]_K\theta(E_3^*-E_4^*)
 &\xrightarrow{\Cop}d\Phi_R(k_3)d\Phi_{\Cop\Sop_4,\rm unres},\\
d\Phi_B[dk_3]_K[dk_4]_K\theta(E_3^*-E_4^*)
 &\xrightarrow{\Cop}d\Phi_B[dk_3]_K d\Phi_{\Cop\Sop_4,\rm unres},
\end{aligned}
\] where \begin{equation}
d\Phi_{\Cop\Sop_4,\rm unres}
=\frac{(E_3^*)^{2-2\epsilon}z^{1-2\epsilon}(1-z)^{-3+2\epsilon}}
       {2(2\pi)^{3-2\epsilon}}\,dz\,
       2^{1-2\epsilon}\eta^{-\epsilon}d\eta\,d\Omega_\perp^*.
\label{eq:3.31}
\end{equation} The integration covers \(0<z<1/2\), \(0<\eta<1\) and the full transverse solid angle, with \(0<E_4^*<E_3^*<E_0^*\).

The local counterterms are \begin{equation}
\begin{aligned}
\Cop\Sop_4F_{gg}&=g_s^2\mathcal C_\epsilon
             \frac{2C_A(1-z)}{z\,k_3\cdot k_4}F_R(k_3),\\
\Cop\Sop_4\Sop_{34}F_{gg}&=g_s^4\mathcal C_\epsilon^2
             \frac{2C_A(1-z)}{z\,k_3\cdot k_4}
                    C_F\mathcal S_{12}(k_3)F_B .
\end{aligned}
\label{eq:3.32}
\end{equation} Here \(k_3\cdot k_4=2(E_3^*)^2z\eta/(1-z)\), and the factor \((1-z)/z=E_3^*/E_4^*\) comes from the single-soft current.

Using \(\int_0^{1/2}dz\,z^{-1-2\epsilon}(1-z)^{-1+2\epsilon}
=-1/(2\epsilon)\) and \(\int_0^1d\eta\,\eta^{-1-\epsilon}=-1/\epsilon\), we obtain \begin{equation}
\begin{gathered}
\begin{aligned}
\int d\Phi_{\Cop\Sop_4,\rm unres}\,\Cop\Sop_4F_{gg}
 &=\mathcal C_{\Cop\Sop_4}(E_3^*,\epsilon)F_R(k_3),\\
\int d\Phi_{\Cop\Sop_4,\rm unres}\,\Cop\Sop_4\Sop_{34}F_{gg}
 &=\mathcal C_{\Cop\Sop_4}(E_3^*,\epsilon)
        g_s^2\mathcal C_\epsilon C_F\mathcal S_{12}(k_3)F_B,
\end{aligned}\\
\mathcal C_{\Cop\Sop_4}(E_3^*,\epsilon)=\frac{g_s^2C_A}{(4\pi)^2}
\left[\frac{2}{\epsilon^2}+\frac{2L_3}{\epsilon}
                              +L_3^2-\zeta_2+O(\epsilon)\right].
\end{gathered}
\label{eq:3.33}
\end{equation} Here \(L_3=\ln[\mu_R^2/(4(E_3^*)^2)]\); the exact kernel is
\begin{equation}
\begin{aligned}
\mathcal C_{\Cop\Sop_4}(E_3^*,\epsilon)
={}&\frac{g_s^2C_A}{(4\pi)^2}
 \left(\frac{\mu_R^2}{4(E_3^*)^2}\right)^\epsilon
 \frac{2e^{\gamma_E\epsilon}}
      {\epsilon^2\Gamma(1-\epsilon)}.
\end{aligned}
\label{eq:cs4-kernel-exact}
\end{equation}
The remaining integrations are over \(d\Phi_R(k_3)\) and \(d\Phi_B[dk_3]_K\), respectively, as for the single-soft terms in \eqref{eq:3.25}.

\Needspace{5\baselineskip}
\subsection{Real-virtual contribution (RV)}\label{jpsicbar-cg-at-one-loop}

The real-virtual contribution is integrated over \(d\Phi_R\), while its soft counterterms use \(d\Phi_B[dk]_K\) with \(0<E_k^*<E_0^*\).
The phase-space parametrisations are the same as for the single-real contribution and are given in appendix~\ref{jpsicbar-cg}.

\Needspace{4\baselineskip}
\subsubsection{Local soft subtraction}\label{local-soft-subtraction}

Applying soft subtraction to the finite remainder and pole term in \eqref{eq:3.34} gives
\begin{equation}
\begin{aligned}
F_{\mathrm{RV}}^{\rm sub}
 &=(1-\Sop_k)F_{\mathrm{RV}}^{\rm fin},\\
F_{\mathrm{RV,SU}}^{\rm sub}
 &=(1-\Sop_k)\left[2a_s\operatorname{Re}Z_{c\bar cg}^{(1)}F_R\right].
\end{aligned}
\label{eq:3.36}
\end{equation}
These terms enter \(\mathrm{RV}_{\mathrm F}\) and \(\mathrm{SU}\), respectively.

The soft limit uses the massive one-loop soft current \cite{ref8,ref9}.
In the normalisation of ref.~\cite{ref6}, \begin{equation}
R_{12}^{(1)}(k)=4C_A
 \left[\frac{\mu_R^2p_1\cdot p_2}
 {2(p_1\cdot k)(p_2\cdot k)}\right]^\epsilon
 \left[\sum_{n=-2}^{1}\epsilon^n R_n+O(\epsilon^2)\right],
\qquad R_n=-\frac12r_n.
\label{eq:3.37}
\end{equation} Here \(r_n\) are the real parts of the time-like coefficients in ref.~\cite{ref9}.
The coefficients \(R_{-2},\ldots,R_1\) and their arguments are given in appendix~\ref{one-loop-soft-current-coefficients}.

The renormalised interference has the soft limit \cite{ref6} \begin{equation}
\begin{gathered}
\Sop_k F_{\mathrm{RV}}=g_s^2\mathcal C_\epsilon C_F\mathcal S_{12}(k)
\Bigl\{F_V^{\rm fin}
 +a_s\bigl[R_{12}^{(1)}(k)+Z_{\alpha_s}^{(1)}
                    +Z_A^{(1)}+2\operatorname{Re}Z_1\bigr]F_B\Bigr\},\\[4pt]
Z_{\alpha_s}^{(1)}=-\frac{\beta_0^{(4)}}{\epsilon},\qquad
Z_A^{(1)}=\frac{\beta_{0,Q}}{\epsilon}
 \left(\frac{\mu_R^2}{m_c^2}\right)^\epsilon
 \Gamma(1+\epsilon)e^{\gamma_E\epsilon}.
\end{gathered}
\label{eq:3.38}
\end{equation} Here \(Z_1\) is the heavy-pair infrared factor in \eqref{eq:2.6}.
The coefficient \(Z_{\alpha_s}^{(1)}\) renormalises the four-flavour \(\overline{\rm MS}\) coupling, and \(Z_A^{(1)}\) is the charm-loop contribution to on-shell gluon-field renormalisation.

Applying the soft limit to \eqref{eq:3.34} gives the counterterm for the finite remainder:
\[
\Sop_kF_{\mathrm{RV}}^{\rm fin}
 =\Sop_kF_{\mathrm{RV}}
  -\Sop_k\!\left[2a_s\operatorname{Re}Z_{c\bar cg}^{(1)}F_R\right].
\]
The two terms are combined on the same soft phase space, retaining their dimensional prefactors until the explicit poles cancel.
The second term is also the counterterm subtracted in \(F_{\mathrm{RV,SU}}^{\rm sub}\).

\Needspace{4\baselineskip}
\subsubsection{Integrated soft counterterm}\label{integrated-soft-counterterm}

The \(d\)-dimensional integral of the full soft limit \eqref{eq:3.38} involves the functions
\begin{equation}
\begin{aligned}
\mathcal S_{\rm int}^{(0)}
&=\mathcal C_\epsilon\int[dk]_K\,\mathcal S_{12}(k)
             =\frac{\mathcal I_{\mathrm S}(\beta,\epsilon,L_S)}{(4\pi)^2},\\
\mathcal R_{\rm int}^{(1)}
&=\mathcal C_\epsilon\int[dk]_K\,\mathcal S_{12}(k)R_{12}^{(1)}(k)\\
&=\frac{4C_A}{(4\pi)^2}e^{2\epsilon L_S}
                  \sum_{n=-3}^{0}\epsilon^n\mathcal R_n(\beta)
                  +O(\epsilon).
\end{aligned}
\label{eq:3.39}
\end{equation} The tree-level integral \(\mathcal I_{\mathrm S}\), valid to all orders in \(\epsilon\), is given in \eqref{eq:A.8}.
The complete coefficients \(\mathcal R_{-3},\ldots,\mathcal R_0\) and the harmonic-polylogarithm definitions are given in appendix~\ref{integrated-one-loop-soft-current-coefficients}.

The integrated counterterm to be added back is \begin{equation}
\begin{aligned}
\int[dk]_K\,\Sop_k F_{\mathrm{RV}}
={}&F_{\mathrm{RV,FR}}+F_{\mathrm{RV,DU}}\\
={}&g_s^2C_F\mathcal S_{\rm int}^{(0)}F_V^{\rm fin}
+g_s^2C_Fa_s\left\{\mathcal R_{\rm int}^{(1)}
+[Z_{\alpha_s}^{(1)}+Z_A^{(1)}
       +2\operatorname{Re}Z_1]\mathcal S_{\rm int}^{(0)}
\right\}F_B .
\end{aligned}
\label{eq:3.40}
\end{equation} The terms proportional to \(F_V^{\rm fin}\) and \(F_B\) define \(F_{\mathrm{RV,FR}}\) and \(F_{\mathrm{RV,DU}}\), respectively.
These enter the finite contributions \(\mathrm{FR}\) and \(\mathrm{DU}\) defined in \eqref{eq:finite-fr} and \eqref{eq:4.12}.

\FloatBarrier

\Needspace{6\baselineskip}
\section{Phase-space parametrisation and numerical integration}\label{common-parametrisations-of-the-real-and-subtraction-terms}

We derive common phase-space parametrisations for the real terms and local counterterms of appendix~\ref{real-emission-subtraction}, starting from the \(d=4-2\epsilon\)-dimensional phase spaces and their soft and collinear projections.
We then give the four-dimensional parametrisations for the finite subtracted integrals.
Within each channel, the spectator momentum \(P\), radiation variables and heavy-quark angles are common to all terms; the heavy-pair momenta and energy ranges depend on the projection.

\Needspace{5\baselineskip}
\subsection{Outer phase space and heavy-pair kinematics}\label{common-kinematics-and-phase-space-factors}

Our phase-space convention is
\begin{equation}
d\Phi_n^{(d)}(L;q_1,\ldots,q_n)
=(2\pi)^d\delta^{(d)}\!\left(L-\sum_{i=1}^nq_i\right)
 \prod_{i=1}^n\frac{d^{d-1}\boldsymbol q_i}{(2\pi)^{d-1}2q_i^0},
\qquad q_i^0=\sqrt{\boldsymbol q_i^2+m_i^2}.
\label{eq:B.measure-definition}
\end{equation}
We denote the solid-angle element on the unit sphere \(S^\nu\) by \(d\Omega_\nu\), analytically continued in \(\nu\), with
\[
d\Omega_\nu=(\sin\vartheta)^{\nu-1}d\vartheta\,d\Omega_{\nu-1},
\qquad
\Omega_\nu\equiv\int d\Omega_\nu
=\frac{2\pi^{(\nu+1)/2}}{\Gamma((\nu+1)/2)}.
\]
\Needspace{7\baselineskip}
For a two-body final state, integrating the momentum-conserving delta function in the rest frame of \(L\) gives
\begin{equation}
d\Phi_2^{(d)}(L;q_a,q_b)
=\frac{\lambda(L^2,m_a^2,m_b^2)^{(d-3)/2}}
 {2^{d-1}(2\pi)^{d-2}(L^2)^{(d-2)/2}}\,d\Omega_{d-2},
\label{eq:B.two-body-d}
\end{equation}
where \(\lambda(x,y,z)=x^2+y^2+z^2-2xy-2xz-2yz\).

For the outer factor \(Q\to P+K\) in \eqref{eq:3.4}, we take the beam as the polar axis and integrate over the common orientation of the final state about that axis:
\begin{equation}
\begin{aligned}
d\Phi_{\rm out}^{(d)}
 &\equiv\frac{dK^2}{2\pi}
        \int_{S^{d-3}}d\Phi_2^{(d)}(Q;P,K)\\
 &=\frac{\Omega_{d-3}\lambda(s,M_\psi^2,K^2)^{(d-3)/2}}
 {2^{d-1}(2\pi)^{d-1}s^{(d-2)/2}}
 (\sin^2\theta)^{(d-4)/2}\,dK^2\,d\cos\theta.
\end{aligned}
\label{eq:B.outer-d}
\end{equation}
The range is \(4m_c^2<K^2<(\sqrt{s}-M_\psi)^2\).

For the heavy pair, let \(T=p_1+p_2\) and let \(\boldsymbol n_T\in S^{d-2}\) be the quark direction in the rest frame of \(T\), with solid-angle element \(d\Omega_T\).
The momenta and two-body phase space are
\begin{equation}
\begin{gathered}
p_{1,2}^{(T)}
 =\left(\frac{\sqrt{T^2}}{2},
       \ \pm\frac{\sqrt{T^2-4m_c^2}}{2}\boldsymbol n_T\right),\\
d\Phi_2^{(d)}(T)\equiv d\Phi_2^{(d)}(T;p_1,p_2)
=\frac{(T^2)^{(d-4)/2}}{2^{d-1}(2\pi)^{d-2}}
 \left(1-\frac{4m_c^2}{T^2}\right)^{(d-3)/2}d\Omega_T.
\end{gathered}
\label{eq:B.pair-d}
\end{equation}
We boost \(p_{1,2}^{(T)}\) to the \(K\) rest frame with velocity \(\boldsymbol T^*/T^{*0}\), then all momenta to the centre-of-mass frame with velocity \(\boldsymbol K/K^0\), using rotation-free Lorentz boosts.
At Born kinematics, \(T=K\), so the first boost is the identity.
The Born phase space is \(d\Phi_B^{(d)}=d\Phi_{\rm out}^{(d)}d\Phi_2^{(d)}(K)\).

For the finite Born integral, we set \(\epsilon=0\) and omit the superscript \((d)\).
The outer phase-space factor becomes
\begin{equation}
d\Phi_{\rm out}
=\left.d\Phi_{\rm out}^{(d)}\right|_{d=4}
=\frac{\sqrt{\lambda(s,M_\psi^2,K^2)}}{32\pi^2s}
 \,dK^2\,d\cos\theta.
\label{eq:B.1}
\end{equation}
With the beam along the \(z\) axis, we choose the \(xz\) plane to contain \(\boldsymbol P\):
\begin{equation}
\begin{gathered}
E_\psi=\frac{s+M_\psi^2-K^2}{2\sqrt{s}},\qquad
\lvert\boldsymbol P\rvert=\frac{\sqrt{\lambda(s,M_\psi^2,K^2)}}{2\sqrt{s}},\\
P=(E_\psi,\lvert\boldsymbol P\rvert\sin\theta,0,\lvert\boldsymbol P\rvert\cos\theta),\qquad
K=Q-P.
\end{gathered}
\label{eq:B.2}
\end{equation}
The corresponding range is \(M_\psi<E_\psi<(s+M_\psi^2-4m_c^2)/(2\sqrt{s})\); its upper endpoint is \(\sqrt{s}/2\) for \(M_\psi=2m_c\).

The heavy-quark direction and solid-angle element in the pair rest frame are
\begin{equation}
\begin{gathered}
\boldsymbol n_T\equiv\boldsymbol n(\theta_T,\phi_T)
 =(\sin\theta_T\cos\phi_T,\sin\theta_T\sin\phi_T,
                                      \cos\theta_T),\\
d\Omega_T=d\cos\theta_T\,d\phi_T.
\end{gathered}
\label{eq:B.3}
\end{equation}
The two-body phase space reduces to
\begin{equation}
d\Phi_2(T)\equiv d\Phi_2(T;p_1,p_2)
 =\frac{1}{32\pi^2}\sqrt{1-\frac{4m_c^2}{T^2}}\,d\Omega_T.
\label{eq:B.4}
\end{equation}

The common outer variables are mapped to \(u_i\in[0,1]\) by
\begin{equation}
\begin{gathered}
K^2=4m_c^2+\Delta_Ku_1,\qquad \cos\theta=2u_2-1,\qquad
\Delta_K=(\sqrt{s}-M_\psi)^2-4m_c^2,\\
d\Phi_{\rm out}
 =\frac{\Delta_K\sqrt{\lambda(s,M_\psi^2,K^2)}}{16\pi^2s}
       \,du_1\,du_2.
\end{gathered}
\label{eq:B.5}
\end{equation}
For the Born phase space, set \(T=K\), \(\cos\theta_T=2u_3-1\) and \(\phi_T=2\pi u_4\), giving
\begin{equation}
d\Phi_B=d\Phi_{\rm out}\,d\Phi_2(K)=J_B\,d^4u,\qquad
J_B=\frac{\Delta_K\sqrt{\lambda(s,M_\psi^2,K^2)}\,\beta}
          {128\pi^3s}.
\label{eq:B.6}
\end{equation}

\Needspace{5\baselineskip}
\subsection{Single-gluon emission}\label{jpsicbar-cg}

For \(0<E_k^*<E_0^*\), the real and soft-counterterm phase spaces are
\begin{equation}
\begin{aligned}
d\Phi_R^{(d)}
 &=d\Phi_{\rm out}^{(d)}\,d\Phi_3^{(d)}(K;p_1,p_2,k)
 =d\Phi_{\rm out}^{(d)}\,d\Phi_2^{(d)}(K-k)\,d\Pi^{(d)}(k),\\
d\Phi_{\Sop}^{(d)}
 &\equiv d\Phi_B^{(d)}[dk]_K
 =d\Phi_{\rm out}^{(d)}\,d\Phi_2^{(d)}(K)\,d\Pi^{(d)}(k),\\
d\Pi^{(d)}(k)
 &=\frac{(E_k^*)^{d-3}}{2(2\pi)^{d-1}}
       \,dE_k^*\,d\Omega_k^*.
\end{aligned}
\label{eq:B.7}
\end{equation}
Here \([dk]_K=d\Pi^{(d)}(k)\theta(E_0^*-E_k^*)\).
The solid-angle element \(d\Omega_k^*\) covers \(S^{d-2}\) in the \(K\) frame.
For a unit direction \(\boldsymbol n_k^*\), the gluon momentum and heavy-pair recoils are
\begin{equation}
\begin{gathered}
k^*=E_k^*(1,\boldsymbol n_k^*),\\
T_R=K-k,\qquad T_R^2=W^2-2WE_k^*,\qquad
T_{\Sop}=K,\qquad T_{\Sop}^2=W^2.
\end{gathered}
\label{eq:B.8}
\end{equation}
Combining \eqref{eq:B.pair-d} and \eqref{eq:B.7} gives, for \(X=R,\Sop\),
\begin{equation}
d\Phi_X^{(d)}=d\Phi_{\rm out}^{(d)}\,
 \frac{(E_k^*)^{d-3}(T_X^2)^{(d-4)/2}}{2^d(2\pi)^{2d-3}}
 \left(1-\frac{4m_c^2}{T_X^2}\right)^{(d-3)/2}
 dE_k^*\,d\Omega_k^*\,d\Omega_T.
\label{eq:B.single-d}
\end{equation}

For the finite single-real integrals, we set \(\epsilon=0\).
With \(d\Omega_k^*=d\cos\theta_k^*\,d\phi_k^*\), the phase space becomes
\begin{equation}
d\Phi_X=d\Phi_{\rm out}\,
 \frac{E_k^*}{512\pi^5}\sqrt{1-\frac{4m_c^2}{T_X^2}}\,
       dE_k^*\,d\Omega_k^*\,d\Omega_T.
\label{eq:B.9}
\end{equation}
At fixed \(K^2\) and \(\cos\theta\), the five variables
\(E_k^*,\theta_k^*,\phi_k^*,\theta_T,\phi_T\) specify both configurations.
The gluon direction is \(\boldsymbol n_k^*=\boldsymbol n(\theta_k^*,\phi_k^*)\), as in \eqref{eq:B.3}, and the heavy-quark momenta follow from \eqref{eq:B.pair-d} with \(T=T_R\) or \(T_{\Sop}\).

We combine the outer map \eqref{eq:B.5} with
\begin{equation}
\begin{gathered}
E_k^*=E_0^*u_3,\qquad
\cos\theta_k^*=2u_4-1,\qquad \phi_k^*=2\pi u_5,\\
\cos\theta_T=2u_6-1,\qquad \phi_T=2\pi u_7.
\end{gathered}
\label{eq:B.10}
\end{equation}
Then \(d\Phi_X=J_X^g\,d^7u\), with
\begin{equation}
\begin{gathered}
J_0^g=\frac{\Delta_K\sqrt{\lambda(s,M_\psi^2,K^2)}
                     E_k^*E_0^*}{512\pi^5s},\\
J_R^g=J_0^g\sqrt{1-\frac{4m_c^2}{T_R^2}},
\qquad J_{\Sop}^g=J_0^g\beta.
\end{gathered}
\label{eq:B.11}
\end{equation}
At fixed \(K\) and angles, the mapped heavy momenta satisfy \(p_{i,R}\to p_{i,\Sop}\) and the Jacobians satisfy \(J_R^g/J_{\Sop}^g\to1\) as \(E_k^*\to0\).

The finite real contribution is
\begin{equation}
R_{\mathrm F}
 =\mathcal N\int_{[0,1]^7}d^7u\,\Bigl[J_R^gF_R(P;p_{1,R},p_{2,R},k)
 -J_{\Sop}^g(\Sop_kF_R)(P;p_{1,\Sop},p_{2,\Sop},k)\Bigr].
\label{eq:B.12}
\end{equation}
For \(\mathrm{RV}_{\mathrm F}\), replace \(F_R\) by
\(F_{\mathrm{RV}}^{\rm fin}\) in \eqref{eq:B.12}.
For \(\mathrm{SU}\) in \eqref{eq:finite-su}, the real map applies to the combined RR and RV insertion, and the soft map to its \(\Sop_k\) counterterm.

\Needspace{5\baselineskip}
\subsection{Light-quark-pair emission}\label{jpsicbar-cqbar-q}

Separating the two massless momenta from the heavy-pair phase space gives
\begin{equation}
\begin{aligned}
d\Phi_{\mathrm{RR}}^{q\bar q,(d)}
 &=d\Phi_{\rm out}^{(d)}\,d\Phi_4^{(d)}(K;p_1,p_2,k_3,k_4)\\
 &=d\Phi_{\rm out}^{(d)}\,d\Phi_2^{(d)}(K-k_3-k_4)
       \,d\Pi^{(d)}(k_3)\,d\Pi^{(d)}(k_4).
\end{aligned}
\label{eq:B.13}
\end{equation}
We use \(E^*=E_3^*+E_4^*\), \(z=E_3^*/E^*\) and
\(\eta=(1-\cos\theta_{34}^*)/2\).
Let \(\boldsymbol n_3^*\in S^{d-2}\) be the direction of \(k_3\), and let \(\boldsymbol e_\perp^*\) be a unit vector in the transverse subspace orthogonal to \(\boldsymbol n_3^*\).
The daughter momenta and their collinear parent \(p=k_3/z\) are
\begin{equation}
\begin{gathered}
\boldsymbol n_4^*=(1-2\eta)\boldsymbol n_3^*
                  +2\sqrt{\eta(1-\eta)}\,\boldsymbol e_\perp^*,\\
k_3^*=zE^*(1,\boldsymbol n_3^*),\qquad
k_4^*=(1-z)E^*(1,\boldsymbol n_4^*),\qquad
p^*=E^*(1,\boldsymbol n_3^*).
\end{gathered}
\label{eq:B.16}
\end{equation}
The solid-angle elements \(d\Omega_3^*\) and \(d\Omega_\perp^*\) cover \(S^{d-2}\) and the transverse sphere \(S^{d-3}\), respectively, as in \eqref{eq:3.13}.
The changes of variables give
\begin{equation}
\begin{gathered}
dE_3^*\,dE_4^*=E^*\,dE^*\,dz,\\
d\Omega_4^*=2^{1-2\epsilon}[\eta(1-\eta)]^{-\epsilon}
                 d\eta\,d\Omega_\perp^*,\\
d\Pi^{(d)}(k_3)\,d\Pi^{(d)}(k_4)
 =\frac{2^{1-2\epsilon}(E^*)^{3-4\epsilon}[z(1-z)]^{1-2\epsilon}}
        {4(2\pi)^{6-4\epsilon}}
 [\eta(1-\eta)]^{-\epsilon}
 dE^*\,dz\,d\Omega_3^*\,d\eta\,d\Omega_\perp^*.
\end{gathered}
\label{eq:B.14}
\end{equation}
Here \(0<z,\eta<1\), with full angular ranges.

The daughter-energy bounds \(zE^*<E_0^*\) and
\((1-z)E^*<E_0^*\) define the common upper limit \(E_{\max}^*\).
The double-real heavy-pair threshold determines the physical upper limit \(E_{\rm phys}^*\):
\begin{equation}
\begin{gathered}
(K-k_3-k_4)^2=W^2-2WE^*+4z(1-z)(E^*)^2\eta\geq4m_c^2,\\
E_{\max}^*(z)=\frac{E_0^*}{\max(z,1-z)},\qquad
E_0^*\leq E_{\rm phys}^*\leq E_{\max}^*,\\
E_{\rm phys}^*(z,\eta)=
 \frac{W\beta^2}{1+\sqrt{1-4\beta^2z(1-z)\eta}}.
\end{gathered}
\label{eq:B.17}
\end{equation}
The physical upper bound is the smaller root of the threshold equation, for which the heavy-pair momentum is future-directed.
Table~\ref{tab:5} lists the corresponding heavy-pair momenta and energy ranges.

\begin{table}[!htbp]
\centering
\normalsize
\renewcommand{\arraystretch}{1.18}
\begin{tabular}{@{}lcc@{}}
\toprule
Term \(X\) & Heavy-pair momentum \(T_X\) & Energy range \\
\midrule
\(\mathrm{RR}\) & \(K-k_3-k_4\) & \(0<E^*<E_{\rm phys}^*\) \\
\(\Cop\) & \(K-p\) & \(0<E^*<E_0^*\) \\
\(\Sop_{34},\ \Cop\Sop_{34}\) & \(K\) & \(0<E^*<E_{\max}^*\) \\
\bottomrule
\end{tabular}
\caption{Heavy-pair recoils and energy ranges for the four \(q\bar q\) terms.}\label{tab:5}
\end{table}

\Needspace{9\baselineskip}
The soft projection retains the full radiation phase space in \eqref{eq:B.14}.
For \(\Cop\) and \(\Cop\Sop_{34}\), the angular projection of \eqref{eq:3.13} sets
\((1-\eta)^{-\epsilon}\to1\), retaining \(\eta^{-\epsilon}\).
Thus, within the domains in table~\ref{tab:5},
\begin{equation}
d\Phi_X^{q\bar q,(d)}
=d\Phi_{\rm out}^{(d)}d\Phi_2^{(d)}(T_X)
 \,d\Pi^{(d)}(k_3)d\Pi^{(d)}(k_4)
 \begin{cases}
 1,&X=\mathrm{RR},\Sop_{34},\\
 (1-\eta)^\epsilon,&X=\Cop,\Cop\Sop_{34}.
 \end{cases}
\label{eq:B.qqbar-d}
\end{equation}
The spin-correlated kernels \eqref{eq:3.14} and \eqref{eq:3.17} use
\(e_\perp^{*\mu}=(0,\boldsymbol e_\perp^*)\) and
\(k_\perp^\mu k_\perp^\nu/k_\perp^2=-e_\perp^\mu e_\perp^\nu\);
the reduced single-real squared matrix element in \(\Cop\) is evaluated at \(p\).

For the subtracted integral \(\mathrm{RR}_{\mathrm F}(q\bar q)\), we now set \(\epsilon=0\).
The full and projected relative angular factors both reduce to
\(2\,d\eta\,d\phi_\perp^*\), giving
\begin{equation}
d\Phi_X^{q\bar q}
 =d\Phi_{\rm out}\,
   \frac{z(1-z)(E^*)^3}{4096\pi^8}
      \sqrt{1-\frac{4m_c^2}{T_X^2}}\,
      dE^*\,dz\,d\Omega_3^*\,d\eta\,d\phi_\perp^*\,d\Omega_T.
\label{eq:B.18}
\end{equation}
Here \(d\Omega_3^*=d\cos\theta_3^*\,d\phi_3^*\), and
\(\phi_3^*,\phi_\perp^*\in[0,2\pi)\).

To construct the four-dimensional momenta, set
\(\boldsymbol n_3^*=\boldsymbol n(\theta_3^*,\phi_3^*)\).
Choose \(\boldsymbol a^*\) from the three Cartesian basis vectors to minimise
\(|\boldsymbol a^*\cdot\boldsymbol n_3^*|\), and define
\begin{equation}
\begin{gathered}
\boldsymbol e_1^*
 =\frac{\boldsymbol a^*
      -(\boldsymbol a^*\cdot\boldsymbol n_3^*)\boldsymbol n_3^*}
          {\sqrt{1-(\boldsymbol a^*\cdot\boldsymbol n_3^*)^2}},
\qquad
\boldsymbol e_2^*=\boldsymbol n_3^*\times\boldsymbol e_1^*,\\
\boldsymbol e_\perp^*
 =\cos\phi_\perp^*\,\boldsymbol e_1^*
       +\sin\phi_\perp^*\,\boldsymbol e_2^*.
\end{gathered}
\label{eq:B.15}
\end{equation}
The daughter momenta follow from \eqref{eq:B.16}, and the heavy-quark momenta from \eqref{eq:B.pair-d} with the corresponding \(T_X\) in table~\ref{tab:5}.

We map the common interval \(0<E^*<E_{\max}^*(z)\) to \(0<u_3<1\), imposing the narrower term-dependent bounds through step functions in the Jacobians.
Together with \eqref{eq:B.5}, the map is
\begin{equation}
\begin{gathered}
E^*=E_{\max}^*(z)u_3,\qquad z=u_4,\qquad \eta=u_7^2,\\
\cos\theta_3^*=2u_5-1,\qquad \phi_3^*=2\pi u_6,\qquad
\phi_\perp^*=2\pi u_8,\\
\cos\theta_T=2u_9-1,\qquad \phi_T=2\pi u_{10}.
\end{gathered}
\label{eq:B.19}
\end{equation}
With \(dE^*\,dz=E_{\max}^*(z)\,du_3\,du_4\) and
\(d\eta=2u_7\,du_7\), the phase space is
\(d\Phi_X^{q\bar q}=J_X^{q\bar q}\,d^{10}u\), where
\begin{equation}
\begin{gathered}
J_0^{q\bar q}
 =\frac{\Delta_K\sqrt{\lambda(s,M_\psi^2,K^2)}}{2048\pi^7s}\,
             z(1-z)(E^*)^3E_{\max}^*(z)\,2u_7,\\
J_X^{q\bar q}=J_0^{q\bar q}\sqrt{1-4m_c^2/T_X^2}
 \,\theta(u_{3,\max}^{X}-u_3),\\
u_{3,\max}^{\mathrm{RR}}=\frac{E_{\rm phys}^*}{E_{\max}^*},
\qquad u_{3,\max}^{\Cop}=\max(z,1-z),\qquad
u_{3,\max}^{\Sop_{34}}=u_{3,\max}^{\Cop\Sop_{34}}=1.
\end{gathered}
\label{eq:B.20}
\end{equation}
Each term is evaluated only where its step function is nonzero.
The common coordinates give \(k_3+k_4\to p\) as \(\eta\to0\) and \(T_X\to K\) as \(E^*\to0\).

The finite contribution is
\begin{equation}
\begin{aligned}
\mathrm{RR}_{\mathrm F}(q\bar q)
 &=\mathcal N\sum_{q=u,d,s}\int_{[0,1]^{10}}d^{10}u\,\Bigl[
 J_{\mathrm{RR}}^{q\bar q}F_{q\bar q}-J_{\Sop_{34}}^{q\bar q}(\Sop_{34}F_{q\bar q})\\
 &\qquad
 -J_{\Cop}^{q\bar q}(\Cop F_{q\bar q})+J_{\Cop\Sop_{34}}^{q\bar q}(\Cop\Sop_{34}F_{q\bar q})\Bigr].
\end{aligned}
\label{eq:B.21}
\end{equation}

\FloatBarrier

\Needspace{5\baselineskip}
\subsection{Ordered two-gluon emission}\label{jpsicbar-cgg}

In the ordered sector \(E_4^*<E_3^*\), the \(J/\psi+c\bar c+gg\) phase space is
\begin{equation}
\begin{aligned}
d\Phi_{\mathrm{RR}}^{gg,(d)}
 &=d\Phi_{\rm out}^{(d)}\,d\Phi_4^{(d)}(K;p_1,p_2,k_3,k_4)
                              \,\theta(E_3^*-E_4^*)\\
 &=d\Phi_{\rm out}^{(d)}\,d\Phi_2^{(d)}(K-k_3-k_4)\,
          d\Pi^{(d)}(k_3)\,d\Pi^{(d)}(k_4)\,\theta(E_3^*-E_4^*).
\end{aligned}
\label{eq:B.22}
\end{equation}
We use \(E^*=E_3^*+E_4^*\), \(z=E_4^*/E^*\),
and \(\eta=(1-\cos\theta_{34}^*)/2\).
With the \(d\)-dimensional directions in \eqref{eq:B.16}, the gluon and parent momenta are
\begin{equation}
k_3^*=(1-z)E^*(1,\boldsymbol n_3^*),\qquad
k_4^*=zE^*(1,\boldsymbol n_4^*),\qquad
p^*=E^*(1,\boldsymbol n_3^*).
\label{eq:B.23}
\end{equation}
The two-gluon phase space follows from \eqref{eq:B.14}, now restricted to \(0<z<1/2\);
the parent is \(p=k_3/(1-z)\).

The harder-gluon bound \((1-z)E^*<E_0^*\) gives the common
upper limit \(E_{\max}^*(z)=E_0^*/(1-z)\); the double-real
threshold gives \(E_{\rm phys}^*(z,\eta)\) in \eqref{eq:B.17}.
Table~\ref{tab:6} gives the heavy-pair momenta and energy ranges for all eight terms.
The reduced single-real squared matrix element is evaluated at \(k_3\) in \(\Sop_4\) and \(\Cop\Sop_4\), and at \(p\) in \(\Cop\).

\begin{table}[!htbp]
\centering
\normalsize
\renewcommand{\arraystretch}{1.18}
\begin{tabular}{@{}lcc@{}}
\toprule
Term \(X\) & Heavy-pair momentum \(T_X\) & Energy range \\
\midrule
\(\mathrm{RR}\) & \(K-k_3-k_4\) & \(0<E^*<E_{\rm phys}^*\) \\
\(\Cop\) & \(K-p\) & \(0<E^*<E_0^*\) \\
\(\Sop_4,\ \Cop\Sop_4\) & \(K-k_3\) & \(0<E^*<E_{\max}^*\) \\
\(\Sop_{34},\ \Sop_4\Sop_{34},\ \Cop\Sop_{34},\ \Cop\Sop_4\Sop_{34}\) & \(K\) & \(0<E^*<E_{\max}^*\) \\
\bottomrule
\end{tabular}
\caption{Heavy-pair recoils and energy ranges for the eight ordered \(gg\) terms.}\label{tab:6}
\end{table}
\FloatBarrier

All terms containing \(\Cop\) use the projected relative angular factor, while the other terms retain the full angular dependence.
For the \(\Cop\Sop_4\) terms, combining \eqref{eq:3.31} with \([dk_3]_K\) and using \(E_3^*=(1-z)E^*\) gives the same projected radiation phase space in \((E^*,z)\).
Within the domains in table~\ref{tab:6}, the \(d\)-dimensional phase spaces are
\begin{equation}
\begin{aligned}
d\Phi_X^{gg,(d)}
 &=d\Phi_{\rm out}^{(d)}d\Phi_2^{(d)}(T_X)
       \,d\Pi^{(d)}(k_3)d\Pi^{(d)}(k_4)\\
 &\quad\times
 \begin{cases}
 1,&X=\mathrm{RR},\Sop_4,\Sop_{34},\Sop_4\Sop_{34},\\
 (1-\eta)^\epsilon,&X=\Cop,\Cop\Sop_4,\Cop\Sop_{34},\Cop\Sop_4\Sop_{34}.
 \end{cases}
\end{aligned}
\label{eq:B.gg-d}
\end{equation}

For the subtracted integral \(\mathrm{RR}_{\mathrm F}(gg)\), we set \(\epsilon=0\), obtaining
\begin{equation}
d\Phi_X^{gg}
 =d\Phi_{\rm out}\,
   \frac{z(1-z)(E^*)^3}{4096\pi^8}
      \sqrt{1-\frac{4m_c^2}{T_X^2}}\,
      dE^*\,dz\,d\Omega_3^*\,d\eta\,d\phi_\perp^*\,d\Omega_T.
\label{eq:B.24}
\end{equation}
The four-dimensional directions are constructed as in \eqref{eq:B.15}, and the heavy-quark momenta follow from \eqref{eq:B.pair-d} with the corresponding \(T_X\).

The ordered-sector map combines \eqref{eq:B.5} with
\begin{equation}
\begin{gathered}
E^*=E_{\max}^*(z)u_3,\qquad z=\frac{u_4}{2},\qquad \eta=u_7^2,\\
\cos\theta_3^*=2u_5-1,\qquad \phi_3^*=2\pi u_6,\qquad
\phi_\perp^*=2\pi u_8,\\
\cos\theta_T=2u_9-1,\qquad \phi_T=2\pi u_{10}.
\end{gathered}
\label{eq:B.25}
\end{equation}
Here \(dE^*\,dz=\tfrac12E_{\max}^*(z)\,du_3\,du_4\)
and \(d\eta=2u_7\,du_7\), so
\(d\Phi_X^{gg}=J_X^{gg}\,d^{10}u\), with
\begin{equation}
\begin{gathered}
J_0^{gg}
 =\frac{\Delta_K\sqrt{\lambda(s,M_\psi^2,K^2)}}{4096\pi^7s}\,
             z(1-z)(E^*)^3E_{\max}^*(z)\,2u_7,\\
J_X^{gg}=J_0^{gg}\sqrt{1-4m_c^2/T_X^2}
 \,\theta(u_{3,\max}^{X}-u_3),\\
u_{3,\max}^{\mathrm{RR}}=\frac{E_{\rm phys}^*}{E_{\max}^*},
\qquad u_{3,\max}^{\Cop}=1-z,\\
u_{3,\max}^{X}=1,
\qquad X=\Sop_4,\Cop\Sop_4,\Sop_{34},\Sop_4\Sop_{34},\Cop\Sop_{34},\Cop\Sop_4\Sop_{34}.
\end{gathered}
\label{eq:B.26}
\end{equation}
The single-soft limit \(z\to0\) gives \(k_4\to0\) and \(p\to k_3\), so the real and collinear heavy-pair momenta approach those of their single-soft counterterms.

Evaluating each term on the phase space specified in table~\ref{tab:6} and grouping terms with the same Jacobian gives
\begin{equation}
\begin{aligned}
\mathrm{RR}_{\mathrm F}(gg)
 &=\mathcal N\int_{[0,1]^{10}}d^{10}u\,\Bigl[
 J_{\mathrm{RR}}^{gg}F_{gg}-J_{\Cop}^{gg}(\Cop F_{gg})
 -J_{\Sop_4}^{gg}(\Sop_4F_{gg}-\Cop\Sop_4F_{gg})\\
 &\qquad
 -J_{\Sop_{34}}^{gg}\bigl(\Sop_{34}F_{gg}-\Sop_4\Sop_{34}F_{gg}
 -\Cop\Sop_{34}F_{gg}+\Cop\Sop_4\Sop_{34}F_{gg}\bigr)\Bigr].
\end{aligned}
\label{eq:B.27}
\end{equation}

\Needspace{5\baselineskip}
\subsection{Numerical integration and spectrum construction}\label{numerical-integration-and-spectra}

\paragraph{Virtual quadrature.}
We integrate \(V_{\mathrm F}\) and \(\mathrm{VV}_{\mathrm F}\) on a fixed \(15\times15\) Gauss--Kronrod grid at the virtual-input energy \(\sqrt{s_V}\), with outer variable \(E_\psi\) and inner invariant \(s_2=P\cdot p_1\).
For either photon projection, the integrand is independent of the overall orientation of the final state.
Integrating these angles in the Born phase space of appendix~\ref{common-kinematics-and-phase-space-factors} gives
\begin{equation}
\begin{gathered}
d\Phi_B=\frac{dE_\psi\,ds_2}{32\pi^3\sqrt{s_V}},
\qquad M_\psi<E_\psi<\frac{\sqrt{s_V}}{2},
\qquad s_{2,-}<s_2<s_{2,+},\\
s_{2,\pm}(E_\psi)
 =\frac{\sqrt{s_V}E_\psi-M_\psi^2}{2}
 \pm\frac{\beta}{4}\sqrt{\lambda(s_V,M_\psi^2,K^2)},
\end{gathered}
\label{eq:virtual-phase-space}
\end{equation}
where \(K^2=s_V+M_\psi^2-2\sqrt{s_V}E_\psi\).
The quadrature uses the dimensionless variables \(E_\psi/m_c\) and \(s_2/m_c^2\).

At each point, the \(\epsilon^0\) coefficient is estimated by averaging the finite interference or squared finite remainder at \(\epsilon=\pm h\), with \(h=10^{-3}\).
After pole cancellation, symmetric averaging removes odd powers of \(\epsilon\), leaving an \(O(h^2)\) truncation error.

\paragraph{Quadrature error estimates.}
For each contribution and photon projection, let \(f_{ij}\) denote the integrand values, including the phase-space prefactor and the factor \(\mathcal N\), at outer and inner nodes \(i,j=1,\ldots,15\).
With mapped Kronrod and Gauss weights \(w^K,w^G\), and the Gauss subset \(\mathcal I_7=\{2,4,\ldots,14\}\), define
\begin{equation}
\begin{aligned}
q_i^K&=\sum_{j=1}^{15}w_{ij}^K f_{ij},
& q_i^G&=\sum_{j\in\mathcal I_7}w_{ij}^G f_{ij},\\
I_{KK}&=\sum_{i=1}^{15}w_i^K q_i^K,
& I_{GK}&=\sum_{i\in\mathcal I_7}w_i^G q_i^K.
\end{aligned}
\label{eq:virtual-gk-sums}
\end{equation}
The embedded error estimate combines the outer difference with the weighted inner differences:
\begin{equation}
\Delta_{\rm emb}
=\lvert I_{KK}-I_{GK}\rvert
+\sum_{i=1}^{15}\lvert w_i^K\rvert\,\lvert q_i^K-q_i^G\rvert.
\label{eq:virtual-gk-error}
\end{equation}
The stored nodes differ slightly from the standard Gauss--Kronrod nodes because of rationalisation and small shifts of several inner midpoints.
Recomputing interpolatory weights at the actual nodes changes \(\Delta_{\rm emb}\) by at most \(0.11\%\).

Table~\ref{tab:virtual-quadrature} gives the estimates for the individual virtual contributions and their linear sum.

\begin{table}[!htbp]
\centering
\small
\renewcommand{\arraystretch}{1.18}
\begin{tabular}{lrr}
\toprule
Contribution & \(\Delta_{\rm emb}(\sigma)\) & \(\Delta_{\rm emb}(\sigma_L)\) \\
\midrule
\(V_{\mathrm F}\) & 0.078 & 0.020 \\
\(\mathrm{VV}_{20}\) & 0.79 & 0.033 \\
\(\mathrm{VV}_{11}\) & 0.32 & 0.014 \\
\midrule
Linear sum & 1.2 & 0.067 \\
\bottomrule
\end{tabular}
\caption{Embedded Gauss--Kronrod error estimates for the virtual contributions at \(\mu_R=2m_c\) and \(\mu_\Lambda=m_c\), in fb.}\label{tab:virtual-quadrature}
\end{table}

\paragraph{Energy bins and interpolation.}
We evaluate bin averages of \(d\sigma/dE_\psi\) using 23 bins of width \(0.10\ {\rm GeV}\) over \(3.00\leq E_\psi\leq5.30\ {\rm GeV}\).
For \(\mathrm{RV}_{\mathrm F}\) and \(\mathrm{FR}\), separate Monte Carlo samples are used for the spectrum and the total cross section.
Within each contribution, correlations between bins are included when rebinning or integrating the spectrum.
The resulting standard errors are added linearly across contributions to obtain conservative Monte Carlo uncertainty estimates.
For \(V_{\mathrm F}\) and \(\mathrm{VV}_{\mathrm F}\), integration over \(s_2\) gives the spectrum at 15 energy nodes.
We interpolate these values with a not-a-knot cubic spline, extrapolate it to the physical endpoints \(E_\psi=3.00\) and \(5.25\ {\rm GeV}\) using the polynomials on its first and last intervals, and integrate over the part of each bin within this range.

\paragraph{Momentum spectrum.}
The quarkonium momentum magnitude in the centre-of-mass frame is \(\lvert\boldsymbol P\rvert=\sqrt{E_\psi^2-M_\psi^2}\).
The differential distributions satisfy
\begin{equation}
\frac{d\sigma}{d\lvert\boldsymbol P\rvert}
=\frac{\lvert\boldsymbol P\rvert}{E_\psi}\frac{d\sigma}{dE_\psi}.
\label{eq:5.3}
\end{equation}
For the binned spectrum, we convert the energy-bin boundaries to momentum and divide the cross section in each bin by the corresponding momentum-bin width.
Evaluating \(V_{\mathrm F}\) and \(\mathrm{VV}_{\mathrm F}\) at \(\sqrt{s_V}\) gives them a kinematic endpoint at \(\lvert\boldsymbol P\rvert\simeq4.31\ {\rm GeV}\), compared with \(4.37\ {\rm GeV}\) for the other contributions.

\Needspace{6\baselineskip}
\section{Auxiliary functions and infrared coefficients}\label{integrated-kernels-and-one-loop-soft-currents}

We give the analytic coefficients used in the counterterms of appendix~\ref{real-emission-subtraction} and the virtual infrared factors of appendix~\ref{virtual-amplitudes-and-infrared-matching}.

\Needspace{10\baselineskip}
\subsection{Double-soft functions}\label{double-soft-functions}

The functions in \eqref{eq:3.11} and \eqref{eq:3.23} use the normalisation \(\mathcal A_S\) in \eqref{eq:double-soft-normalisation}.
The labels \(AA\) and \(AB\) denote equal and distinct emitters, respectively.
The double-soft functions have the Laurent expansions
\begin{equation}
\begin{aligned}
f_{ij}^{q\bar q}(\beta,\epsilon)
 &=\sum_{n=-2}^{0}\epsilon^n f_{ij,n}^{q\bar q}(\beta)+O(\epsilon),\\
f_{ij}^{gg}(\beta,\epsilon)
 &=\sum_{n=-3}^{0}\epsilon^n f_{ij,n}^{gg}(\beta)+O(\epsilon).
\end{aligned}
\label{eq:A.9}
\end{equation} The coefficients below are taken from ref.~\cite{ref7} and its ancillary file.
They are expressed in terms of Goncharov polylogarithms, defined by
\begin{equation}
\begin{gathered}
G(a_1,\ldots,a_w;x)=\int_0^x\frac{dt}{t-a_1}
                           G(a_2,\ldots,a_w;t),\\
G(;x)=1,\qquad
G(\underbrace{0,\ldots,0}_{w};x)=\frac{\ln^w x}{w!}.
\end{gathered}
\label{eq:A.10}
\end{equation} The recursive integral definition applies when at least one index is nonzero.
We use \(G_{a_1,\ldots,a_w}=G(a_1,\ldots,a_w;\beta)\), with letters \(a_i\in\{0,\pm1,\pm3\}\) and the real branch for \(0<\beta<1\).

\Needspace{8\baselineskip}
\subsubsection{Double-soft quark-pair coefficients}\label{quark-pair-emission}

The self-correlations are
\begin{align}
f_{AA,-2}^{q\bar q}={}&- \frac{1}{4},\notag\\[3pt]
f_{AA,-1}^{q\bar q}={}&- \frac{2 \ln2 - 3}{2}
 + \frac{1}{4 \beta}(- G_{-1} + G_{1}).
\label{eq:A.11}\\[3pt]
f_{AA,0}^{q\bar q}={}&8 \ln2 - \zeta_2 - 6
 + \frac{8 \ln2 - 11}{4 \beta}(- G_{-1} + G_{1})
 + \frac{1}{2 \beta}(G_{-1,-1} - G_{1,1})\notag\\
&\quad + G_{-1,1} + G_{1,-1}
 - \frac{2 \beta + 1}{2 \beta}G_{0,-1}
 - \frac{2 \beta - 1}{2 \beta}G_{0,1}.
\label{eq:A.12}
\end{align}

\Needspace{7\baselineskip}
The correlations between distinct emitters are
\begin{align}
f_{AB,-2}^{q\bar q}={}&- \frac{1}{12}
 + \frac{\beta^{2} + 1}{12 \beta}(- G_{-1} + G_{1}),\notag\\[3pt]
f_{AB,-1}^{q\bar q}={}&- \frac{12 \ln2 - 17}{36}
 + \frac{- 37 \beta^{2} + 24 \ln2 (\beta^{2} + 1) - 43}{72 \beta}(- G_{-1} + G_{1})\notag\\
&\quad + \frac{\beta^{2} + 1}{6 \beta}(G_{-1,-1} - G_{-1,1} + G_{1,-1} - G_{1,1})
 + \frac{\beta^{2} + 1}{3 \beta}(- G_{0,-1} + G_{0,1}).
\label{eq:A.13}\displaybreak[3]\\[3pt]
f_{AB,0}^{q\bar q}={}&- \frac{- 192 \ln2 + 18 \zeta_2 + 161}{54}
 + \frac{G_{-1}+G_1}{12}\notag\\*
&\quad + \frac{154 \beta^{2} - 24 \ln2 (10 \beta^{2} + 7) + 36 \zeta_2 (\beta^{2} + 1) + 115}{108 \beta}(G_1-G_{-1})\notag\\
&\quad + \frac{- 37 \beta^{2} + 24 \ln2 (\beta^{2} + 1) - 43}{36 \beta}(G_{-1,-1} - G_{1,1})\notag\\
&\quad + \frac{- 17 \beta^{2} + 12 \ln2 (\beta^{2} + 1) - 23}{18 \beta}(G_{1,-1}-G_{-1,1})\notag\\
&\quad + \frac{- 71 \beta^{2} + 48 \ln2 (\beta^{2} + 1) - 89}{36 \beta}(G_{0,1}-G_{0,-1})\notag\\
&\quad + \frac13(G_{-1,1}+G_{1,-1}-G_{0,-1}-G_{0,1})\notag\\
&\quad + \frac{\beta^{2} + 1}{3 \beta}\bigl[-G_{-1,-1,-1}+G_{-1,-1,1}+G_{-1,1,-1}+G_{-1,1,1}
 -G_{1,-1,-1}\notag\\*
&\qquad -G_{1,-1,1}-G_{1,1,-1}+G_{1,1,1}-2G_{-1,0,1}+2G_{0,-1,-1}\notag\\*
&\qquad -2G_{0,0,-1}+2G_{0,0,1}-2G_{0,1,1}+2G_{1,0,-1}\bigr].
\label{eq:A.14}
\end{align}

\Needspace{7\baselineskip}
\subsubsection{Double-soft gluon coefficients}\label{gluon-emission}

The self-correlations are
\begin{align}
f_{AA,-3}^{gg}={}&- \frac{1}{8},\notag\\[3pt]
f_{AA,-2}^{gg}={}&\frac{1}{4}
 + \frac{1}{4 \beta}(- G_{-1} + G_{1}),\notag\\[3pt]
f_{AA,-1}^{gg}={}&- \frac{4 \ln2 - 1}{2}
 + \frac{3}{4 \beta}(G_{-1} - G_{1})
 + \frac{\beta - 1}{4 \beta}(- G_{-1,-1} + G_{-1,1})\notag\\
&\quad + \frac{1}{2 \beta}(- G_{0,-1} + G_{0,1})
 + \frac{\beta + 1}{4 \beta}(G_{1,-1} - G_{1,1}).
\label{eq:A.15}\\[3pt]
f_{AA,0}^{gg}={}&12 \ln2 - 2 \zeta_2 - 3
 - \frac{8 \beta \ln2 + \beta + 2}{2 \beta^{2}}G_{-1}
 + \frac{8 \beta \ln2 + \beta - 2}{2 \beta^{2}}G_{1}\notag\\
&\quad + \frac{3}{2}(G_{-1,-1} + G_{1,1})
 + \frac{\beta + 3}{2 \beta}G_{-1,1}
 + \frac{\beta - 3}{2 \beta}G_{1,-1}\notag\\
&\quad - \frac{4 \beta - 3}{2 \beta}G_{0,-1}
 - \frac{4 \beta + 3}{2 \beta}G_{0,1}
 + \frac{3 \beta - 1}{4 \beta}(G_{-1,-1,-1} - G_{-1,-1,1})\notag\\
&\quad + \frac{\beta - 1}{2 \beta}(- G_{-1,0,-1} + G_{-1,0,1})
 + \frac{\beta + 1}{4 \beta}(- G_{-1,1,-1} + G_{-1,1,1})\notag\\
&\quad + \frac{2 \beta - 1}{2 \beta}(- G_{0,-1,-1} + G_{0,-1,1})
 + \frac{1}{\beta}(- G_{0,0,-1} + G_{0,0,1})\notag\\
&\quad + \frac{2 \beta + 1}{2 \beta}(G_{0,1,-1} - G_{0,1,1})
 + \frac{\beta - 1}{4 \beta}(G_{1,-1,-1} - G_{1,-1,1})\notag\\
&\quad + \frac{\beta + 1}{2 \beta}(G_{1,0,-1} - G_{1,0,1})
 + \frac{3 \beta + 1}{4 \beta}(- G_{1,1,-1} + G_{1,1,1}).
\label{eq:A.16}
\end{align}

\Needspace{7\baselineskip}
The correlations between distinct emitters are
\begin{align}
f_{AB,-3}^{gg}={}&\frac{3}{8}
 + \frac{\beta^{2} + 1}{4 \beta}(- G_{-1} + G_{1}),\notag\\[3pt]
f_{AB,-2}^{gg}={}&- \frac{4}{3}
 + \frac{13 \beta^{2} + 31}{24 \beta}(G_{-1} - G_{1})\notag\\
&\quad + \frac{(\beta^{2} + 1)^{2}}{4 \beta^{2}}(- G_{-1,-1} + G_{-1,1} + G_{1,-1} - G_{1,1})
 + \frac{\beta^{2} + 1}{2 \beta}(- G_{0,-1} + G_{0,1}),\notag\\[3pt]
f_{AB,-1}^{gg}={}&\frac{15 \ln2 - 13}{9}
 + \frac{- 62 \beta^{2} + 66 \ln2 (\beta^{2} + 1) + 25}{36 \beta}(- G_{-1} + G_{1})\notag\\
&\quad + \frac{12 \beta^{4} - \beta^{3} + 9 \beta^{2} + 2 \beta + 24}{12 \beta^{2}}(G_{-1,-1} - G_{-1,1})
 + \frac{4 \beta^{2} + 13}{6 \beta}(G_{0,-1} - G_{0,1})\notag\\
&\quad + \frac{12 \beta^{4} + \beta^{3} + 9 \beta^{2} - 2 \beta + 24}{12 \beta^{2}}(- G_{1,-1} + G_{1,1})\notag\\
&\quad + \frac{(\beta^{2} + 1) (\beta^{2} + \beta + 1)}{\beta^{2}}(G_{-1,-1,-1} - G_{-1,-1,1})\notag\\
&\quad + \frac{3 (\beta^{2} + 1)^{2}}{4 \beta^{2}}(- G_{-1,0,-1} + G_{-1,0,1} + G_{1,0,-1} - G_{1,0,1})\notag\\
&\quad + \frac{\beta^{2} + 1}{\beta}(- G_{-1,1,-1} + G_{-1,1,1} - G_{0,0,-1} + G_{0,0,1}
 - G_{1,-1,-1} + G_{1,-1,1})\notag\\
&\quad + \frac{(\beta^{2} + 1)^{2}}{\beta^{2}}(- G_{0,-1,-1} + G_{0,-1,1} + G_{0,1,-1} - G_{0,1,1})\notag\\
&\quad + \frac{(\beta^{2} + 1) (\beta^{2} - \beta + 1)}{\beta^{2}}(- G_{1,1,-1} + G_{1,1,1}).
\label{eq:A.17}\\[3pt]
f_{AB,0}^{gg}={}&\frac{- 1740 \ln2 + 180 \zeta_2 + 1787}{108}
 - \frac{11\beta^2+12}{12\beta^2}(G_{-1}+G_1)\notag\\*
&\quad + \frac{1769\beta^2-12\ln2(199\beta^2+259)+396\zeta_2(\beta^2+1)+2573}{216\beta}(G_1-G_{-1})\notag\\
&\quad + \frac{(\beta - 1) (\beta + 1)^{2} (\beta + 3)}{4 \beta^{2}}
 (-2G_{-3,-1} + G_{-3,1})\notag\\
&\quad - \frac{\beta^4-24\beta^2+8\ln2(3\beta^2-1)+29}{4\beta^2}(G_{-1,-1}+G_{1,1})\notag\\
&\quad + \frac{23\beta^2+6\ln2(\beta^2+4)-55}{9\beta}(G_{-1,-1}-G_{1,1})\notag\\
&\quad + \frac{3\beta^4-43\beta^2+12\ln2(3\beta^2-1)+39}{6\beta^2}(G_{-1,1}+G_{1,-1})\notag\\
&\quad + \frac{13\beta^2-24\ln2(\beta^2+4)+43}{36\beta}(G_{-1,1}-G_{1,-1})\notag\\
&\quad + \frac{-209\beta^2+192\ln2(\beta^2+1)-47}{36\beta}(G_{0,1}-G_{0,-1})
 + \frac53(G_{0,-1}+G_{0,1})\notag\\
&\quad + \frac{(\beta - 3) (\beta - 1)^{2} (\beta + 1)}{4 \beta^{2}}
 (G_{3,-1} - 2G_{3,1})\notag\\
&\quad + \frac{(\beta + 1)^{2} (2 \beta - 1)}{2 \beta^{2}}
 (-2G_{-1,-3,-1} + G_{-1,-3,1})\notag\\
&\quad + \frac{- 48 \beta^{4} + 2 \beta^{3} - 3 \beta^{2} - 73 \beta + 12 \ln2 (\beta + 1)^{2} (\beta^{2} + 1) - 108}{12 \beta^{2}}G_{-1,-1,-1}\notag\\
&\quad - \frac{- 48 \beta^{4} - 28 \beta^{3} - 75 \beta^{2} - 91 \beta + 12 \ln2 (\beta + 1)^{2} (\beta^{2} + 1) - 84}{12 \beta^{2}}G_{-1,-1,1}\notag\\
&\quad - \frac{- 6 \beta^{4} + 4 \beta^{3} - 11 \beta^{2} + 3 \beta + 2 \ln2 (\beta^{2} + 1)^{2} - 10}{2 \beta^{2}}G_{-1,0,-1}\notag\\
&\quad + \frac{- 18 \beta^{4} - 10 \beta^{3} - 33 \beta^{2} - 13 \beta + 6 \ln2 (\beta^{2} + 1)^{2} - 30}{6 \beta^{2}}G_{-1,0,1}\notag\\
&\quad + \frac{40 \beta^{3} - 75 \beta^{2} + 97 \beta + 12 \ln2 (\beta - 1)^{2} (\beta^{2} + 1) + 24}{12 \beta^{2}}G_{-1,1,-1}\notag\\
&\quad - \frac{8 \beta^{3} - 21 \beta^{2} + 77 \beta + 12 \ln2 (\beta - 1)^{2} (\beta^{2} + 1) + 6}{12 \beta^{2}}G_{-1,1,1}\notag\\
&\quad + \frac{(\beta - 1) (\beta + 1)}{2 \beta}
 (G_{-1,3,-1}-G_{1,-3,1}-2G_{-1,3,1}+2G_{1,-3,-1})\notag\\
&\quad + \frac{24 \beta^{4} + 4 \beta^{3} + 18 \beta^{2} + 7 \beta + 48}{6 \beta^{2}}G_{0,-1,-1}\notag\\
&\quad - \frac{8 \beta^{4} - 2 \beta^{3} + 6 \beta^{2} - \beta + 16}{2 \beta^{2}}G_{0,-1,1}
 + \frac{4 \beta^{2} + 13}{3 \beta}(G_{0,0,-1} - G_{0,0,1})\notag\\
&\quad - \frac{8 \beta^{4} + 2 \beta^{3} + 6 \beta^{2} + \beta + 16}{2 \beta^{2}}G_{0,1,-1}\notag\\
&\quad + \frac{24 \beta^{4} - 4 \beta^{3} + 18 \beta^{2} - 7 \beta + 48}{6 \beta^{2}}G_{0,1,1}\notag\\
&\quad - \frac{- 8 \beta^{3} - 21 \beta^{2} - 77 \beta + 12 \ln2 (\beta + 1)^{2} (\beta^{2} + 1) + 6}{12 \beta^{2}}G_{1,-1,-1}\notag\\
&\quad + \frac{- 40 \beta^{3} - 75 \beta^{2} - 97 \beta + 12 \ln2 (\beta + 1)^{2} (\beta^{2} + 1) + 24}{12 \beta^{2}}G_{1,-1,1}\notag\\
&\quad + \frac{- 18 \beta^{4} + 10 \beta^{3} - 33 \beta^{2} + 13 \beta + 6 \ln2 (\beta^{2} + 1)^{2} - 30}{6 \beta^{2}}G_{1,0,-1}\notag\\
&\quad - \frac{- 6 \beta^{4} - 4 \beta^{3} - 11 \beta^{2} - 3 \beta + 2 \ln2 (\beta^{2} + 1)^{2} - 10}{2 \beta^{2}}G_{1,0,1}\notag\\
&\quad - \frac{- 48 \beta^{4} + 28 \beta^{3} - 75 \beta^{2} + 91 \beta + 12 \ln2 (\beta - 1)^{2} (\beta^{2} + 1) - 84}{12 \beta^{2}}G_{1,1,-1}\notag\\
&\quad + \frac{- 48 \beta^{4} - 2 \beta^{3} - 3 \beta^{2} + 73 \beta + 12 \ln2 (\beta - 1)^{2} (\beta^{2} + 1) - 108}{12 \beta^{2}}G_{1,1,1}\notag\\
&\quad + \frac{(\beta - 1)^{2} (2 \beta + 1)}{2 \beta^{2}}
 (-G_{1,3,-1} + 2G_{1,3,1})\notag\\
&\quad + \frac{\beta^2+1}{2\beta^2}\biggl[
 \frac{(\beta+1)^2}{2}(2G_{-1,-1,-3,-1}-2G_{1,-1,-3,-1}-G_{-1,-1,-3,1}+G_{1,-1,-3,1})\notag\\*
&\qquad -(7\beta^2+10\beta+7)G_{-1,-1,-1,-1}
 +(5\beta^2+6\beta+5)G_{-1,-1,-1,1}\notag\\
&\qquad +(5\beta^2+4\beta+5)(G_{-1,-1,0,-1}-G_{-1,-1,0,1})\notag\\
&\qquad +4\beta\bigl(G_{-1,-1,1,-1}+G_{-1,1,-1,1}-G_{0,-1,1,-1}+G_{0,-1,1,1}
 -G_{0,0,0,-1}\notag\\*
&\qquad\quad +G_{0,0,0,1}-G_{0,1,-1,-1}+G_{0,1,-1,1}-G_{1,-1,1,-1}-G_{1,1,-1,1}\bigr)\notag\\
&\qquad +\frac{3\beta^2-2\beta+3}{2}(G_{-1,-1,1,1}-G_{1,-1,1,1})\notag\\
&\qquad +(7\beta^2+8\beta+7)G_{-1,0,-1,-1}-(5\beta^2+8\beta+5)G_{-1,0,-1,1}\notag\\
&\qquad +3(\beta^2+1)(-G_{-1,0,0,-1}+G_{-1,0,0,1}+G_{1,0,0,-1}-G_{1,0,0,1})\notag\\
&\qquad -(3\beta^2+8\beta+3)G_{-1,0,1,-1}+(\beta^2+8\beta+1)G_{-1,0,1,1}\notag\\
&\qquad +\frac{3\beta^2+2\beta+3}{2}(-G_{-1,1,-1,-1}+G_{1,1,-1,-1})\notag\\
&\qquad +(\beta^2+4\beta+1)(-G_{-1,1,0,-1}+G_{-1,1,0,1})
 +3(\beta+1)^2G_{-1,1,1,-1}\notag\\
&\qquad -(\beta^2+10\beta+1)G_{-1,1,1,1}
 +\frac{(\beta-1)^2}{2}(G_{-1,1,3,-1}-G_{1,1,3,-1})\notag\\
&\qquad +(\beta-1)^2(-G_{-1,1,3,1}+G_{1,1,3,1})\notag\\
&\qquad +4(2\beta^2+\beta+2)(G_{0,-1,-1,-1}-G_{0,-1,-1,1})\notag\\
&\qquad +6(\beta^2+1)(-G_{0,-1,0,-1}+G_{0,-1,0,1}+G_{0,1,0,-1}-G_{0,1,0,1})\notag\\
&\qquad +8(\beta^2+1)(-G_{0,0,-1,-1}+G_{0,0,-1,1}+G_{0,0,1,-1}-G_{0,0,1,1})\notag\\
&\qquad +4(2\beta^2-\beta+2)(-G_{0,1,1,-1}+G_{0,1,1,1})\notag\\
&\qquad -(\beta^2-10\beta+1)G_{1,-1,-1,-1}+3(\beta-1)^2G_{1,-1,-1,1}\notag\\
&\qquad +(\beta^2-4\beta+1)(G_{1,-1,0,-1}-G_{1,-1,0,1})\notag\\
&\qquad +(\beta^2-8\beta+1)G_{1,0,-1,-1}-(3\beta^2-8\beta+3)G_{1,0,-1,1}\notag\\
&\qquad -(5\beta^2-8\beta+5)G_{1,0,1,-1}+(7\beta^2-8\beta+7)G_{1,0,1,1}\notag\\
&\qquad +(5\beta^2-4\beta+5)(-G_{1,1,0,-1}+G_{1,1,0,1})\notag\\*
&\qquad +(5\beta^2-6\beta+5)G_{1,1,1,-1}-(7\beta^2-10\beta+7)G_{1,1,1,1}\biggr].
\label{eq:A.18}
\end{align}

\Needspace{5\baselineskip}
\subsection{Eikonal integrals for single-soft subtraction}\label{single-soft-insertion-with-a-surviving-gluon}

The kernel \(\mathcal K_{\Sop_4}\) in \eqref{eq:3.25} is expressed in terms of the eikonal integrals~\cite{ref6,alioli_powheg}
\begin{equation}
\begin{aligned}
\mathcal J(a,b;E_3^*)
&=\mathcal C_\epsilon\int[dk]_K\,\theta(E_3^*-u_K\cdot k)
               \frac{a\cdot b}{(a\cdot k)(b\cdot k)}\\
&=-\frac1{(4\pi)^2}\biggl[
\frac{I_{ab}^{(-1)}}{\epsilon^2}
+\frac{I_{ab}^{(0)}+L_3 I_{ab}^{(-1)}}{\epsilon}\\
&\qquad+I_{ab}^{(1)}+L_3 I_{ab}^{(0)}
+\left(\frac{L_3^2}{2}-\frac32\zeta_2\right)I_{ab}^{(-1)}
\biggr]+O(\epsilon),
\end{aligned}
\label{eq:A.19}
\end{equation}
where \(L_3=\ln[\mu_R^2/(4(E_3^*)^2)]\).

\Needspace{9\baselineskip}
For two distinct massive emitters, \(a^2=b^2=m_c^2\), let
\(\boldsymbol u^*=\boldsymbol a^*/(u_K\cdot a)\) and
\(\boldsymbol w^*=\boldsymbol b^*/(u_K\cdot b)\) be their velocities in the \(K\) rest frame.
With \(v_{ab}=\sqrt{1-m_c^4/(a\cdot b)^2}\), the coefficients are
\begin{equation}
\begin{aligned}
I_{ab}^{(-1)}&=0,\qquad
I_{ab}^{(0)}=\frac1{v_{ab}}\ln\frac{1+v_{ab}}{1-v_{ab}},\\
I_{ab}^{(1)}&=\frac{1-\boldsymbol u^*\cdot\boldsymbol w^*}
                  {\sqrt{\Delta_{ab}^2-\rho_{ab}^2}}\,[\mathcal K(z_2)-\mathcal K(z_1)].
\end{aligned}
\label{eq:A.22}
\end{equation}
The auxiliary variables in \(I_{ab}^{(1)}\) are
\begin{equation}
\begin{aligned}
\Delta_{ab}&=|\boldsymbol u^*-\boldsymbol w^*|,&
\rho_{ab}^2&=(\boldsymbol u^*)^2(\boldsymbol w^*)^2-(\boldsymbol u^*\cdot\boldsymbol w^*)^2,\\
X_1&=(\boldsymbol u^*)^2-\boldsymbol u^*\cdot\boldsymbol w^*,&
X_2&=(\boldsymbol w^*)^2-\boldsymbol u^*\cdot\boldsymbol w^*,\\
z_1&=\sqrt{X_1^2+\rho_{ab}^2}-X_1,&
z_2&=\sqrt{X_2^2+\rho_{ab}^2}+X_2.
\end{aligned}
\label{eq:A.20}
\end{equation}
With \(z_\pm=\Delta_{ab}\pm\sqrt{\Delta_{ab}^2-\rho_{ab}^2}\), the function \(\mathcal K\) is
\begin{equation}
\begin{aligned}
\mathcal K(y)={}&
-2\operatorname{Li}_2\!\left(
 \frac{2z_-(z_+-y)}{(z_+-z_-)(z_-+y)}\right)
-2\operatorname{Li}_2\!\left(
 \frac{-2z_+(z_-+y)}{(z_+-z_-)(z_+-y)}\right)\\
&-\frac12\ln^2\!\left[
 \frac{(y-z_-)(z_+-y)}{(y+z_-)(z_++y)}\right].
\end{aligned}
\label{eq:A.21}
\end{equation}
For the Born momenta \(p_1+p_2=K\), with \(\beta=\sqrt{1-4m_c^2/K^2}\), one obtains
\[
I_{p_1p_2}^{(1)}
 =\frac{1+\beta^2}{\beta}\left[
 \operatorname{Li}_2\!\left(\frac{2\beta}{1+\beta}\right)
 -\operatorname{Li}_2\!\left(-\frac{2\beta}{1-\beta}\right)\right].
\]

For \(a=b=p_i\), the coefficients depend only on the speed
\(\kappa_i=\sqrt{1-m_c^2/(u_K\cdot p_i)^2}\):
\begin{equation}
I_{p_ip_i}^{(-1)}=0,\qquad
I_{p_ip_i}^{(0)}=2,\qquad
I_{p_ip_i}^{(1)}=-\frac2{\kappa_i}
                           \ln\frac{1-\kappa_i}{1+\kappa_i}.
\label{eq:A.23}
\end{equation}

For a massive--massless pair \((p_i,k_3)\), define
\(\xi_i=(p_i\cdot k_3)/[(u_K\cdot p_i)(u_K\cdot k_3)]\).
The coefficients are
\begin{equation}
\begin{aligned}
I_{p_ik_3}^{(-1)}&=-1,\qquad
I_{p_ik_3}^{(0)}=\ln\frac{\xi_i^2}{1-\kappa_i^2},\\
I_{p_ik_3}^{(1)}&=-2\biggl[
\frac14\ln^2\frac{1-\kappa_i}{1+\kappa_i}
+\ln\frac{\xi_i}{1+\kappa_i}\ln\frac{\xi_i}{1-\kappa_i}\\
&\qquad
+\operatorname{Li}_2\!\left(1-\frac{\xi_i}{1+\kappa_i}\right)
+\operatorname{Li}_2\!\left(1-\frac{\xi_i}{1-\kappa_i}\right)\biggr].
\end{aligned}
\label{eq:A.24}
\end{equation}

\Needspace{9\baselineskip}
\subsection{Local one-loop soft-current coefficients}\label{one-loop-soft-current-coefficients}

For two outgoing massive emitters, we define
\begin{equation}
\begin{gathered}
\alpha_1=\frac{m_c^2p_2\cdot k}
                 {2(p_1\cdot p_2)(p_1\cdot k)},\qquad
\alpha_2=\frac{m_c^2p_1\cdot k}
                 {2(p_1\cdot p_2)(p_2\cdot k)},\\
d_i=1-2\alpha_i,\qquad
v=\sqrt{1-4\alpha_1\alpha_2}=v_{12},\qquad
v_\pm=\frac{1\pm v}{2},\qquad x=\frac{v_-}{v_+}.
\end{gathered}
\label{eq:A.25}
\end{equation}
At Born kinematics, \(v=2\beta/(1+\beta^2)\).
We use the logarithms
\begin{equation}
L_{\alpha_i}=\ln\frac{\alpha_i}{v_+},\qquad
\bar L_{\alpha_i}=\ln\!\left(1-\frac{\alpha_i}{v_+}\right),\qquad
L_+=\ln v_+,\qquad L_v=\ln v.
\label{eq:A.26}
\end{equation}

The coefficients of the poles in \eqref{eq:3.37} are~\cite{ref9}
\begin{equation}
R_{-2}=-\frac12,\qquad
R_{-1}=-\frac12\left[L_+-\frac{v_-}{v}
                              (L_{\alpha_1}+L_{\alpha_2})\right].
\label{eq:A.27}
\end{equation}
The finite and \(O(\epsilon)\) coefficients are
\begin{equation}
\begin{aligned}
R_0
={}&
-\frac12
\Bigg\{
\frac{1}{v}
\bigg[
\frac{
(\alpha_1v_+-\alpha_2v_-)L_{\alpha_1}^2
+(\alpha_2v_+-\alpha_1v_-)L_{\alpha_2}^2
}{d_1+d_2}
\\
&\qquad
+(L_{\alpha_1}+L_{\alpha_2})(v_+L_+-L_v)
-\operatorname{Li}_2(x)
\bigg]
\\
&\quad 
+\frac12L_+^2
+\zeta_2\left(\frac7v-\frac{19}{2}\right)
\Bigg\}.
\end{aligned}
\label{eq:A.28}
\end{equation}
\begin{align}
R_1={}&-\frac12\Biggl\{
\frac{1}{d_1+d_2}\Biggl[
(1-d_1-d_2)\biggl(
\bar L_{\alpha_1}L_{\alpha_1}^2+\bar L_{\alpha_2}L_{\alpha_2}^2
\notag\\
&\qquad\quad
+2L_{\alpha_1}\operatorname{Li}_2\!\left(\frac{\alpha_1}{v_+}\right)
+2L_{\alpha_2}\operatorname{Li}_2\!\left(\frac{\alpha_2}{v_+}\right)
-\operatorname{Li}_2(x)(L_{\alpha_1}+L_{\alpha_2})
\notag\\
&\qquad\quad
+2\operatorname{Li}_3(x)
-2\operatorname{Li}_3\!\left(\frac{\alpha_1}{v_+}\right)
-2\operatorname{Li}_3\!\left(\frac{\alpha_2}{v_+}\right)
+2\zeta_3\biggr)
-7\zeta_2(L_{\alpha_1}+L_{\alpha_2})
\notag\\
&\qquad
+\frac1v\biggl[
\bigl((\alpha_2v_+-\alpha_1v_-)L_{\alpha_1}^2
+(\alpha_1v_+-\alpha_2v_-)L_{\alpha_2}^2\bigr)L_+
\notag\\
&\qquad\quad
+(\alpha_1-\alpha_2)(L_{\alpha_1}^2-L_{\alpha_2}^2)L_v
-(d_1L_{\alpha_1}+d_2L_{\alpha_2})
\bigl(\operatorname{Li}_2(x)-7\zeta_2\bigr)\biggr]\Biggr]
\notag\\
&\quad
+\frac1v\Biggl[
\biggl(L_+\left(\frac{3+v}{4}L_+-L_v\right)
-\frac{9v_-}{2}\zeta_2\biggr)(L_{\alpha_1}+L_{\alpha_2})
-\frac{v_-}{6}(L_{\alpha_1}^3+L_{\alpha_2}^3)
\notag\\
&\qquad
+2\operatorname{Li}_3(1-x)+\operatorname{Li}_3(x)
-\biggl(\operatorname{Li}_2(x)+\zeta_2\left(5+\frac{19}{2}v\right)\biggr)L_+
+12\zeta_2L_v\Biggr]
\notag\\*
&\quad
+\frac16L_+^3-\left(\frac73+\frac1v\right)\zeta_3\Biggr\}.
\label{eq:A.29}
\end{align}

\Needspace{11\baselineskip}
\subsection{Integrated one-loop soft-current coefficients}\label{integrated-one-loop-soft-current-coefficients}

The integrated coefficients \(\mathcal R_n\) in \eqref{eq:3.39} depend on the heavy-quark speed at Born kinematics,
\(\beta=\sqrt{1-4m_c^2/K^2}\).
We write \(H_{a_1,\ldots,a_w}=H_{a_1,\ldots,a_w}(\beta)\) for the harmonic polylogarithms defined by
\begin{equation}
\begin{gathered}
H_\emptyset(t)=1,\qquad
H_{\underbrace{0,\ldots,0}_{w}}(t)=\frac{\ln^w t}{w!},\\
H_{a,\vec b}(t)=\int_0^t du\,f_a(u)H_{\vec b}(u),
\qquad (a,\vec b)\ne(0,\ldots,0),\\
f_{-1}(u)=\frac1{1+u},\qquad f_0(u)=\frac1u,\qquad
f_1(u)=\frac1{1-u}.
\end{gathered}
\label{eq:A.30}
\end{equation}
For \(a_j\in\{0,\pm1\}\), \(H_{a_1,\ldots,a_w}(\beta)=(-1)^{N_+}G(a_1,\ldots,a_w;\beta)\),
where \(N_+\) counts the entries equal to \(+1\).

The coefficients are \cite{ref6}
\begin{equation}
\begin{aligned}
\mathcal R_{-3}(\beta)={}&-1+\frac{1+\beta^2}{2\beta}(H_{-1}+H_1),
\end{aligned}
\label{eq:A.31}
\end{equation}
\begin{equation}
\begin{aligned}
\mathcal R_{-2}(\beta)={}&1
+\frac{(1+\beta^2)^2}{4\beta^2}(H_{-1}+H_1)^2
-\frac2\beta(H_{-1}+H_1)\\
&+\frac{1+\beta^2}{\beta}(H_{0,-1}+H_{0,1}),
\end{aligned}
\label{eq:A.32}
\end{equation}
\begin{align}
\mathcal R_{-1}(\beta)={}&
\left(11-\frac{3(1+\beta^2)}{\beta}\right)\zeta_2
-\frac{(1+\beta^2)(3\beta^2+2\beta+3)}{12\beta^2}H_{-1}^3\notag\\
&+\frac{(1+\beta^2)(3\beta^2-2\beta+3)}{12\beta^2}
 (H_1^3+3H_{-1}H_1^2)
-\frac{2+\beta+\beta^3}{2\beta^2}(H_1^2+2H_{-1}H_1)\notag\\
&+H_{-1}^2\biggl[
\frac{-2+\beta+\beta^3}{2\beta^2}
+\frac{(\beta-1)^2(1+\beta^2)}{4\beta^2}H_1\biggr]
-\frac4\beta(H_{0,-1}+H_{0,1})\notag\\
&+(H_{-1}+H_1)\biggl[
\frac2\beta+\frac{(1+\beta^2)(3\beta^2-11\beta+3)}{2\beta^2}\zeta_2\notag\\*
&\qquad\qquad
+\frac{(1+\beta^2)^2}{\beta^2}(H_{0,-1}+H_{0,1}-H_{-1,1})\biggr]\notag\\*
&+\frac{2(1+\beta^2)}{\beta}(H_{-1,1}+H_{0,0,-1}+H_{0,0,1}).
\label{eq:A.33}
\end{align}
\begin{align}
\mathcal R_0(\beta)={}&-4
+\biggl[\frac{6(1+\beta^2)}{\beta}(1-2\ln2)-11\biggr]\zeta_2
+\frac{14}{3}\zeta_3
\notag\\
&+\frac{1+\beta^2}{12\beta^2}
 \bigl[(3\beta^2+4\beta+3)H_{-1}^4+(3\beta^2-4\beta+3)H_1^4\bigr]
\notag\\
&+\left(\frac{13}{6}-\frac{3}{2\beta^2}-2\beta\right)H_1^3
+H_{-1}^3\biggl[-\frac{13}{6}+\frac{3}{2\beta^2}-2\beta
+\frac{4(1+\beta^2)}{3\beta}H_1\biggr]
\notag\\
&+\frac{3(1+\beta^2)\zeta_2}{\beta^2}H_0
 \bigl[(1+\beta^2)(H_{-1}+H_1)-2\beta\bigr]
+\frac{(1+\beta^2)^2}{\beta^2}(H_{0,-1}+H_{0,1})^2
\notag\\
&+\biggl[\frac4\beta
 +\frac{(1+\beta^2)(3\beta^2-11\beta+3)}{\beta^2}\zeta_2
 -\frac{2(1+\beta^2)^2}{\beta^2}H_{-1,1}\biggr](H_{0,-1}+H_{0,1})
\notag\\
&+H_1^2\biggl[-1+\frac2{\beta^2}+\beta
 +\frac{(1+\beta^2)(\beta^2-12\beta+1)}{4\beta^2}\zeta_2
\notag\\*
&\qquad
 -\frac{2(1+\beta^2)(\beta^2-\beta+1)}{\beta^2}H_{-1,1}
 +\frac{(1+\beta^2)(3\beta^2-2\beta+3)}{2\beta^2}(H_{0,-1}+H_{0,1})\biggr]
\notag\\
&+H_{-1}^2\biggl[-1+\frac2{\beta^2}-\beta
 -\frac{(1+\beta^2)(23\beta^2+12\beta+23)}{4\beta^2}\zeta_2
\notag\\*
&\qquad
 +\left(\frac72-\frac{3}{2\beta^2}+\frac2\beta-4\beta\right)H_1
 +\frac{(\beta-1)^2(1+\beta^2)}{2\beta^2}H_1^2
 -\frac{6(1+\beta^2)}{\beta}H_{-1,1}
\notag\\*
&\qquad
 -\frac{(1+\beta^2)(3\beta^2+2\beta+3)}{2\beta^2}(H_{0,-1}+H_{0,1})\biggr]
\notag\\
&-4\beta H_{-1,1}
 -\frac4\beta\bigl[(\beta-1)^2H_{-1,-1,1}+(\beta+1)^2H_{-1,1,1}\bigr]
\notag\\
&+\left(8-\frac{2(1+\beta^2)}{\beta}\right)(H_{0,-1,-1}+H_{0,1,-1})
 -\frac8\beta(H_{0,0,-1}+H_{0,0,1})
\notag\\
&+2\left(4+\frac{1+\beta^2}{\beta}\right)(H_{0,-1,1}+H_{0,1,1})
\notag\\
&+(H_{-1}+H_1)\biggl[
 \left(6-\frac6{\beta^2}+\frac{6(1+\beta^2)^2\ln2}{\beta^2}\right)\zeta_2
 -\frac{7(1+\beta^2)}{3\beta}\zeta_3
\notag\\*
&\qquad
 +\frac{(1+\beta^2)(\beta^2-4\beta+1)}{\beta^2}(H_{0,-1,-1}+H_{0,1,-1})
 +\frac{2(1+\beta^2)^2}{\beta^2}(H_{0,0,-1}+H_{0,0,1})
\notag\\*
&\qquad
 -\frac{(1+\beta^2)(\beta^2+4\beta+1)}{\beta^2}(H_{0,-1,1}+H_{0,1,1})\biggr]
\notag\\
&+\frac{2H_1}{\beta^2}\biggl[
 \beta(8-3\beta^2)\zeta_2
 +(2\beta^3-3\beta^2+2\beta+3)H_{-1,1}
 -(2+\beta+\beta^3)(H_{0,-1}+H_{0,1})
\notag\\*
&\qquad
 +(1+\beta^2)
 \bigl[(\beta-1)^2H_{-1,-1,1}+(\beta+1)^2H_{-1,1,1}\bigr]\biggr]
\notag\\
&+H_{-1}\biggl[
 \left(\frac{28}{\beta}+6\beta\right)\zeta_2
 +\left(\frac{13}{2}-\frac9{2\beta^2}-6\beta\right)H_1^2
 +\frac{(1+\beta^2)(3\beta^2-4\beta+3)}{3\beta^2}H_1^3
\notag\\*
&\qquad
 +\left(-14+\frac6{\beta^2}\right)H_{-1,1}
 +\frac{2(-2+\beta+\beta^3)}{\beta^2}(H_{0,-1}+H_{0,1})
\notag\\*
&\qquad
 +\frac{2(1+\beta^2)}{\beta^2}
 \bigl[(\beta^2+10\beta+1)H_{-1,-1,1}
       +(\beta^2-10\beta+1)H_{-1,1,1}\bigr]\biggr]
\notag\\
&+H_{-1}H_1\biggl[
 -2+\frac4{\beta^2}+2\beta
 -\frac{(1+\beta^2)(11\beta^2+12\beta+11)}{2\beta^2}\zeta_2
\notag\\*
&\qquad
 -\frac{2(1+\beta^2)(\beta^2-4\beta+1)}{\beta^2}H_{-1,1}
 +\frac{(\beta-1)^2(1+\beta^2)}{\beta^2}(H_{0,-1}+H_{0,1})\biggr]
\notag\\
&+\frac{24(1+\beta^2)}{\beta}
 \bigl(-H_{-1,-1,-1,1}+H_{-1,-1,1,1}-H_{-1,1,1,1}\bigr)
\notag\\
&+\frac{16(1+\beta^2)}{\beta}
 \bigl(H_{0,-1,-1,-1}+H_{0,-1,-1,1}+H_{0,-1,1,-1}+H_{0,-1,1,1}
\notag\\*
&\qquad\qquad
 +H_{0,1,-1,-1}+H_{0,1,-1,1}+H_{0,1,1,-1}+H_{0,1,1,1}\bigr)
\notag\\*
&+\frac{4(1+\beta^2)}{\beta}(H_{0,0,0,-1}+H_{0,0,0,1}).
\label{eq:A.34}
\end{align}

\Needspace{7\baselineskip}
\subsection{Two-loop massive cusp anomalous dimension}\label{two-loop-massive-cusp-anomalous-dimension}

For the outgoing heavy-quark pair with relative velocity \(v=v_{12}\), let
\(\chi=\tfrac12\ln[(1+v)/(1-v)]\).
The time-like two-loop cusp coefficient entering \(Z_2\) in \eqref{eq:2.6} is~\cite{becher_neubert,ferroglia_massive,ref6}
\begin{align}
\gamma_{{\rm cusp},Q}^{(1)}(v)
={}&\frac{\gamma_{\rm cusp}^{(1)}}{v}(\chi-i\pi)
 +8C_A\Biggl\{\zeta_3-\frac{5\pi^2}{6}+\chi^2\notag\\*
&+\frac1{v^2}\left[
  \frac{\chi^3}{3}+\chi\operatorname{Li}_2(e^{-2\chi})
  -\frac{5\pi^2\chi}{6}+\operatorname{Li}_3(e^{-2\chi})-\zeta_3
 \right]\notag\\
&+\frac1v\left[
  \frac{5\pi^2}{6}(1+\chi)-\chi^2
  -2\chi\ln(1-e^{-2\chi})-\frac{\chi^3}{3}
  +\operatorname{Li}_2(e^{-2\chi})\right]\notag\\
&+i\pi\Biggl[\frac1{v^2}\left(
  \frac{\pi^2}{6}-\chi^2-\operatorname{Li}_2(e^{-2\chi})\right)
  -2\chi\notag\\*
&\qquad\quad+\frac1v\left(
  -\frac{\pi^2}{6}+2\ln(1-e^{-2\chi})+2\chi+\chi^2
  \right)\Biggr]\Biggr\},
\label{eq:A.35}
\end{align}
where the massless cusp coefficient is
\[
\gamma_{\rm cusp}^{(1)}
=\left(\frac{268}{9}-\frac{4\pi^2}{3}\right)C_A
 -\frac{80}{9}T_Fn_l,\qquad n_l=3.
\]

\FloatBarrier

\end{document}